\documentclass[sigconf,10pt,screen]{acmart}

\usepackage{amsmath}

\usepackage{algorithm}  
\usepackage{algorithmic} 

\usepackage{array}      
\usepackage{booktabs}   
\usepackage{enumerate}  

\usepackage{float}      
\usepackage[T1]{fontenc} 
\usepackage{graphicx}

\usepackage[utf8]{inputenc}   
\usepackage{listings}   
\usepackage{longtable}  
\usepackage{lscape}     

\usepackage{mathtools}  
\usepackage{microtype}  
\usepackage{multicol}   
\usepackage{multirow}   
\usepackage{orcidlink}  
\usepackage{siunitx}    
\usepackage{subcaption} 
\usepackage{placeins}   
\usepackage{tabularx}   
\usepackage{textcomp}   

\DeclareRobustCommand{\earlog}{\mbox{\textit{EarLog}}}
\makeatletter
\renewcommand{\orcid}[1]{\unskip\ignorespaces\expandafter\gdef\csname
    typeset@author\the\num@authors\endcsname##1{##1\,\orcidlink{#1}}}
\makeatother

\newif\ifEarLogSharedAffiliationLayout
\EarLogSharedAffiliationLayouttrue
\ifEarLogSharedAffiliationLayout
\makeatletter
\def\@mkauthors@iii{\author@bx@wd=\textwidth\relax
  \advance\author@bx@wd by -\author@bx@sep\relax
  \ifnum\@ACM@authorsperrow>0\relax
    \divide\author@bx@wd by \@ACM@authorsperrow\relax
  \else
    \ifcase\num@authorgroups
    \relax
    \or
    \or \divide\author@bx@wd by \num@authorgroups\relax
    \or \divide\author@bx@wd by \num@authorgroups\relax
    \or \divide\author@bx@wd by 2\relax
    \else \divide\author@bx@wd by 3\relax
    \fi
  \fi
  \advance\author@bx@wd by -\author@bx@sep\relax
  \gdef\@currentauthors{}\gdef\@currentaffiliation{}\gdef\@sharedtitleaffiliation{}\def\@author##1{\ifx\@currentauthors\@empty
    \gdef\@currentauthors{\par##1}\else
    \g@addto@macro\@currentauthors{\par##1}\fi
  \gdef\and{}}\def\email##1##2{\ifx\@currentaffiliation\@empty
    \gdef\@currentaffiliation{\bgroup\normalsize
      \urlstyle{same}\nolinkurl{##2}\egroup}\else
    \g@addto@macro\@currentaffiliation{\par\bgroup\normalsize
      \urlstyle{same}\nolinkurl{##2}\egroup}\fi}\def\affiliation##1##2{\ifx\@sharedtitleaffiliation\@empty
      \gdef\@sharedtitleaffiliation{\setkeys{@ACM@affiliation@}{obeypunctuation=false}\setkeys{@ACM@affiliation@}{##1}\@ACM@resetaffil
        ##2\@ACM@checkaffil}\fi
    \global\let\and\@typeset@author@bx
  }\hsize=\textwidth
  \global\setbox\mktitle@bx=\vbox{\noindent
    \unvbox\mktitle@bx\par\medskip\leavevmode
    \lineskip=1pc\relax\centering\hspace*{-1em}\addresses\let\and\@typeset@author@bx\and\par
    \ifx\@sharedtitleaffiliation\@empty\else
      \medskip
      {\centering\@affiliationfont\@sharedtitleaffiliation\par}\fi
    \bigskip}}
\makeatother
\fi

\makeatletter
\newcommand{\EarLogTableCaption}[2]{\begingroup
    \let\caption@type@warning\relax
    \captionof{table}{#1}\phantomsection\label{#2}\endgroup}
\makeatother

\newlength{\EarLogSavedEmergencyStretch}

\usepackage{xurl}

\definecolor{NIPSTeal}{HTML}{00FF00}  
\definecolor{NIPSRed}{HTML}{FF0000}   
\definecolor{NIPSUrl}{HTML}{0B63B6}   

\AtBeginDocument{\hypersetup{
    linkcolor=NIPSRed,    
    citecolor=NIPSTeal,   
    urlcolor=NIPSUrl,     
    breaklinks=true
  }\citestyle{acmnumeric}  
  \setcitestyle{numbers, square, comma, sort&compress}  
}

\acmYear{2026}
\copyrightyear{2026}
\setcopyright{cc}
\setcctype[4.0]{by}
\acmConference[MobiCom '26]{The 32nd Annual International Conference on Mobile Computing and Networking}{October 26--30, 2026}{Austin, TX, USA}
\acmBooktitle{The 32nd Annual International Conference on Mobile Computing and Networking (MobiCom '26), October 26--30, 2026, Austin, TX, USA}
\acmDOI{10.1145/3795866.3844456}
\acmISBN{979-8-4007-2505-0/26/10}

\apptocmd{\bibfont}{\baselineskip=10pt plus 1pt\relax}{}{}
\AtBeginEnvironment{thebibliography}{\setlength{\emergencystretch}{2em}}

\begin{document}

\title{\earlog{}: Opportunistic Ear Physiology Sensing toward Personalized Hearing Risk Awareness}

\author{Yongzhi Huang}
\orcid{0000-0003-3728-3502}
\email{huangyongzhi@email.szu.edu.cn}
\affiliation{\department{Data Science and Analytics Thrust (DSA), Information Hub (INF)}
  \institution{The Hong Kong University of Science and Technology (Guangzhou)}
  \city{Guangzhou}
  \country{China}
}

\author{Jiayi Zhao}
\orcid{0009-0001-8121-0063}
\email{jzhao238@connect.hkust-gz.edu.cn}
\affiliation{\department{Data Science and Analytics Thrust (DSA), Information Hub (INF)}
  \institution{The Hong Kong University of Science and Technology (Guangzhou)}
  \city{Guangzhou}
  \country{China}
}

\author{Kaishun Wu}
\orcid{0000-0003-2216-0737}
\correspondingauthor
\email{wuks@hkust-gz.edu.cn}

\affiliation{\department{Data Science and Analytics Thrust (DSA), Information Hub (INF)}
  \institution{The Hong Kong University of Science and Technology (Guangzhou)}
  \city{Guangzhou}
  \country{China}
}

\renewcommand{\shortauthors}{Huang et al.}

\begin{abstract}
Many headphone users risk irreversible hearing damage during high-load listening, yet current protection still relies on sound-pressure limits. Clinical tests detect damage after it becomes measurable, leaving the risk-accumulating process unobserved. The acoustic reflex (AR), an involuntary response to high acoustic load, marks when the ear enters a protective state. AR traces are weak, and existing measurements require dedicated probes and controlled conditions that are difficult to maintain during everyday audio.
We present \earlog{}, which turns playback, the dominant interference, into an opportunistic probe. \earlog{} introduces content-conditioned residual sensing to suppress variations caused by nonstationary content, individual ear acoustics, and device response. It identifies weak AR through loudness-dependent changes and localized spectral irregularities in the residual. A resonance-constrained sparse finite impulse response (FIR) model captures the short-window response. Dual-microphone observations, mild-distortion recovery, and cross-session reference updates keep detection usable under low signal-to-noise ratio (SNR), fit changes, and environmental drift.
\earlog{} runs in real time on an embedded controller using a low-cost dual-microphone add-on. We evaluate it with 81 participants, 16 headphone models, and more than 400 hours of recordings. \earlog{} achieves 99.0\% sensitivity and 98.7\% specificity on a clinically referenced benchmark. \earlog{} records listener-specific AR events during everyday listening, providing data for future longitudinal studies of hearing risk.

\end{abstract}




\baselineskip=12pt plus 1pt\relax

\setlength{\EarLogSavedEmergencyStretch}{\emergencystretch}
\setlength{\emergencystretch}{1em}
\maketitle
\setlength{\emergencystretch}{\EarLogSavedEmergencyStretch}

\section{Introduction}

\begin{figure}[t]
    \centering
    \includegraphics[width=\linewidth]{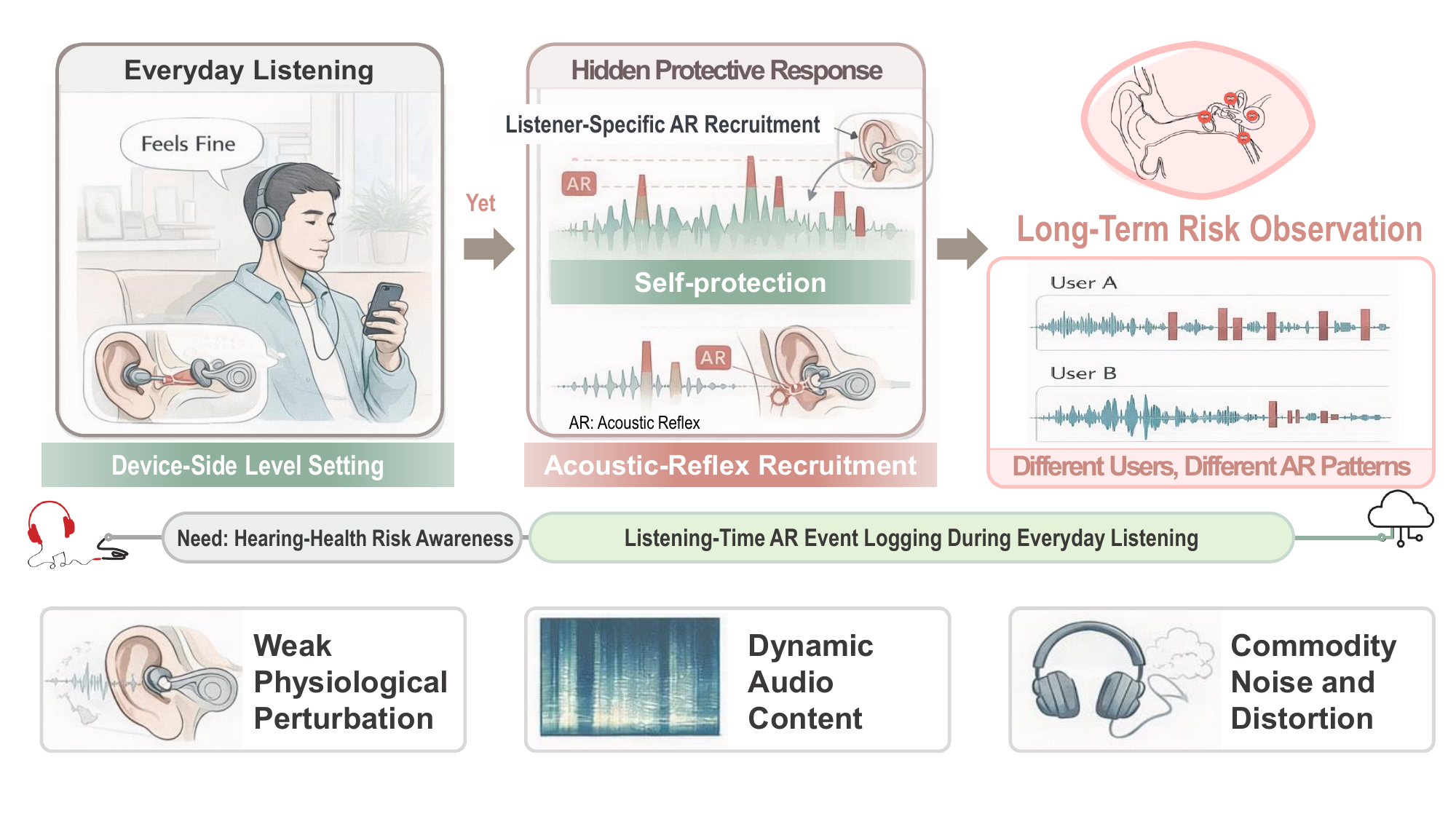}
    \caption{\textbf{\earlog{} overview.} \earlog{} turns everyday playback into an in-ear probe and records listener-specific AR events that remain invisible to device-side volume rules.}
    \Description{Three-part overview of the \earlog{} application. Everyday listening at a device-recommended level can still cross a listener-specific AR threshold. \earlog{} uses the playback itself as an in-ear probe and records AR events over time. The bottom row highlights three sensing obstacles: the weak physiological perturbation, dynamic broadband content, and commodity-hardware noise and distortion.}
    \label{project_introduction}
\end{figure}

More than one billion young people are at risk of irreversible hearing damage from unsafe headphone listening~\citep{WHOITU2019_SafeListening}. Current protection covers two endpoints: device-side control of sound pressure level (SPL) and listening exposure~\citep{WHOITU2019_SafeListening,turchet2024hearing,knoetze2021sound,chen2023hearing}, and clinical hearing tests that detect damage after it occurs~\citep{Hill2016_HairCellSynaptopathy,WangRen2012Repeated,Kujawa2009_TTSsynap,Fernandez2015_AgingAfterNoise,Liberman2016Differential}. Between these endpoints, how risk accumulates during high-load listening remains a blind spot in hearing protection.

This blind spot matters because hearing risk varies across listeners and from moment to moment. Physiological tolerance differs widely, and brief high-energy peaks in modern audio can push the auditory system into overload even when the volume setting appears safe~\citep{reynolds2023music,Causon2020_AR,Feeney2023_WBART}. The ear signals these moments through its own protective response. When the load reaches a listener-specific threshold, the brainstem contracts the stapedius muscle in the middle ear. This contraction stiffens the sound-conduction path and reduces the acoustic energy transmitted to the cochlea~\citep{Trevino2023_MEMR,Valero2016_MEMR}. This involuntary protective response, known as the acoustic reflex (AR), also leaves a minute perturbation in ear-canal reflections that a headphone can capture. These activations can occur even at ordinary listening levels. Recording these AR activations during everyday listening would allow hearing protection to track how each ear responds over time.

However, this stable measurement setup is unavailable during everyday listening on commodity headphones. Clinical systems use a dedicated probe microphone, a fixed probe tone, a separate elicitor, and a stationary subject. Together, these controls provide a stable acoustic environment for measuring small changes in admittance~\citep{titan-manual,InteracousticsTitan,Feeney2017_WAI_Norms}. Everyday listening provides far less stability and can both trigger AR and mask its minute trace. Changes driven by its dynamic content can be much larger than the reflex trace and shift as the content evolves. Ear-canal multipath, headphone frequency response, and fit changes introduce additional variations at the same scale as AR. Reliable clinical AR labels are also scarce in this scenario, making it difficult to train an end-to-end model directly from raw signals.

\earlog{} turns playback, the dominant interference, into the probe. It uses the known input to build a no-AR baseline for the expected in-ear response. \earlog{} then adapts the baseline using playback energy in each frequency band, allowing the estimate to follow changes in content. Because AR typically lasts 100--200 ms, \earlog{} estimates the response within a short window to avoid temporal smearing of this transient response~\citep{Qiu1998_ARlatency,Borg1982_timecourse,Gorga1983_temporal}. \earlog{} uses ear-canal resonances to constrain the plausible delays of reflected sound. A sparse finite impulse response (FIR) model then captures the in-ear response within the short window. With this baseline, \earlog{} applies content-conditioned residual sensing. The measured response is compared with the baseline to suppress variations caused by nonstationary content, individual ear acoustics, and device response. Loudness-dependent changes and localized spectral irregularities in the residual serve as evidence of the weak AR trace. Fit changes, motion, leakage, and measurement saturation can still distort the residual. \earlog{} therefore uses dual-microphone observations to reject unreliable windows, recovers from mild distortion, and updates the baseline across sessions. These designs let \earlog{} record AR events during everyday listening without interrupting playback, forming a longitudinal record of each ear's protective response (Fig.~\ref{project_introduction}).

{\setlength{\emergencystretch}{1em}
We implement \earlog{} as an approximately \$5 dual-microphone add-on. The module follows the inward- and outward-facing microphone layout of a simplified active-noise-cancellation (ANC) front end, and the full pipeline runs in real time on an embedded controller. We evaluate \earlog{} with 81 participants, 16 headphone models, and more than 400 hours of recordings. On a clinically referenced benchmark, \earlog{} achieves 99.0\% sensitivity and 98.7\% specificity.\par}

This paper makes three contributions.

\textbf{A new sensing paradigm for everyday hearing protection.} \earlog{} moves observation from device output and episodic hearing screening to AR events during everyday listening. These event logs provide a process-level record that could support individualized, long-term hearing-protection research.

{\setlength{\emergencystretch}{1em}
\textbf{Opportunistic sensing under dynamic playback.} \earlog{} turns the user's playback into the probe and introduces content-conditioned residual sensing to separate weak AR traces from observations shaped jointly by nonstationary content, individual ear acoustics, and device response. This formulation makes AR during dynamic playback a measurable earable sensing target even though ordinary playback does not provide per-window AR labels.\par}

{\setlength{\emergencystretch}{1em}
\textbf{A complete sensing system for commodity headphones.} \earlog{} combines short-window modeling constrained by ear-canal resonance with dual-microphone observation, recovery from mild distortion, and cross-session reference updates. These components keep weak AR detection usable under low SNR, dynamic content, fit changes, and environmental drift. The system uses low-cost commercial components and is validated across users, headphone models, everyday behaviors, and cross-day conditions.\par}

\section{Related Work}
\label{sec:related_work}

At a broader level, commodity mobile sensing has used vibration and optical responses to infer latent physical states, including liquid identity and gradual quality changes~\citep{huang2021vi,huang2021portable,huang2021li,huang2023beverage}. In mobile health, pulmonary-function systems have extended spirometry from clinical instruments to phones and earphones, using breathing, coughing, and speech to reduce testing burden~\citep{goel2016spirocall,song2020spirosonic,yin2022aware,yin2023ptease,xie2023earspiro,xu2025easyspiro,cao2025espiro}. Earphones and hearables have also become important physiological sensing interfaces. Existing systems use in-ear microphones, active-noise-cancellation (ANC) paths, and repurposed headphone transducers to sense heart rate, heart sounds, respiration, and other weak physiological signals~\citep{butkow2023heart,sun2022earmonitor,cao2023earace,fan2023apg,cao2025heartprint,fan2021headfi,chen2024exploring}. Related commodity-device research has expanded to voluntary middle-ear activity, ear-condition screening, cochlear responses, otoacoustic emissions, and tympanometry~\citep{roddiger2021earrumble,jin2022earhealth,chan2022cochlea,chan2022off,chan2022performing,chan2023wireless}. Software-defined and interference-aware platforms further support acoustic sensing across devices and in the presence of competing sound~\citep{cai2023asdp,cai2023acute}. These studies establish earphones as practical interfaces for periodic, repeatable, or task-elicited physiological sensing. Most closely related, EarCSI reconstructs user-specific ear-canal channels from passive broadband audio for fine-grained sensing of tympanic-membrane changes~\citep{huang2025reconstructing}. \earlog{} targets weak, brief, and unpredictable non-periodic events that arise opportunistically during arbitrary dynamic playback, where periodic structure, preset tasks, clear raw observables, and scalable label collection offer limited help.

Clinical AR measurement, by contrast, detects the reflex with a dedicated probe microphone, a fixed probe tone, a separate elicitor, and stable measurement conditions~\citep{titan-manual,InteracousticsTitan,Feeney2017_WAI_Norms}. These conditions disappear during arbitrary playback on commodity headphones. The next section shows why clinical AR criteria and direct classification of raw in-ear signals do not readily transfer to this setting.

\section{Problem Formulation}
\label{sec:problem_formulation}
\label{sec:why_formulation}

\subsection{Why Direct AR Sensing Fails}
\label{sec:cost}

\begin{figure}[thb]
    \centering
    \includegraphics[width=\linewidth]{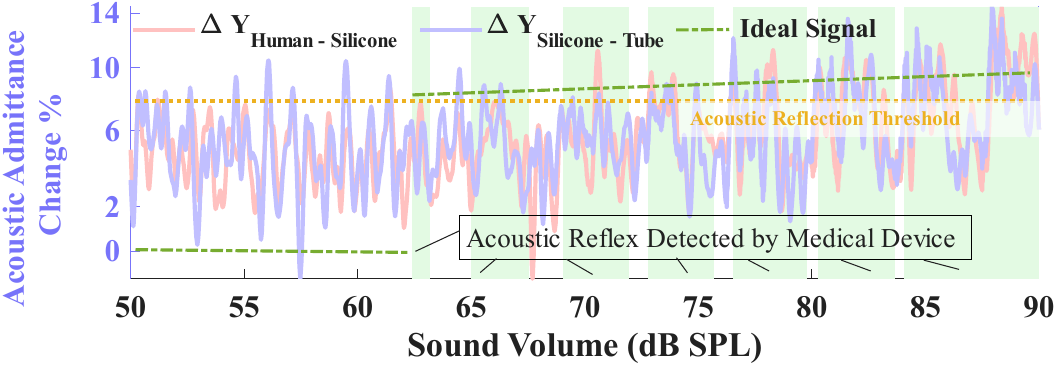}
    \caption{\textbf{The clinical AR criterion does not transfer to commodity headphones.} The contrast between two passive fixtures crosses the criterion although neither can generate AR.}
    \Description{Response changes from 50 to 90 dB SPL. The pink curve is the human--silicone contrast, and the blue curve is the silicone--IEC 60318-4 ear-simulator contrast. Green bands mark Titan-detected AR ranges, the orange line marks the Titan admittance-change criterion, and the dashed green curve fits the Titan measurements. The passive contrast varies with sound level and crosses the Titan criterion.}
    \label{fig:titan_mic}
\end{figure}

\begin{figure*}[!t]
    \centering

    \begin{minipage}[t]{0.28\textwidth}
        \vspace{0pt}
        \centering
        \includegraphics[
            width=\linewidth,
            keepaspectratio
        ]{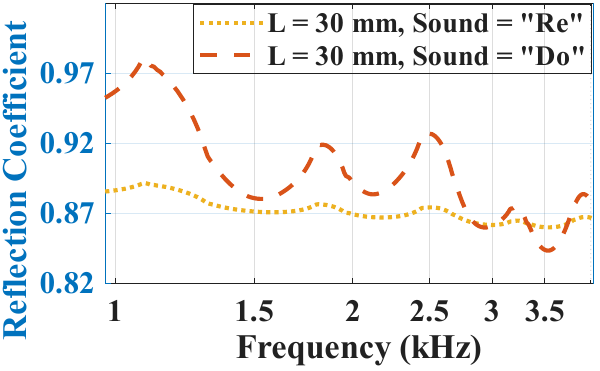}
        \captionof{figure}{
            Reflection coefficient curves for two notes
            (Do and Re) in the same ear canal.
        }
        \Description{Reflection-coefficient curves for the Do and Re notes in the same ear canal, showing that the measured response changes with playback content.}
        \label{fig:ideal_CSI_diff}
    \end{minipage}
    \hfill
    \begin{minipage}[t]{0.67\textwidth}
        \vspace{0pt}
        \centering

        \refstepcounter{figure}
        \setcounter{subfigure}{0}

        \begin{subfigure}[t]{0.49\linewidth}
            \centering
            \includegraphics[width=\linewidth,keepaspectratio]{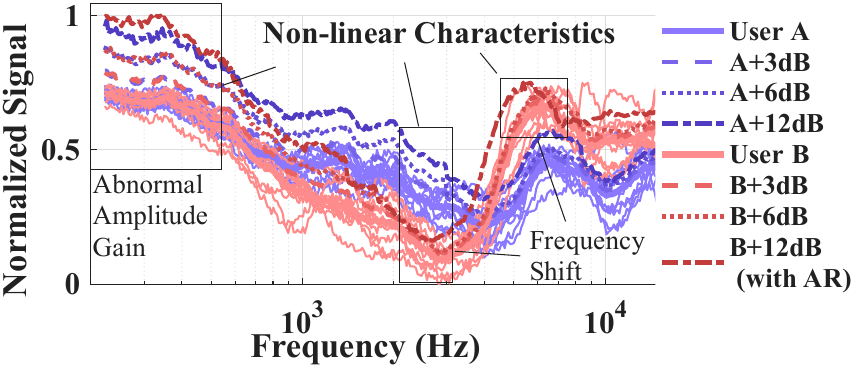}
            \caption{Responses at different sound levels.}
            \label{fig:amplitude_response}
        \end{subfigure}
        \hfill
        \begin{subfigure}[t]{0.49\linewidth}
            \centering
            \includegraphics[width=\linewidth,keepaspectratio]{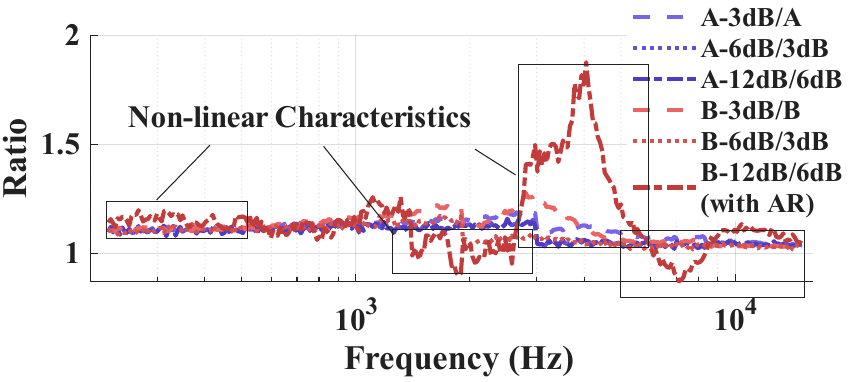}
            \caption{Signal enhancement ratio.}
            \label{fig:boost_ratio}
        \end{subfigure}

        \addtocounter{figure}{-1}

        \par\vspace{3pt}

        \captionof{figure}{
            Frequency shifts and nonlinear effects associated
            with acoustic reflex.
        }
        \Description{Two panels compare responses across sound levels and the corresponding signal-enhancement ratio, highlighting localized frequency shifts and nonlinear changes associated with acoustic reflex.}
        \label{fig:comparison}
    \end{minipage}
\end{figure*}

Clinical acoustic-reflex tests infer AR from changes in absolute acoustic admittance~\citep{Qiu1998Characterization,Kujawa2009Adding}. This criterion does not transfer to commodity headphones. In Fig.~\ref{fig:titan_mic}, the silicone artificial ear and IEC 60318-4 ear simulator cannot generate AR, yet their response contrast varies with sound pressure level and crosses the Titan clinical criterion. The change must therefore come from device response and acoustic propagation. Clinical systems avoid this ambiguity with high-precision instrumentation, a dedicated probe, and fixed excitation. Commodity headphones lack this stable reference, so changes in absolute admittance or reflection cannot provide direct evidence of AR. Appendix~\ref{app:clinical_admittance} describes the control setup and contrast calculation.

A fixed threshold calibrated for commodity headphones also fails because the in-ear response changes with playback even when AR is absent. As Fig.~\ref{fig:ideal_CSI_diff} shows, the Do and Re notes produce different reflection curves in the same ear canal under nominally identical conditions. The inward microphone receives overlapping ear-canal propagation paths whose contributions depend on the current spectrum and playback level. A fixed threshold or one-time reference therefore becomes stale as the content changes. \earlog{} needs a content-conditioned baseline that follows the current playback. Appendix~\ref{app:content_confound} shows how playback content and ear-canal multipath jointly shape the observation.

\subsection{\earlog{}'s Sensing Core}
\label{sec:sensing_core}

Dynamic playback shifts the baseline, but it does not erase the local structure that AR leaves in the response. Figure~\ref{fig:comparison} shows that, in the absence of AR, response changes across adjacent loudness levels remain relatively smooth. When AR engages, some bands exhibit localized gain enhancement, slight shifts in spectral peaks, and irregular gain changes across adjacent levels. Removing the expected response for the current content reveals these local deviations that are otherwise masked by the much stronger playback-driven response. This observation motivates content-conditioned residual sensing. \earlog{} predicts the no-AR response to the current playback, compares this baseline with the measured response, and identifies weak AR from loudness-dependent changes and localized spectral irregularities in the residual. Sections~\ref{sec:baseline} and~\ref{sec:verification} describe the baseline model and residual features, respectively.

\begin{figure*}[!t]
    \centering
    \includegraphics[width=0.88\linewidth]{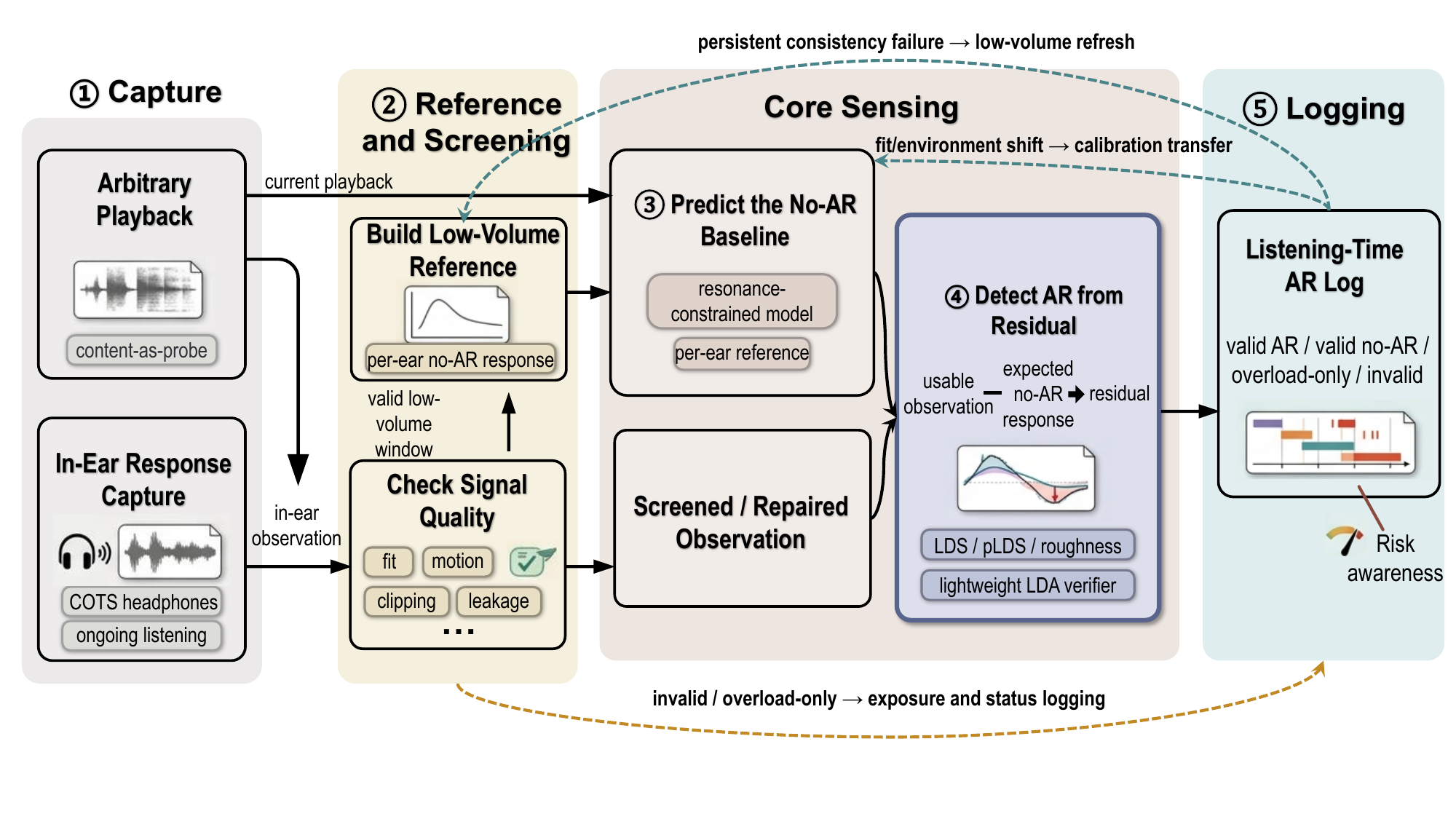}
    \caption{\textbf{\earlog{} workflow.} \earlog{} screens each window, predicts the no-AR baseline for the current playback, detects AR from the residual, and records the resulting window state.}
    \Description{Workflow from playback and dual-microphone capture to signal screening, no-AR baseline prediction, residual-based AR detection, and window logging, with baseline maintenance across sessions.}
    \label{fig:workflow}
\end{figure*}

{\setlength{\emergencystretch}{1em}
Figure~\ref{fig:workflow} summarizes \earlog{}'s window-level pipeline. The system first checks whether the playback and dual-microphone observations support reliable AR detection. Windows that fail this check are logged as \emph{invalid} or \emph{overload-only} and excluded from AR detection. For each valid window, \earlog{} predicts the no-AR baseline response for the current playback (Sec.~\ref{sec:baseline}). When the observation contains mild clipping or recoverable noise, this baseline also guides signal repair (Sec.~\ref{sec:guardrails}). \earlog{} then compares the measured response with the baseline to form the residual (Sec.~\ref{sec:baseline_construction}) and examines the residual for loudness-dependent features and localized spectral irregularities (Sec.~\ref{sec:verification}). A lightweight verifier outputs \emph{valid AR} or \emph{valid no-AR}, and the system records the result. Across sessions, \earlog{} checks baseline consistency and updates the reference when fit or environmental conditions shift (Sec.~\ref{sec:baseline_maintenance}). The resulting log records AR events during playback together with each window's sensing status.\par}

\section{\earlog{} Design}
\label{sec:improve} \label{sec:design}

\subsection{Predicting the No-AR Baseline}
\label{sec:baseline}
\label{sec:ear_canal_model}

Section~\ref{sec:why_formulation} shows that the absolute in-ear response alone does not provide reliable evidence of AR during dynamic playback. \earlog{} therefore asks what response the same ear would produce under the current playback if AR were absent. This expected response defines the no-AR baseline. Because the current window may already contain AR, \earlog{} learns the no-AR response from low-volume segments and predicts the baseline for the current playback condition.

\subsubsection{Building the Low-Volume Reference}
\label{sec:Parameter}

Low-volume playback provides the reference needed by \earlog{}. In this regime, non-silent audio still excites the in-ear acoustic response, while the response follows the smooth no-AR pattern illustrated in Fig.~\ref{fig:comparison}. \earlog{} therefore builds its per-ear no-AR reference from non-silent low-volume segments without requiring a clinical WB-ART measurement. Silent blocks provide an independent negative control, but without excitation, they cannot be used to identify the acoustic response.

The digital playback samples are known, but the in-ear observation is not a copy of the source. The headphone transducer, insertion depth, seal, and ear-canal multipath transform the sound before the inward-facing microphone records it. \earlog{} pairs the known playback with the corresponding inward-microphone response from the same ear and device configuration. Over a short, low-volume segment, this input--output relation can be approximated as locally linear.

A single nominal volume is insufficient for grouping these calibration segments. Two audio windows with the same overall level can distribute their energy across different frequency bands and produce different ear-canal responses. \earlog{} therefore computes a playback-energy vector over \(B\) analysis bands: \(\mathbf{V}=[E_1,\ldots,E_B]\), where \(E_b\) is the playback energy in band \(b\), and \(B\) is the number of analysis bands. Windows with similar \(\mathbf{V}\) enter the same band-energy bucket. Each bucket represents a local no-AR operating condition under similar content and gain, instead of a single reference intended to cover all playback.

We use \(V\) to denote the bucket associated with one such band-energy pattern. Within that bucket, \earlog{} models the local no-AR input--output relation with a finite impulse response (FIR):
\[
y^{(V)}(n)
=
\sum_{k=0}^{M-1}h_k^{(V)}x(n-k)
+
\eta(n).
\]
Here, \(x(n)\) is the \(n\)-th digital playback sample, \(y^{(V)}(n)\) is the inward-microphone observation, \(h_k^{(V)}\) is the \(k\)-th FIR coefficient for bucket \(V\), \(M\) is the FIR length, and \(\eta(n)\) is additive noise. \earlog{} estimates the bucket-specific coefficients with the least mean squares (LMS) algorithm. This step identifies the local no-AR response. AR detection occurs later. Bucketing reduces content variation, but a short calibration segment still cannot reliably estimate a generic high-order FIR with many free coefficients. \earlog{} next constrains the tap locations with ear-canal physics.

\subsubsection{Resonance-Constrained Sparse FIR Modeling}
\label{sec:Sensitivity}
\label{sec:SNR}

The response model must be estimated within a short window. A longer window provides more samples for FIR estimation, but it also spreads a short AR transition across unrelated content and causes temporal smearing~\citep{Qiu1998_ARlatency,Borg1982_timecourse,Gorga1983_temporal}. A generic high-order FIR therefore has too many free coefficients for the available window. Ear-canal reflections provide the missing constraint because their main arrival times occupy a limited set of propagation delays.

\earlog{} approximates the ear canal as a quarter-wave resonator. Its lowest credible resonance \(f_0\) provides a coarse effective propagation length \(L\simeq c/(4f_0)\), where \(c\) is the speed of sound. The corresponding one-way delay is \(t_0=L/c\). \earlog{} restricts nonzero FIR taps to narrow neighborhoods around the direct arrival and successive round-trip delays \(k_i\simeq2it_0f_s\), where \(f_s\) is the sampling rate and \(i\) is the reflection order. The effective length serves as a structural prior on tap locations. Constraining the FIR support preserves the main propagation components while reducing the number of coefficients that must be estimated in each short window.

Each band-energy bucket maintains its own LMS state. Let \(e^{(V)}(n)\) be the difference between the inward-microphone sample and the bucket-specific FIR prediction at sample \(n\). For a tap index \(k\) inside the resonance-constrained support \(\mathcal K_V\), the online update is
\begin{equation}
h_k^{(V)}(n+1)
=
h_k^{(V)}(n)
+
\mu e^{(V)}(n)x(n-k),
\qquad k\in\mathcal K_V,
\label{eq:sparse_lms_update}
\end{equation}
where \(\mu\) is the LMS step size. Buckets update independently, and coefficients outside \(\mathcal K_V\) remain zero. The support constraint therefore limits adaptation to propagation delays supported by the ear-canal model rather than allowing an unconstrained filter to absorb the short-window observation.

After sparse FIR identification, a gammatone filterbank maps the response into the same \(B\) analysis bands used by the playback-energy vector. The band-wise baseline parameters and residual features in Sec.~\ref{sec:verification} use this same frequency representation. The shared band structure connects low-volume calibration, baseline prediction, and residual-based AR detection without repeatedly converting between incompatible representations.

\subsubsection{Predicting the Baseline for Current Playback}
\label{sec:baseline_construction}

The low-volume buckets cover only part of the content and energy states encountered during everyday playback. Once playback leaves those states, the measured response may already contain AR and can no longer serve as its own no-AR reference. \earlog{} therefore predicts the baseline for the current playback from the responses learned during low-volume calibration.

For each analysis band \(b\), \earlog{} represents the bucket response with a baseline parameter \(\theta_b(V)\). Here, \(\theta_b(V)\) is the log-magnitude of the identified no-AR transfer response in band \(b\) for playback state \(V\). \earlog{} assumes that this parameter varies smoothly across nearby playback conditions and fits a low-order model:
\[
\theta_b(V)
=
\sum_{m=0}^{P}a_{b,m}\phi_m(V).
\]
Here, \(\phi_m(V)\) is the \(m\)-th basis function, \(a_{b,m}\) is its fitted coefficient, and \(P\) is the model order. This low-order relation avoids a separate calibration for every content--level combination and limits the variation introduced by the finite low-volume dataset.

{\setlength{\emergencystretch}{1em}
At runtime, \earlog{} computes the current playback state \(V_t\) and reconstructs the corresponding no-AR baseline \(H_{\mathrm{baseline}}(f,V_t)\). Comparing it with the observed in-ear response \(H_{\mathrm{obs}}(f,t)\) yields the log-magnitude residual:
\[
\Delta\log H(f,t)
=
\log|H_{\mathrm{obs}}(f,t)|
-
\log|H_{\mathrm{baseline}}(f,V_t)|.
\]
This subtraction removes most of the variation explained by the current playback and the modeled in-ear response. The residual may still contain baseline mismatch, fit drift, leakage, noise, or distortion. The next subsection therefore tests whether the residual shows the loudness dependence and localized spectral structure associated with AR.\par}

Removing the physics-constrained no-AR baseline reduces accuracy from 98.71\% to 90.35\%, a loss of 8.36 percentage points (Table~\ref{tab:ablation_results}). The 8.36-point loss confirms the role of baseline prediction in residual-based AR detection. Appendix~\ref{ap:fir_details_main} gives bucketed identification details and runtime construction, while Appendix~\ref{ap:FIR_Tap} gives the resonance and sparse-tap derivation.
\subsection{Detecting AR from the Residual}
\label{sec:verification}

Subtracting the no-AR baseline removes most of the response explained by the current playback, but a nonzero residual alone is insufficient evidence of AR. Baseline mismatch, fit drift, cross-band content coupling, and recovery artifacts can all produce a nonzero residual. The observations in Sec.~\ref{sec:sensing_core} suggest two tests that are more specific to AR. When AR occurs, the residual is expected to become loudness-dependent in a subset of bands and to exhibit a localized spectral irregularity rather than a smooth shift across frequency. \earlog{} uses three residual cues for these tests. The Loudness Differential Signature (LDS) measures how the residual changes with loudness, the partial Loudness Differential Signature (pLDS) accounts for cross-band contributions before estimating the local slope, and local spectral roughness describes whether the remaining change is localized in frequency.

\subsubsection{Loudness Differential Signature}
\label{sec:LDS}

The first cue tests whether the residual varies with playback loudness. Let \(r_b(t)\) denote the residual log response in analysis band \(b\), derived from \(\Delta\log H(f,t)\). Let \(L_{s,b}(t)\) denote the source-side log loudness in the same band. The Loudness Differential Signature (LDS) is the local sensitivity
\[
\mathrm{LDS}_b
=
\frac{\partial r_b(t)}
{\partial L_{s,b}(t)}.
\]

The no-AR baseline already accounts for the ordinary level-dependent response expected from the current playback. Without AR, the residual therefore remains approximately flat with respect to source-side loudness, and \(\mathrm{LDS}_b\) stays close to zero. When AR engages, stapedius contraction changes middle-ear transmission. Near ear-canal resonances, this transmission change appears as a new local slope in a subset of bands. LDS measures the loudness-dependent response that remains after baseline subtraction, without relying on the absolute response magnitude used by clinical thresholds.

A slope measured in one band, however, may include variation from neighboring bands. Dynamic audio excites many frequencies at once, and harmonics, ear-canal multipath, and filterbank leakage couple observations across bands. \earlog{} therefore accounts for these cross-band contributions before interpreting the local slope.

\subsubsection{Removing Cross-Band Coupling with pLDS}
\label{sec:pLDS}

To separate the band-local response from cross-band content variation, let \(B\) denote the number of analysis bands. For frames \(t=t_1,\ldots,t_W\) in a short estimation window, \earlog{} jointly models the residuals as
\[
\mathbf r(t)
=
\mathbf A\mathbf L_s(t)
+
\mathbf c
+
\boldsymbol\epsilon(t).
\]
Here, \(\mathbf r(t)\in\mathbb R^B\) is the residual log-response vector, \(\mathbf L_s(t)\in\mathbb R^B\) is the source-side loudness vector, \(\mathbf A\in\mathbb R^{B\times B}\) is the cross-band coefficient matrix, \(\mathbf c\) is an intercept, and \(\boldsymbol\epsilon(t)\) is the remaining modeling error. Dynamic playback often produces strongly correlated loudness trajectories across bands, so \earlog{} estimates the coefficients with
\begin{equation}
\left(\widehat{\mathbf A},\widehat{\mathbf c}\right)
=
\arg\min_{\mathbf A,\mathbf c}
\sum_{t=t_1}^{t_W}
\left\|
\mathbf r(t)
-
\mathbf A\mathbf L_s(t)
-
\mathbf c
\right\|_2^2
+
\lambda_A\|\mathbf A\|_F^2,
\label{eq:plds_ridge}
\end{equation}
where \(\lambda_A\geq0\) controls the ridge penalty.

Each entry \(\widehat A_{ij}\) describes how source-side loudness in band \(j\) contributes to the residual in band \(i\). The off-diagonal coefficients account for cross-band contributions, while the diagonal coefficient estimates the band-local loudness-to-residual relation after conditioning on the other excited bands, giving \(\mathrm{pLDS}_b=\widehat A_{bb}\). pLDS therefore reduces the cross-band contamination that makes a single-band LDS sensitive to rich broadband content unrelated to the band-local AR change. Ridge regularization stabilizes the joint estimate when source-band loudness trajectories are correlated. Because the current playback does not reliably excite every band, \earlog{} computes pLDS only where playback energy is sufficient. This support rule prevents leakage, noise, or numerical instability in weakly excited bands from being interpreted as a local loudness response.

pLDS estimates the band-local loudness dependence, but a local slope can still arise from smooth gain drift, mild fit variation, or baseline extrapolation error. These effects differ from AR mainly in how they vary across frequency.

\subsubsection{Local Spectral Roughness}
\label{sec:roughness}

AR most strongly perturbs a subset of bands near ear-canal resonances. The residual can therefore contain localized gain changes, small spectral-peak shifts, or abrupt differences between adjacent bands, whereas device gain drift and mild fit changes more often vary smoothly across frequency. For each window, \earlog{} robustly normalizes the residual as
\begin{equation}
\tilde r_b(t)
=
\frac{
r_b(t)-\operatorname{median}_{b'}r_{b'}(t)
}{
\operatorname{MAD}_{b'}(\mathbf r(t))+\epsilon
},
\qquad \epsilon>0.
\label{eq:roughness_normalization}
\end{equation}
The median removes a global offset. The median absolute deviation (MAD), computed across the \(B\) bands around that median, supplies a robust within-window scale; \(\epsilon\) keeps the normalization defined when the residual is nearly flat. This centering and scaling prevents a broadband offset or uniform gain change from appearing as localized structure. \earlog{} then computes
\begin{equation}
\begin{aligned}
\mathrm{TV}(t)
&=
\sum_{b=1}^{B-1}
\left|\tilde r_{b+1}(t)-\tilde r_b(t)\right|,
\\
\mathrm{Curv}(t)
&=
\sum_{b=2}^{B-1}
\left|
\tilde r_{b+1}(t)-2\tilde r_b(t)+\tilde r_{b-1}(t)
\right|,
\\
J_{\max}(t)
&=
\max_{1\leq b<B}
\left|\tilde r_{b+1}(t)-\tilde r_b(t)\right|.
\end{aligned}
\label{eq:roughness_descriptors}
\end{equation}
The three descriptors capture complementary aspects of the normalized residual. Total variation accumulates adjacent-band changes, curvature responds to local changes in spectral slope, and \(J_{\max}\) isolates the strongest adjacent-band discontinuity. Smooth gain or fit drift generally keeps all three small, whereas a localized AR perturbation can increase them. LDS and pLDS test whether the residual changes with loudness. The roughness descriptors test whether that change is concentrated in a subset of bands rather than appearing as a smooth global shift.

\subsubsection{Lightweight AR Verifier}
\label{sec:LDA}
\label{sec:System Design}

Each valid window yields
\begin{equation}
\begin{aligned}
\mathbf z(t)
=
\big[
&\mathrm{LDS}(t)^\top,\,
\mathrm{pLDS}(t)^\top,\,
\mathrm{TV}(t),\\[-2pt]
&\mathrm{Curv}(t),\,
J_{\max}(t)
\big]^\top.
\end{aligned}
\label{eq:verifier_features}
\end{equation}
The implementation normalizes \(\mathbf z(t)\) using z-score statistics computed from the training participants. A shrinkage-regularized linear discriminant analysis (LDA) model learns \(\mathbf w\) and \(b\) from the clinical labels and produces
\begin{equation}
\begin{aligned}
s(t)
&=
\mathbf w^\top\mathbf z(t)+b,
\\
\bar s(t)
&=
\lambda_s\bar s(t-1)
+
(1-\lambda_s)s(t),
\qquad \lambda_s\in[0,1).
\end{aligned}
\label{eq:verifier_score}
\end{equation}
Covariance shrinkage stabilizes the discriminant under the available labels. Score smoothing suppresses frame-level jitter; it does not replace the residual evidence in \(\mathbf z(t)\).

The z-score statistics, \(\mathbf w\), and \(b\) are learned from the training participants and remain fixed during online inference. Runtime processing therefore consists of constructing \(\mathbf z(t)\), evaluating the linear score, and updating the smoothed score for eligible windows. This separation prevents the verifier from adapting its decision boundary to a transient residual or to the current user's event history.

A two-threshold hysteresis rule converts \(\bar s(t)\) into the window state. \earlog{} enters the AR state only after the score stays above the higher threshold for \(K_+\) consecutive valid windows, and exits only after it stays below the lower threshold for \(K_-\) consecutive valid windows. Invalid and overload-only windows receive no AR decision and do not advance the AR state or event count, although they remain in the listening-exposure and signal-quality log. Consecutive valid windows assigned to the AR state are merged into one event.

The ablation results separate the roles of pLDS and roughness. Replacing the multiband pLDS representation with single-band LDS reduces accuracy from 98.71\% to 86.55\%, a loss of 12.16 percentage points. Removing the roughness descriptors and roughness-based gating reduces accuracy to 92.20\%, a loss of 6.51 percentage points (Table~\ref{tab:ablation_results}). The 12.16-point loss reflects the value of cross-band decoupling under dynamic broadband playback. The 6.51-point loss reflects the additional separation between localized AR structure and smooth drift.

Appendix~\ref{app:residual_derivations} derives the residual relation and records pLDS window construction, regularization details, training-time feature normalization, and the hysteresis parameters. The residual cues above assume that the playback and microphone observations remain physically meaningful. Large motion, leakage, saturation, and environmental noise can invalidate that assumption or distort the residual. The next subsection explains how \earlog{} rejects unusable windows and repairs mild artifacts before verification.

\subsection{Screening and Repairing In-Ear Observations}
\label{sec:Interference}
\label{sec:guardrails}

Sections~\ref{sec:baseline} and~\ref{sec:verification} assume that the inward-facing microphone still captures a meaningful response to the known playback under the current headphone fit and ear-canal condition. Everyday disturbances violate this assumption in two different ways. Severe leakage, large motion, abrupt fit changes, and hard saturation can break the playback--response relation. Mild clipping and localized background noise corrupt only part of the response. As shown in Fig.~\ref{fig:workflow}, \earlog{} routes these cases through screening and repair before computing residual features. Unusable windows are excluded from the AR-detection path and are logged as \emph{invalid} or \emph{overload-only}. Recoverable windows are repaired using the current no-AR baseline as a constraint before residual detection. The repair is deliberately conservative because the verifier relies on loudness-dependent changes and local spectral roughness. Preprocessing must preserve these patterns without introducing artifacts that resemble them.

\subsubsection{Checking Window Validity}
\label{sec:nonlinear}

\earlog{} begins by checking whether the current window can be compared reliably with its no-AR baseline. Four complementary indicators screen for different failure modes. The clipped-sample ratio \(r_{\mathrm{clip}}(t)\) detects saturation in the microphone or analog front end. The leakage ratio \(\lambda_{\mathrm{leak}}(t)\) compares outer- and inner-microphone energy to reveal seal loss or abnormal leakage of external sound. The source-coherence score \(c_{\mathrm{src}}(t)\) checks whether the inward observation still follows the known playback, exposing large motion or interference unrelated to the playback source. The baseline-consistency score \(m_{\mathrm{fit}}(t)\) checks whether the current response is consistent with the recent no-AR baseline under the present fit. These indicators determine whether the captured playback--response relation remains interpretable.

A window remains eligible for further processing only while all four indicators stay within their calibrated ranges. To prevent rapid switching near a boundary, the gate changes state only after either a violation or a recovery persists across consecutive windows. Windows dominated by hard front-end saturation are logged as \emph{overload-only}, because the playback--response relation has been destroyed and cannot constrain a reconstruction. Other windows that cannot be interpreted reliably are logged as \emph{invalid}. Windows in either state are excluded from AR detection and do not contribute to the AR event count, although they remain in the listening-exposure and signal-quality records described in Sec.~\ref{sec:setup}. A transient baseline inconsistency causes the current window to be rejected. Persistent inconsistency is handled in Sec.~\ref{sec:baseline_maintenance}, which adapts or refreshes the no-AR reference.

The gate removes observations whose input--output relation has been lost. Only locally corrupted windows that retain an interpretable playback--response relation are repaired. For mild clipping, \earlog{} uses the current no-AR baseline to reconstruct the dominant linear response and applies residual compensation only in bands whose reconstruction error remains elevated. For background interference, it suppresses only components with low correlation to the known playback. Both operations are constrained to preserve the local structure used by LDS, pLDS, and roughness, and candidate AR windows never update the repair parameters. Removing software artifact/interference mitigation reduces accuracy from 98.71\% to 75.90\%, a loss of 22.81 percentage points (Table~\ref{tab:ablation_results}). These mechanisms address short-lived corruption within individual windows; Sec.~\ref{sec:baseline_maintenance} instead handles path drift that persists across windows.

\subsection{Maintaining the Baseline over Time}
\label{sec:baseline_maintenance}

The preceding sections handle content variation and transient observation failures. A slower source of error remains: drift in the no-AR baseline. Refitting changes insertion depth and seal, while humidity and prolonged wear gradually alter the acoustic response between the headphone and the ear. A stale baseline can then leave a persistent residual offset even when every window passes the validity checks. \earlog{} accounts for this drift in the no-AR response model and keeps the AR verifier unchanged.

\subsubsection{Calibration Transfer}
\label{sec:baseline_transfer}
\label{sec:guardrails_drift}

Changes in fit, seal, and humidity do not affect all frequency bands independently. Across paired low-volume no-AR observations, their effects recur as a small set of band-wise patterns. \earlog{} extracts the leading patterns and forms a compact drift basis \(\mathbf{\Psi}\). For playback state \(V_t\) and current fit or environmental condition \(u_t\), the adapted baseline parameters are
\begin{equation}
\boldsymbol{\theta}(V_t,u_t)
=
\boldsymbol{\theta}(V_t)
+
\mathbf{\Psi}\boldsymbol{\alpha}_t.
\label{eq:calibration_transfer_model}
\end{equation}
Here, \(\boldsymbol{\theta}(V_t)\) is the baseline predicted for the current playback, the columns of \(\mathbf{\Psi}\) describe recurring band-wise drift patterns, and \(\boldsymbol{\alpha}_t\) gives the current shift in that basis.

Let \(\mathbf q_t\in\mathbb R^B\) be the band-wise average log-transfer observation from the current valid low-volume support windows, expressed in the same representation as \(\boldsymbol{\theta}\). Let \(\mathcal B\) be the stored low-volume baseline buckets, and let \(\mathbf P_{\mathcal A}\) select the stable anchor bands. \earlog{} first matches the current support to a stored reference:
\begin{equation}
\hat V_t
=
\arg\min_{V\in\mathcal B}
\left\|
\mathbf P_{\mathcal A}
\left[
\mathbf q_t-\boldsymbol{\theta}(V)
\right]
\right\|_2^2.
\label{eq:calibration_anchor_matching}
\end{equation}

After matching, \earlog{} estimates the drift coefficients from the remaining band-wise mismatch:
\begin{equation}
\hat{\boldsymbol{\alpha}}_t
=
\arg\min_{\boldsymbol{\alpha}}
\left\|
\mathbf W
\left[
\mathbf q_t
-
\boldsymbol{\theta}(\hat V_t)
-
\mathbf{\Psi}\boldsymbol{\alpha}
\right]
\right\|_2^2
+
\lambda_\alpha
\|\boldsymbol{\alpha}\|_2^2,
\label{eq:calibration_online_transfer}
\end{equation}
where \(\mathbf W=\mathbf I-\mathbf P_{\mathcal A}\) selects the non-anchor bands that carry most of the drift, and \(\lambda_\alpha\) controls the ridge penalty. The matched bucket and \(\hat{\boldsymbol{\alpha}}_t\) provide the stored reference and current drift term in Eq.~\ref{eq:calibration_transfer_model}; the resulting baseline replaces the stale reference used by Secs.~\ref{sec:baseline_construction} and~\ref{sec:verification}.

Each transfer is estimated directly against the stored calibration reference. Previous transfer coefficients do not enter the next objective, so transfer error does not accumulate recursively across sessions. Several factors may act together, such as insertion depth, humidity, and seal evolution; their combined effect remains correctable while its band-wise displacement lies within the learned drift basis. Appendix~\ref{app:baseline_transfer} describes paired-difference construction, the drift and anchor sets, support-window formation, frequency-domain transport, and consistency-triggered refresh.

\subsubsection{Consistency Check and Low-Volume Refresh}

\earlog{} retains a transferred baseline only while it remains consistent with current low-volume observations. On valid low-volume support windows, \earlog{} monitors residual energy and the low-volume LDS, pLDS, and roughness statistics. A single mismatch causes the validity gate in Sec.~\ref{sec:guardrails} to reject the current window. A mismatch that persists across consecutive windows indicates that the current shift is no longer explained by the learned drift patterns.

When this occurs, \earlog{} pauses AR detection and refreshes the band-energy buckets and no-AR baseline using naturally occurring low-volume playback. Listening exposure and signal-quality status continue to be recorded during the refresh. AR detection resumes after the new baseline passes the same consistency check.

The seven-day evaluation in Sec.~\ref{sec:eval:crossday_refresh} quantifies how often this fallback is required and how much low-volume playback it consumes. The results show that calibration transfer absorbs most day-to-day drift and that refresh is needed only occasionally.

\section{Implementation}
\label{Module}

\begin{figure*}[!t]
  \centering
  \captionsetup[subfigure]{font=footnotesize,justification=centering}
  \begin{subfigure}[t]{0.46\textwidth}
    \vspace{0pt}
    \centering
    \includegraphics[height=36mm,keepaspectratio]{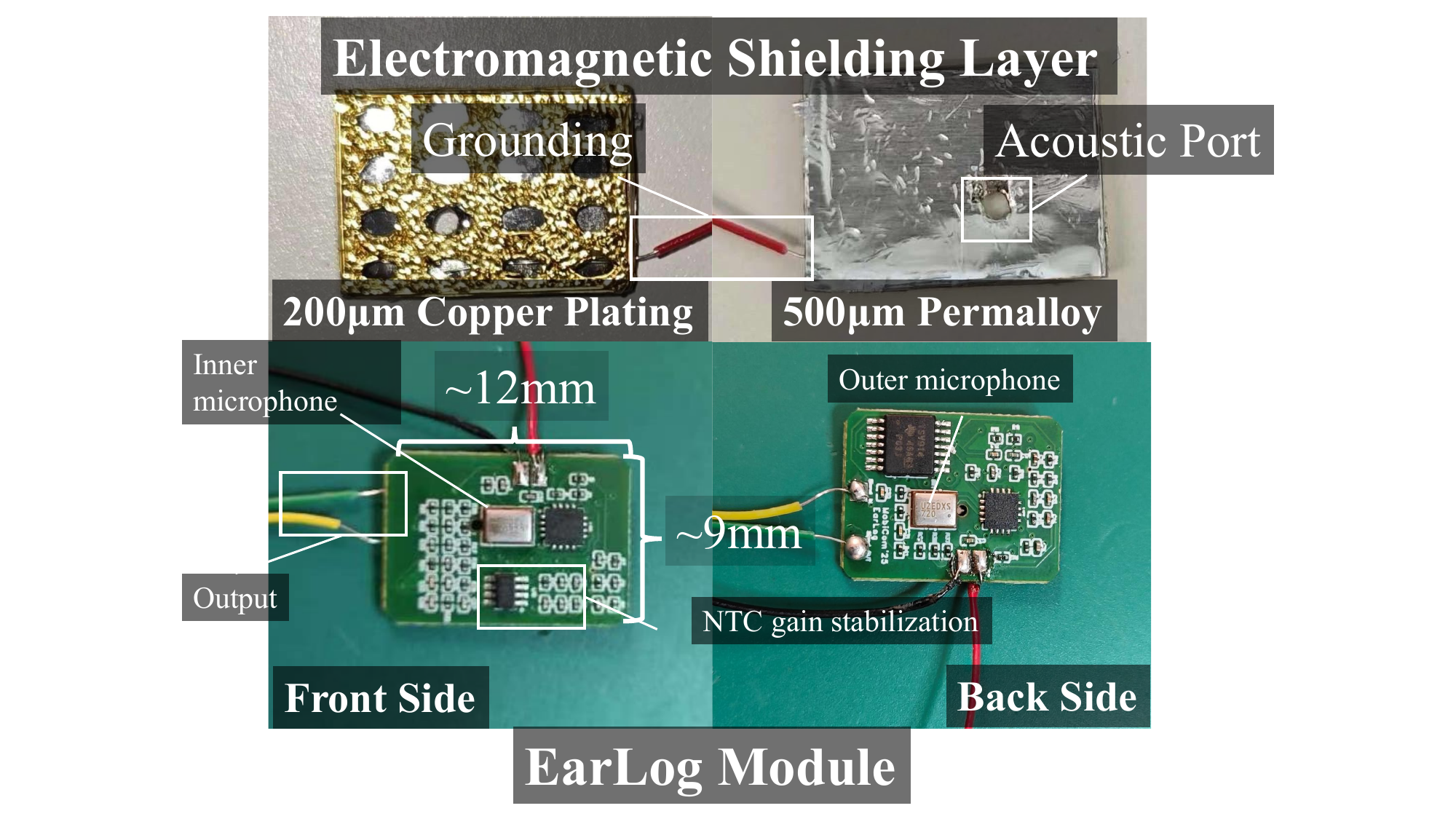}
    \caption{Dual-microphone add-on module.}
    \label{fig:earlog_module_panel}
  \end{subfigure}\hfill
  \begin{subfigure}[t]{0.52\textwidth}
    \vspace{0pt}
    \centering
    \includegraphics[height=36mm,keepaspectratio]{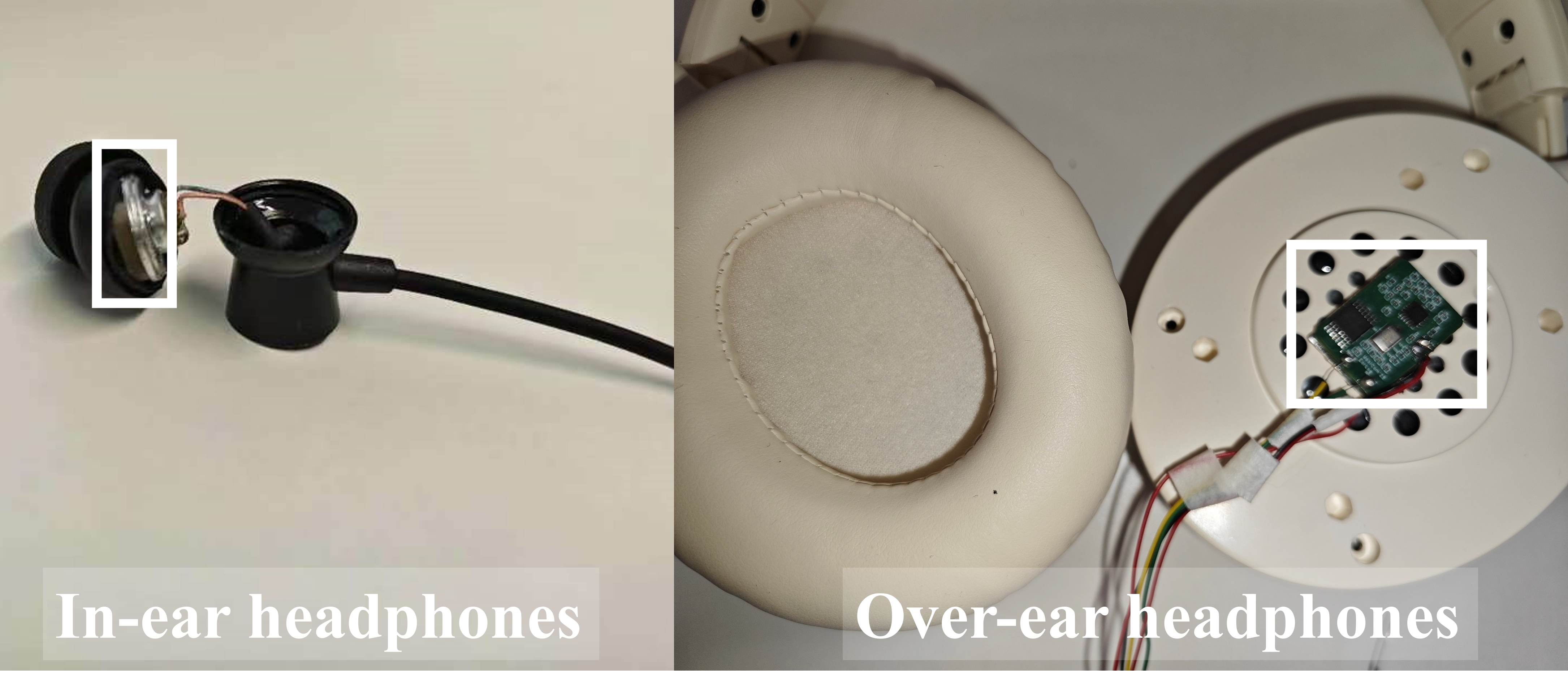}
    \caption{In-ear and over-ear deployment geometry.}
    \label{fig:headphone_display}
  \end{subfigure}
  \caption{\textbf{\earlog{} hardware and deployment.} The \(12\,\mathrm{mm} \times 9\,\mathrm{mm}\) add-on orients the inner microphone toward the ear canal and the outer microphone toward the environment; the deployment view shows how it sits inside representative in-ear and over-ear headphones.}
  \Description{Panel a shows front and back views of the EarLog module, including the ambient-facing and ear-canal-facing microphones, output, shielding, acoustic port, and temperature-sensitive control module. Panel b shows the module mounted inside in-ear and over-ear headphones.}
  \label{fig:earlog_module}
\end{figure*}

Figure~\ref{fig:earlog_module} shows the \earlog{} hardware and its placement. The prototype is a \(12\,\mathrm{mm}\times9\,\mathrm{mm}\) dual-microphone add-on with a bill-of-materials cost of approximately \$5. When mounted in an in-ear assembly, the add-on sits between the headphone driver and the ear-canal entrance. Two analog microelectromechanical systems (MEMS) microphones (ICS-40720) face opposite directions. The inner microphone records the response shaped by playback, headphone fit, and the ear canal, supporting no-AR baseline prediction and residual detection. The outer microphone provides the environmental and leakage reference used by the validity checks in Sec.~\ref{sec:guardrails}. The two channels are conditioned separately and digitized as independent 16-bit streams at 48~kHz. To reduce slow temperature-dependent gain drift in the microphone and analog front end, \earlog{} places an NTC thermistor in the op-amp feedback path. Calibration transfer handles the remaining frequency-dependent changes caused by fit and environment. Appendix~\ref{app:analog_frontend} details the circuit.

A fixed-point implementation on the Arduino Portenta H7 runs the complete pipeline at \(1.05\times\) real time. Relative to audio playback alone, the \earlog{} processing and sensing path adds \(16\,\mathrm{mW}\) on average. This corresponds to approximately \(0.016\,\mathrm{Wh}\) for one hour of listening, or about \(8\%\) of a representative \(0.20\,\mathrm{Wh}\) earbud battery. Appendix~\ref{app:Implementation} separates this incremental overhead from whole-board development-platform power and reports the fallback-refresh measurement.

The prototype provides the four capabilities required by \earlog{} in a uniform form across headphone models: an inward reflection microphone, an outward environmental reference, synchronized access to playback and microphone streams, and stable local signal processing. Many active-noise-cancellation (ANC) headphones already contain comparable inward and outward microphones together with on-board DSP~\citep{fan2023apg}. A product implementation can reuse these components when synchronized raw access is available. For headphones without an accessible reference path, the add-on supplies the same observations and enables controlled evaluation across the 16 headphone models reported in this paper.

\section{Experimental Setup}
\label{sec:setup}

\paragraph{Participant-disjoint evaluation.}
All main results use leave-one-subject-out (LOSO) evaluation. Every window from one participant remains in the same split, so the test participant contributes no window to supervised training. The short low-volume segment used to build that participant's no-AR baseline is an inference-time, per-ear calibration step. All LDA parameters, feature-normalization statistics, and other supervised parameters are learned from the remaining participants. We report sensitivity, specificity, precision, recall, F1 score, and area under the precision-recall curve (AUPRC). Confidence intervals, when reported, use participant-level bootstrapping rather than window-level resampling (Appendix~\ref{app:eval_splits}).

\paragraph{Clinical reference and synchronization.}
During synchronized trials, the clinical reference neither enters nor stimulates the \earlog{} ear. Each participant first undergoes a separate standard examination with an Interacoustics Titan tympanometer. This examination measures the evaluation ear's wideband acoustic-reflex threshold (WB-ART), the lowest wideband stimulus level at which Titan detects reflex activation~\citep{Feeney2017_WAI_Norms,InteracousticsTitan}. In our evaluation, the measured WB-ART is used to schedule clinically verified sub-threshold control blocks and to stratify the cohort in Sec.~\ref{sec:eval:wbart}. Its numerical value is not provided to \earlog{}'s baseline model or AR verifier.

In subsequent trials, the headphone stimulates the \earlog{}-instrumented target ear while Titan monitors time-varying admittance in the contralateral ear. The acoustic-reflex pathway is bilateral, so sufficient stimulation of the target ear can produce a synchronized reflex response in the opposite ear. The Titan trace therefore supplies an independent time-resolved reference without contributing to the headphone-side acoustic observation. The Titan and \earlog{} signals are recorded as separate streams and aligned to the stimulus timeline by cross-correlation. The maximum measured misalignment across participants is 10~ms. The passive Titan-versus-commodity comparison in Sec.~\ref{sec:cost} is a separate bench experiment and does not generate participant labels.

\paragraph{Independent reference labels.}
Reference labels are generated before any \earlog{} inference. An independent script reads only the Titan trace and stimulus timestamps, then marks intervals as AR-active or AR-inactive. A window is reference-positive when at least \(0.5T_{\mathrm{win}}\) overlaps a Titan-active interval. A reference-negative window lies within a Titan-inactive interval, including silent and below-threshold blocks. The overlap rule associates analysis windows with the clinical trace; physiological onset latency is not counted as part of the labeled event duration.

Both silent and non-silent below-threshold blocks provide reference-negative windows. Baseline identification uses only the non-silent blocks because input--output estimation requires a known excitation. No baseline value, residual feature, or \earlog{} prediction enters the reference-label script. The unilateral music-validation trials use the same frozen classifier, stream alignment, and synchronized contralateral Titan reference. The intervals shown in Fig.~\ref{fig:performance_music} follow this protocol.

\paragraph{Window states, events, and coverage.}
Evaluation follows the four runtime states defined in Sec.~\ref{sec:LDA}. Only valid windows enter the clinical confusion matrix. Consecutive valid-AR windows within one uninterrupted AR state are merged into one event, and event rates are normalized by valid listening time.

The valid-window ratio is the fraction of all recorded windows that pass the validity gate and remain eligible for AR detection. It measures sensing coverage, not classification accuracy. Invalid and overload-only windows do not contribute to TP, TN, FP, or FN counts and are never merged into AR events. They remain in the listening-exposure and signal-quality logs. Evaluation therefore reports accepted-window errors together with valid-window coverage when disturbances cause abstention.

\paragraph{Corpus and metric denominators.}
The complete corpus contains one evaluation ear from each of 81 participants, spans 16 headphone models, and exceeds 400 hours. It includes controlled measurements, robustness tests, and ordinary-listening deployments with wideband noise, piano tones, instrument timbres, and music excerpts at 50--90~dB SPL.

Clinical accuracy uses the synchronized and valid subset of this corpus. This benchmark contains 145,800 non-overlapping windows with \(T_{\mathrm{win}}=250\,\mathrm{ms}\), corresponding to 10.1 hours of Titan-synchronized valid time. Only these valid referenced windows enter sensitivity, specificity, precision, recall, F1, and AUPRC. The unilateral music validation is analyzed separately but uses the same frozen classifier and contralateral Titan reference. Ordinary-listening records without synchronized Titan observation support standalone logging analyses and do not enter the clinical accuracy denominator. The study was approved by the institutional review board. Participant statistics, stimulus schedules, device configurations, and robustness protocols appear in Appendix~\ref{app:eval_details}.

\section{Evaluation}
\label{sec:Evaluation}

\subsection{Accuracy Across Listener-Specific Reflex Thresholds}
\label{sec:eval:wbart}

\begin{figure}[!t]
  \centering
  \includegraphics[width=0.96\linewidth]{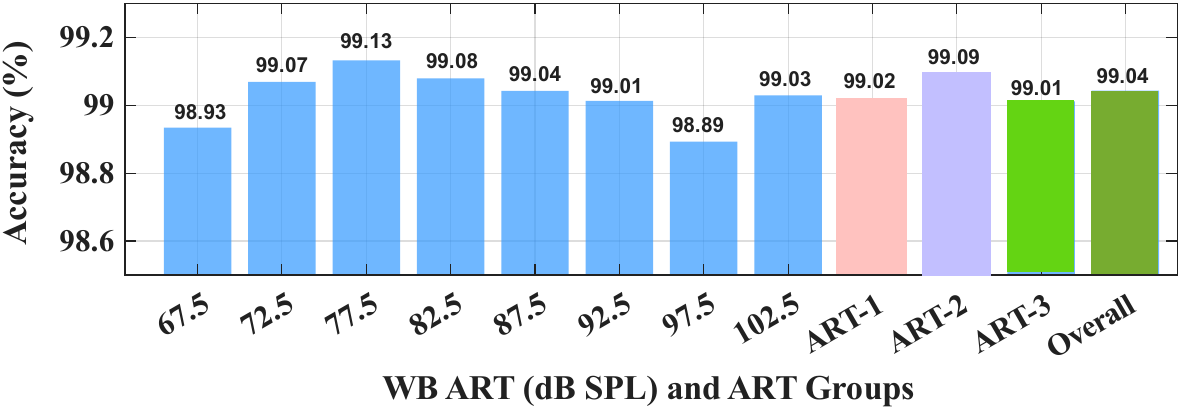}
  \caption{\earlog{} maintains high recognition accuracy across listener-specific WB-ART levels.}
  \Description{Recognition accuracy plotted across listener-specific wideband acoustic-reflex-threshold intervals.}
  \label{fig:overall_accuracy}
\end{figure}

We first verify that \earlog{} operates across listeners with different physiological protection thresholds. All results use leave-one-subject-out evaluation, so no window from the test participant contributes to supervised training. We group the 81 evaluation ears by wideband acoustic-reflex threshold (WB-ART): ART-1 spans 65--75~dB SPL (\(n=16\)), ART-2 spans 75--85~dB SPL (\(n=29\)), and ART-3 is above 85~dB SPL (\(n=36\)). Appendix~\ref{app:eval_participants} gives the cohort distribution.

Figure~\ref{fig:overall_accuracy} shows consistently high recognition accuracy across the measured WB-ART range. The clinically referenced benchmark contains 145,800 valid windows, including 7,883 true positives, 136,036 true negatives, 1,801 false positives, and 80 false negatives. These counts yield \textbf{99.0\% sensitivity, 98.7\% specificity, 81.4\% precision, 89.3\% F1, and 98.71\% accuracy}. The 95\% confidence intervals for sensitivity and specificity are \(\pm0.8\) percentage points.

The same participant-disjoint model therefore covers ears with substantially different reflex thresholds. This result shows that \earlog{} can log listener-specific AR events across a broad range of reflex thresholds rather than operating only within one narrow threshold range. Appendix~\ref{app:eval_wbart_content} provides the interval-wise and cross-content subgroup results.

\subsection{Label Efficiency of the Residual Representation}
\label{sec:eval:generic-baselines}

The clinical benchmark establishes detection accuracy, but it does not show whether a generic classifier could obtain the same result directly from raw spectra. We therefore train a support vector machine (SVM), a decision tree, and a lightweight deep neural network (DNN) under the same participant-disjoint supervision used by \earlog{}.

\begin{figure*}[!t]
  \centering
  \captionsetup[subfigure]{font=footnotesize,justification=centering}
  \begin{subfigure}[t]{0.245\textwidth}
    \vspace{0pt}
    \centering
    \includegraphics[height=25.5mm,keepaspectratio]{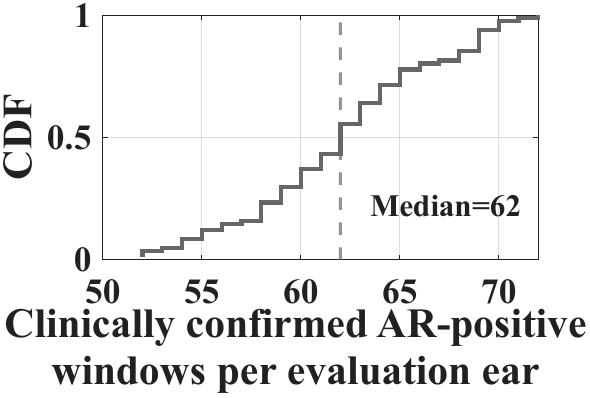}
    \caption{Positive-label availability.}
    \label{fig:eval:generic-baselines-a}
  \end{subfigure}\hfill\begin{subfigure}[t]{0.245\textwidth}
    \vspace{0pt}
    \centering
    \includegraphics[height=25.5mm,keepaspectratio]{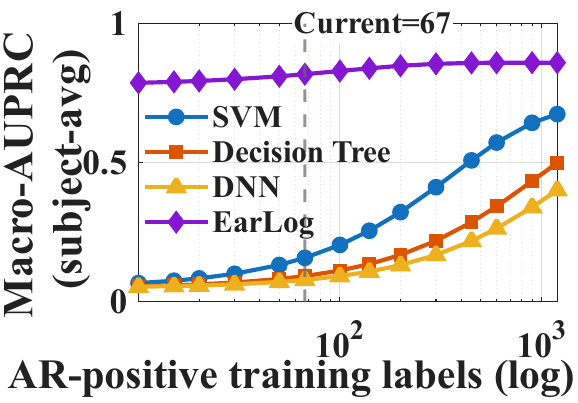}
    \caption{AUPRC versus label budget.}
    \label{fig:eval:generic-baselines-b}
  \end{subfigure}\hfill\begin{subfigure}[t]{0.245\textwidth}
    \vspace{0pt}
    \centering
    \includegraphics[height=25.5mm,keepaspectratio]{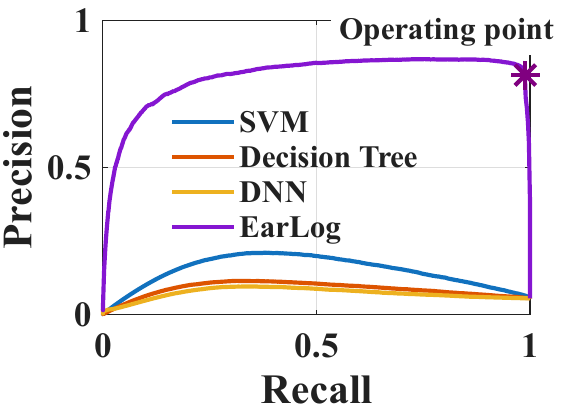}
    \caption{Pooled precision--recall.}
    \label{fig:eval:generic-baselines-c}
  \end{subfigure}\hfill\begin{subfigure}[t]{0.245\textwidth}
    \vspace{0pt}
    \centering
    \includegraphics[height=25.5mm,keepaspectratio]{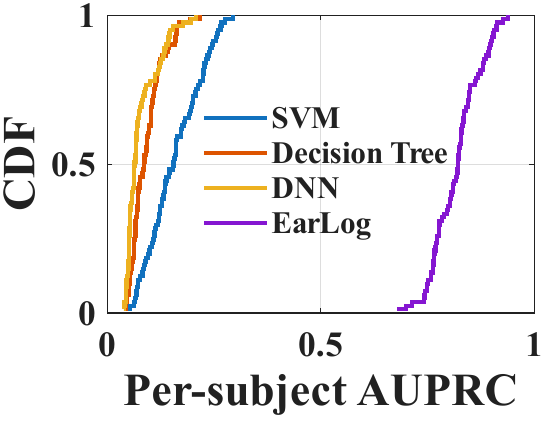}
    \caption{Per-participant AUPRC.}
    \label{fig:eval:generic-baselines-d}
  \end{subfigure}
  \caption{The residual representation improves label efficiency and participant-disjoint precision--recall performance over raw-spectrum classifiers.}
  \Description{Four panels show positive-label availability, AUPRC versus positive-label budget, pooled precision-recall curves, and the cumulative distribution of per-participant AUPRC for EarLog and three raw-spectrum classifiers.}
  \label{fig:eval:generic-baselines}
\end{figure*}

Ordinary playback provides abundant audio, but clinically confirmed AR-positive windows cannot be expanded in the same way. Repeatedly eliciting positive reflexes is constrained by participant safety and ethics. The raw spectral appearance of the same physiological event also changes with playback content, ear acoustics, headphone response, and fit. Under limited supervision, a raw-spectrum model can therefore rely on content or fit correlations. \earlog{} first removes the expected playback and in-ear response, so each clinical label describes a more focused residual difference.

Figure~\ref{fig:eval:generic-baselines-a} shows that the median availability of clinically confirmed positive windows is 62 per evaluation ear. Figure~\ref{fig:eval:generic-baselines-b} varies the AR-positive training-label budget and marks 67 labels as the current operating point.

{\setlength{\emergencystretch}{1em}
At the marked \earlog{} operating point in Fig.~\ref{fig:eval:generic-baselines-c}, sensitivity, specificity, precision, F1, and false-positive rate (FPR) are 98.9\%, 98.7\%, 81.4\%, 89.3\%, and 1.3\%, respectively. Its pooled AUPRC is 0.807, compared with 0.147 for SVM, 0.085 for the decision tree, and 0.071 for the DNN. Figure~\ref{fig:eval:generic-baselines-d} reports the cumulative distribution of per-participant AUPRC. Both views show that \earlog{} improves the precision--recall trade-off and shifts the participant distribution toward higher AUPRC. The advantage comes from the content-conditioned residual representation, which makes each clinically confirmed label more informative.\par}

\subsection{Ablation Study}
\label{sec:eval:controlled-ablations}

We next verify that the measured accuracy follows from the coupled sensing design. The first comparison replaces the sensing formulation or removes the outward reference microphone. The second removes complete processing blocks from the residual pipeline.

\begin{figure*}[!t]
  \centering

  \begin{minipage}[t]{0.25\textwidth}
    \vspace{0pt}
    \centering
    \includegraphics[width=\linewidth,height=27mm,keepaspectratio]{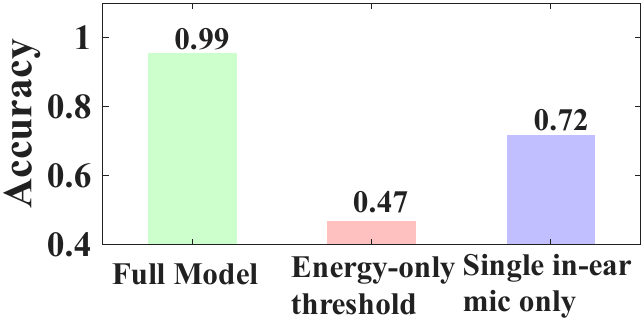}
    \captionsetup{justification=raggedright,singlelinecheck=false}
    \captionof{figure}{Full \earlog{}, energy-only, and single-microphone variants.}
    \Description{A three-bar chart comparing Full EarLog at approximately 0.99 accuracy, an energy-only threshold at 0.47, and a single-microphone EarLog variant at 0.72.}
    \label{fig:accuracy_wb}
  \end{minipage}\hfill
  \begin{minipage}[t]{0.47\textwidth}
    \vspace{0pt}
    \centering
    \scriptsize
    \setlength{\tabcolsep}{2.5pt}
    \renewcommand{\arraystretch}{1.08}
    \begin{tabularx}{\linewidth}{@{}Xrr@{}}
      \toprule
      \textbf{Variant} & \textbf{Acc. (\%)} & \textbf{Change (pp)}\\
      \midrule
      Full \earlog{} & 98.71 & 0.00\\
      Without physics-constrained no-AR baseline & 90.35 & -8.36\\
      Single-band LDS variant & 86.55 & -12.16\\
      Without roughness descriptors and gating & 92.20 & -6.51\\
      Without artifact and interference handling & 75.90 & -22.81\\
      \bottomrule
    \end{tabularx}
    \EarLogTableCaption{Each processing block addresses a distinct source of error in the residual pipeline.}{tab:ablation_results}
  \end{minipage}\hfill
  \begin{minipage}[t]{0.25\textwidth}
    \vspace{0pt}
    \centering
    \scriptsize
    \setlength{\tabcolsep}{2.5pt}
    \renewcommand{\arraystretch}{1.25}
    \begin{tabularx}{\linewidth}{@{}>{\raggedright\arraybackslash}Xr@{}}
      \toprule
      \textbf{Environment} & \textbf{Acc. (\%)}\\
      \midrule
      Quiet dormitory (45 dB) & 99.37\\
      Classroom (55 dB) & 97.78\\
      Office with talking (65 dB) & 96.21\\
      Outdoor road (70 dB) & 95.15\\
      \bottomrule
    \end{tabularx}
    \EarLogTableCaption{\earlog{} retains high accuracy as ambient noise increases.}{tab:environmental_impact}
  \end{minipage}
\end{figure*}

The energy-only baseline uses short-time in-ear energy and a fixed threshold. It does not predict the no-AR response or use LDS, pLDS, and roughness. Figure~\ref{fig:accuracy_wb} shows that this baseline reaches 47\% accuracy. The single-microphone variant retains the inward channel, no-AR baseline, and residual verifier but removes the outward environmental reference. It reaches 72\% accuracy. Full \earlog{} reaches 98.71\%. The comparison shows that the residual formulation handles dynamic playback, while the outward microphone enables leakage, environment, and window-validity checks.

Table~\ref{tab:ablation_results} isolates the internal design blocks. Removing artifact and interference handling reduces accuracy by 22.81 percentage points, showing the value of validity screening, clipping recovery, and selective denoising on commodity hardware. Replacing multiband pLDS with single-band LDS costs 12.16 percentage points, confirming the role of cross-band decoupling under broadband playback. Removing the physics-constrained no-AR baseline costs 8.36 percentage points, while removing roughness descriptors costs 6.51 percentage points. Together, the four ablations isolate content variation, cross-band coupling, smooth drift, and real-world interference as distinct error sources.

\subsection{Generalization Across Everyday Conditions}
\label{sec:eval:realistic}

The preceding experiments establish the core detector and its design dependencies. We further examine whether the same frozen system continues to operate as the headphone, playback content, environment, fit, behavior, and recording day vary. The tests progress from external device and content variation, to active changes in the fitted acoustic path, and finally to seven-day reference drift.

\begin{figure*}[!t]
  \centering
  \captionsetup[subfigure]{font=footnotesize,justification=centering}
  \begin{subfigure}[t]{0.49\textwidth}
    \vspace{0pt}
    \centering
    \includegraphics[height=36mm,keepaspectratio]{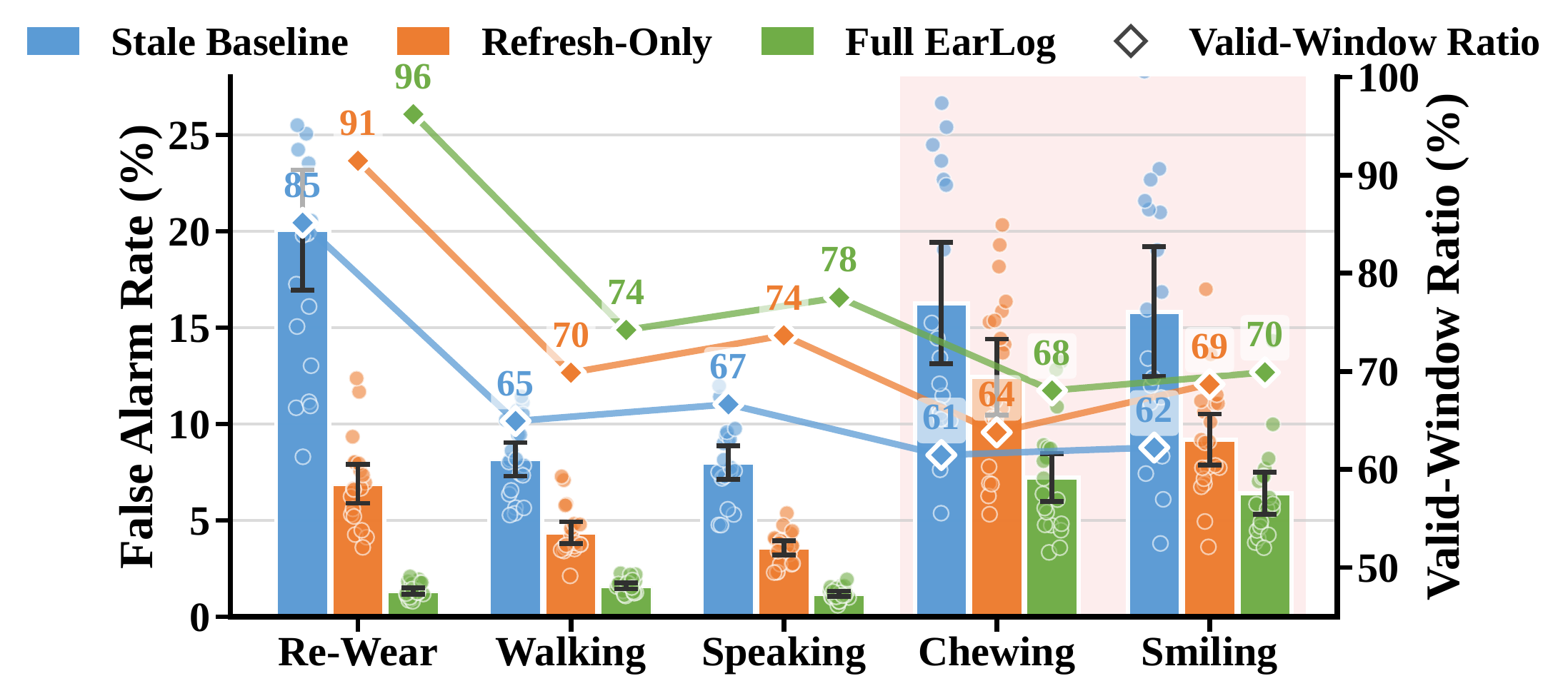}
    \caption{False-alarm rate and retained sensing coverage under refitting and everyday behaviors.}
    \label{fig:realistic_robustness_a}
  \end{subfigure}\hfill
  \begin{subfigure}[t]{0.49\textwidth}
    \vspace{0pt}
    \centering
    \includegraphics[height=36mm,keepaspectratio]{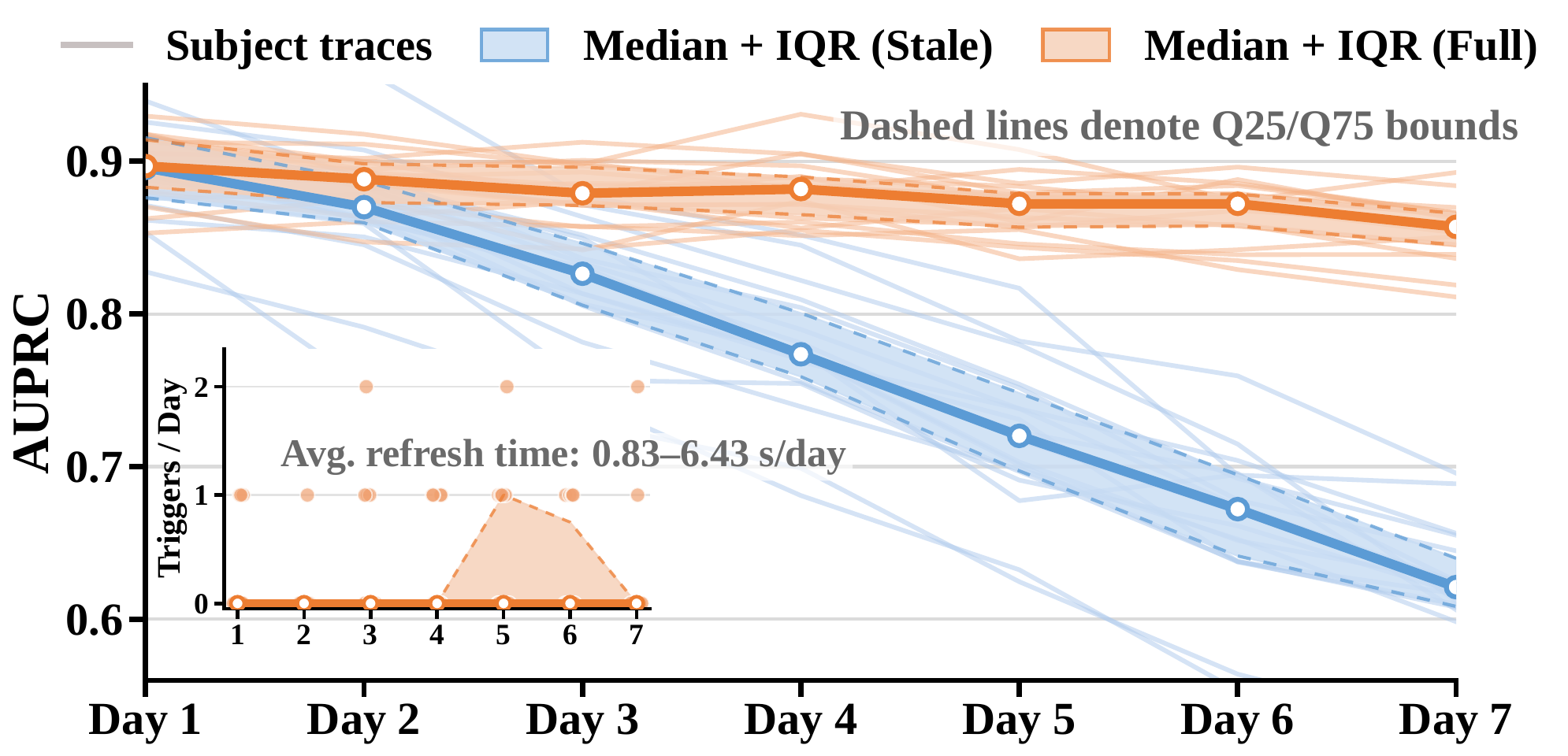}
    \caption{Participant-level reference stability across seven days.}
    \label{fig:realistic_robustness_b}
  \end{subfigure}
  \caption{Quality-aware robustness under refitting, everyday behaviors, and cross-day drift.}
  \Description{Panel a reports false-alarm rate and valid-window ratio for refitting, walking, speaking, chewing, and smiling. Panel b follows participant-level AUPRC trajectories for a stale baseline and the full EarLog reference-maintenance pipeline across seven days.}
  \label{fig:realistic_robustness}
\end{figure*}

\subsubsection{Devices, Ambient Noise, and Unseen Content}
\label{sec:eval:realistic-ood}

To verify hardware and content generalization without retraining, we keep the wideband-trained classifier fixed. We test 16 headphone models and four ambient-noise levels. Appendix~\ref{app:eval_wbart_content} extends the subgroup analysis to piano and instrument playback.

\begin{figure}[!tbp]
  \centering
  \includegraphics[width=0.7\linewidth]{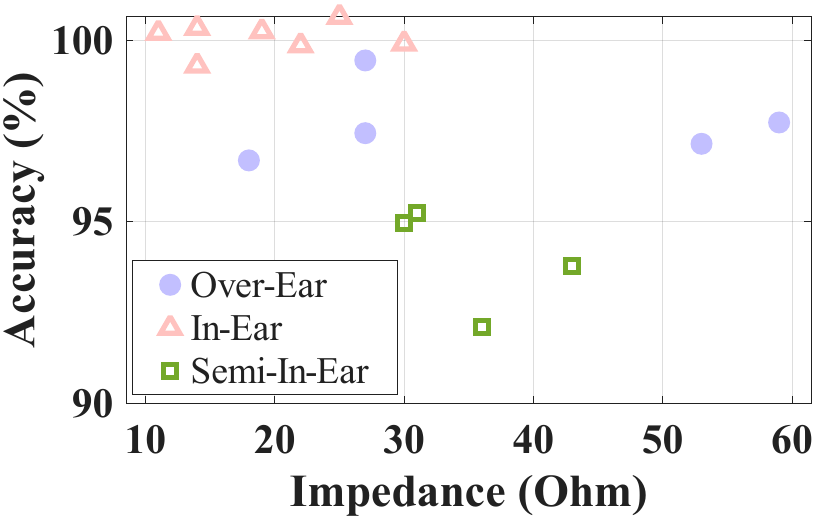}
  \caption{\earlog{} maintains high recognition accuracy across headphone form factors.}
  \Description{A bar chart comparing EarLog recognition accuracy for over-ear, in-ear, and semi-in-ear headphones.}
  \label{fig:different_headphone_impact}
\end{figure}

Across over-ear, in-ear, and semi-in-ear devices, \earlog{} reaches 97.0\%, 100.0\%, and 94.5\% accuracy, respectively, as shown in Fig.~\ref{fig:different_headphone_impact}. Table~\ref{tab:environmental_impact} shows that accuracy remains 95.15\% at 70~dB ambient noise and changes by only 4.22 percentage points across the tested 45--70~dB range. The outward reference, validity gate, and selective noise mitigation preserve a usable playback--response observation across these hardware and environmental conditions.

{\setlength{\emergencystretch}{1em}
Appendix~\ref{app:eval_wbart_content} combines wideband, piano, and instrument playback in a noisy environment. Without additional fine-tuning, accuracy reaches 95.9\% for ART-1 ears, ranges from 88.3\% to 92.5\% across the ART-2 transition range, and reaches 87.7\% for ART-3 ears, preserving the cross-threshold pattern observed in the clinically referenced benchmark.\par}

\subsubsection{Refitting and Everyday Behaviors}
\label{sec:eval:refit_behavior}

The preceding tests preserve the fitted headphone--ear interface. We next deliberately change that interface to verify whether fit and behavior are incorrectly written into the AR log.

Playback remains 10~dB below each ear's WB-ART, and synchronized Titan-inactive intervals define an AR-negative condition. Every positive output in this experiment is therefore a false alarm.

We compare a stale baseline, low-volume refresh alone, and the full calibration-transfer-plus-refresh pipeline. Figure~\ref{fig:realistic_robustness_a} reports false-alarm rate among valid windows together with the valid-window ratio. Reporting both quantities shows how effectively \earlog{} controls erroneous events while retaining sensing coverage.

After re-wearing, a stale baseline produces 20.05\% false alarms. Low-volume refresh reduces this value to 6.89\%, while the full pipeline reaches 1.33\% and retains 96.20\% of the recorded windows. Calibration transfer therefore aligns the new fit while preserving most of the listening period for sensing.

Walking and speaking retain 74.21\% and 77.52\% of recorded windows, with false-alarm rates of 1.60\% and 1.19\%. During chewing and smiling, \earlog{} retains 68.05\% and 69.90\% of windows and limits false alarms to 7.22\% and 6.39\%. The quality-aware output preserves roughly seven in ten observations even when facial motion changes both the seal and canal geometry. Appendix~\ref{app:eval_refit} gives the detailed protocol.

\subsubsection{Cross-Day Reference Maintenance}
\label{sec:eval:crossday_refresh}

{\setlength{\emergencystretch}{1em}
Window-level screening handles short disturbances. We next ask whether the same personalized baseline remains useful across days. Multiple participants repeat the experiment for seven days with the same evaluation ear and headphone model. We compare continued reuse of the Day-1 baseline with the full reference-maintenance pipeline.\par}

With the initial baseline reused unchanged, participant-level median AUPRC changes from 0.896 on Day~1 to 0.621 on Day~7. Full \earlog{} starts at 0.897 and reaches 0.857 on Day~7. Reference maintenance therefore provides a 0.236 AUPRC gain over the stale baseline by the end of the study.

Calibration transfer accounts for most of the observed cross-day stability. The median refresh count remains zero on every day, and the mean ranges from 0.06 to 0.39 triggers per day. Across days, accumulated low-volume refresh time ranges from 0.83 to 6.43~s per day. Each triggered refresh lasts approximately 13.8--16.5~s. The stored reference therefore remains useful through transfer, and refresh is needed only occasionally. Appendix~\ref{app:eval_refresh} gives the session protocol and participant-level aggregation.

\subsection{What the Event Log Reveals}
\label{sec:eval:event-log}

The previous experiments establish accuracy and robustness. We finally use \earlog{} in its intended logging mode to examine whether ordinary playback produces listener-specific recruitment patterns. We keep the classifier frozen, play music unilaterally through the target headphone, and monitor the contralateral ear with Titan. This design provides an independent clinical reference while leaving the headphone-side acoustic observation unchanged.

\begin{figure}[!tbp]
  \centering
  \includegraphics[width=0.90\linewidth]{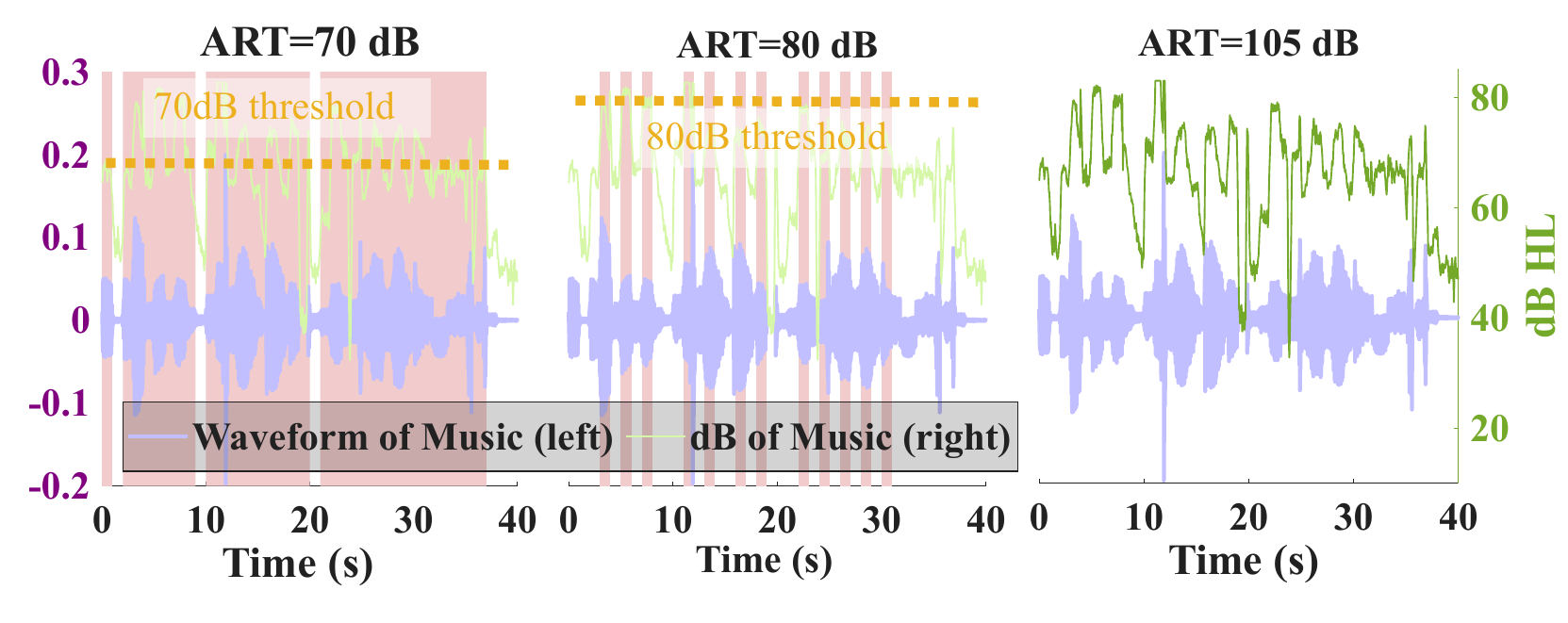}
  \captionsetup{justification=raggedright,singlelinecheck=false}
  \caption{Titan-confirmed AR-event logs reveal listener-specific recruitment under the same device-recommended playback setting.}
  \Description{Representative listening-time logs for one evaluation ear from each WB-ART group. Red intervals mark EarLog detections independently confirmed by synchronized contralateral Titan monitoring.}
  \label{fig:performance_music}
\end{figure}

Figure~\ref{fig:performance_music} shows one representative evaluation ear from each WB-ART range. Every red interval is an \earlog{} detection that overlaps a synchronized contralateral Titan-active interval. Confirmed events cluster around transient high-energy and spectrally intense passages. The same device-recommended playback setting produces different recruitment densities across listeners. Lower-threshold ears show denser confirmed intervals, while higher-threshold ears show sparser recruitment under the same playback regime.

\earlog{} aggregates these momentary events against valid listening time and co-reports sensing coverage. The resulting log describes how each ear recruits protection across content and sessions, revealing listener-specific patterns that remain invisible to a common device setting. Section~\ref{sec:setup} defines the reference-event protocol, and Appendix~\ref{app:eval_details} gives the participant, stimulus, and deployment details.

\section{Conclusion}  \label{sec:Conclusion}

This paper presents \earlog{}, which enables commodity headphones to record acoustic-reflex events without interrupting everyday listening. The challenge is that AR is weak and short-lived, whereas nonstationary content, individual ear acoustics, device response, fit, and environmental disturbances introduce stronger variations in the in-ear observation. \earlog{} turns playback into an opportunistic probe and uses content-conditioned residual sensing to suppress these dominant variations. Resonance-constrained short-window modeling, dual-microphone observations, and cross-session reference updates allow the system to run in real time on commodity hardware. We evaluate \earlog{} with 81 participants, 16 headphone models, and more than 400 hours of recordings. \earlog{} achieves 99.0\% sensitivity and 98.7\% specificity on a clinically referenced benchmark and remains usable across content, headphone models, everyday behaviors, refitting, and cross-day conditions. The resulting logs also show that the same audio played at the same nominally safe volume can produce markedly different AR recruitment patterns across listeners. These listener-specific recruitment logs can support long-term risk observation and future personalized safe-listening strategies.

\begin{acks}
This work was supported by the Guangdong Provincial Key Laboratory of Integrated Communication, Sensing and Computation for Ubiquitous Internet of Things (No. 2023B\allowbreak{}1212\allowbreak{}010007), the National Natural Science Foundation of China (Grant No. 62472366), the Department of Education of Guangdong Province (Nos. 2023\allowbreak{}KCXTD\allowbreak{}042, 2024GCZX\allowbreak{}003, 2024\allowbreak{}KCXTD\allowbreak{}008, and 2025KCXTD\allowbreak{}056), the 111 Center (No. D25008), and the Shenzhen Science and Technology Foundation (No. ZDSYS\allowbreak{}20190902\allowbreak{}092853047).
\end{acks}

\bibliographystyle{ACM-Reference-Format}
\nocite{huang2021vi,huang2021li,huang2021portable,huang2023beverage,huang2025reconstructing,zhao2024chameleon,li2025fedwcm,li2025fedwmsam,dai2018mmhumidity,huang2018oinput,huang2026earlog,li2025fedwmsam_arxiv,li2025fedwcm_arxiv,huang2026smartphone_arxiv,xie2024litecrypt,huang_patent_sensory_substitution_2022,huang_patent_lora_isac_2025,huang_patent_multikeyboard_2025,huang_patent_capillary_2022,huang_patent_bone_input_2021,huang_patent_humidity_2020,huang_patent_wifi_fire_2018,huang_patent_glasses_2025,huang_patent_action_recognition_2021,huang_patent_assistive_eyeglasses_2023}
\bibliography{EarLog,EarLog_Revision_Additions,EarLog_arXiv_Additions}


\begin{thebibliography}{75}

%%% except for \shownote{} and \showURL{}.  The latter does
%%% not use final punctuation, in order to avoid confusing it with
%%% the Web address.

\ifx \showCODEN    \undefined \def \showCODEN     #1{\unskip}     \fi
\ifx \showISBNx    \undefined \def \showISBNx     #1{\unskip}     \fi
\ifx \showISBNxiii \undefined \def \showISBNxiii  #1{\unskip}     \fi
\ifx \showISSN     \undefined \def \showISSN      #1{\unskip}     \fi
\ifx \showLCCN     \undefined \def \showLCCN      #1{\unskip}     \fi
\ifx \shownote     \undefined \def \shownote      #1{#1}          \fi
\ifx \showarticletitle \undefined \def \showarticletitle #1{#1}   \fi
\ifx \showURL      \undefined \def \showURL       {\relax}        \fi
% invisible to TeX
\providecommand\bibfield[2]{#2}
\providecommand\bibinfo[2]{#2}
\providecommand\natexlab[1]{#1}
\providecommand\showeprint[2][]{arXiv:#2}
\makeatletter
\@ifundefined{NAT@parse@date}{}{\let\NAT@parse@date@orig\NAT@parse@date}
\@ifundefined{NAT@parse@date}{}{\def\NAT@parse@date#1#2#3#4#5#6@@{\NAT@parse@date@orig#1#2#3#4#5#6@@\def\NAT@tempyear{0000}\def\NAT@tempexlab{{?}}\ifx\NAT@year\NAT@tempyear\ifx\NAT@exlab\NAT@tempexlab\def\NAT@date{[n.\,d.]}\else\edef\NAT@date{[n.\,d.]\NAT@exlab}\fi\fi}}
\makeatother

\bibitem[Baricevich et~al\mbox{.}(2025)]%
        {baricevich2025frequency}
\bibfield{author}{\bibinfo{person}{Abbie Baricevich}, \bibinfo{person}{Danielle
  Bassett}, \bibinfo{person}{Sophia Chan}, \bibinfo{person}{Shayna Lavi}, {and}
  \bibinfo{person}{Jonathan Siegel}.} \bibinfo{year}{2025}\natexlab{}.
\newblock \showarticletitle{Frequency and level dependence of the middle ear
  acoustic reflex and its decay measured in wideband absorbance with
  contralateral narrowband noise elicitors}.
\newblock \bibinfo{journal}{\emph{Hearing Research}}  \bibinfo{volume}{459}
  (\bibinfo{date}{April} \bibinfo{year}{2025}), \bibinfo{pages}{109225}.
\newblock
\href{https://doi.org/10.1016/j.heares.2025.109225}{doi:\nolinkurl{10.1016/j.heares.2025.109225}}


\bibitem[Borg(1982)]%
        {Borg1982_timecourse}
\bibfield{author}{\bibinfo{person}{Erik Borg}.}
  \bibinfo{year}{1982}\natexlab{}.
\newblock \showarticletitle{Time Course of the Human Acoustic Stapedius Reflex:
  A Comparison of Eight Different Measures in Normal-Hearing Subjects}.
\newblock \bibinfo{journal}{\emph{Scandinavian Audiology}}
  \bibinfo{volume}{11}, \bibinfo{number}{4} (\bibinfo{year}{1982}),
  \bibinfo{pages}{237--242}.
\newblock
\href{https://doi.org/10.3109/01050398209087473}{doi:\nolinkurl{10.3109/01050398209087473}}


\bibitem[Butkow et~al\mbox{.}(2023)]%
        {butkow2023heart}
\bibfield{author}{\bibinfo{person}{Kayla-Jade Butkow}, \bibinfo{person}{Ting
  Dang}, \bibinfo{person}{Andrea Ferlini}, \bibinfo{person}{Dong Ma}, {and}
  \bibinfo{person}{Cecilia Mascolo}.} \bibinfo{year}{2023}\natexlab{}.
\newblock \showarticletitle{{hEARt}: Motion-resilient Heart Rate Monitoring
  with In-ear Microphones}. In \bibinfo{booktitle}{\emph{2023 IEEE
  International Conference on Pervasive Computing and Communications (PerCom)}}
  (Atlanta, GA, USA). \bibinfo{publisher}{IEEE}, \bibinfo{address}{Piscataway,
  NJ, USA}, \bibinfo{pages}{200--209}.
\newblock
\href{https://doi.org/10.1109/PERCOM56429.2023.10099317}{doi:\nolinkurl{10.1109/PERCOM56429.2023.10099317}}


\bibitem[Cai et~al\mbox{.}(2023a)]%
        {cai2023asdp}
\bibfield{author}{\bibinfo{person}{Chao Cai}, \bibinfo{person}{Henglin Pu},
  \bibinfo{person}{Menglan Hu}, \bibinfo{person}{Rong Zheng}, {and}
  \bibinfo{person}{Jun Luo}.} \bibinfo{year}{2023}\natexlab{a}.
\newblock \showarticletitle{Acoustic Software Defined Platform: A Versatile
  Sensing and General Benchmarking Platform}.
\newblock \bibinfo{journal}{\emph{IEEE Transactions on Mobile Computing}}
  \bibinfo{volume}{22}, \bibinfo{number}{2} (\bibinfo{date}{Feb.}
  \bibinfo{year}{2023}), \bibinfo{pages}{647--660}.
\newblock
\href{https://doi.org/10.1109/TMC.2021.3093259}{doi:\nolinkurl{10.1109/TMC.2021.3093259}}


\bibitem[Cai et~al\mbox{.}(2023b)]%
        {cai2023acute}
\bibfield{author}{\bibinfo{person}{Chao Cai}, \bibinfo{person}{Henglin Pu},
  \bibinfo{person}{Liyuan Ye}, \bibinfo{person}{Hongbo Jiang}, {and}
  \bibinfo{person}{Jun Luo}.} \bibinfo{year}{2023}\natexlab{b}.
\newblock \showarticletitle{Active Acoustic Sensing for ``{Hearing}''
  Temperature Under Acoustic Interference}.
\newblock \bibinfo{journal}{\emph{IEEE Transactions on Mobile Computing}}
  \bibinfo{volume}{22}, \bibinfo{number}{2} (\bibinfo{date}{Feb.}
  \bibinfo{year}{2023}), \bibinfo{pages}{661--673}.
\newblock
\href{https://doi.org/10.1109/TMC.2021.3096792}{doi:\nolinkurl{10.1109/TMC.2021.3096792}}


\bibitem[Cao et~al\mbox{.}(2025a)]%
        {cao2025heartprint}
\bibfield{author}{\bibinfo{person}{Yetong Cao}, \bibinfo{person}{Chao Cai},
  \bibinfo{person}{Fan Li}, \bibinfo{person}{Zhe Chen}, {and}
  \bibinfo{person}{Jun Luo}.} \bibinfo{year}{2025}\natexlab{a}.
\newblock \showarticletitle{Enabling Passive User Authentication via Heart
  Sounds on In-Ear Microphones}.
\newblock \bibinfo{journal}{\emph{IEEE Transactions on Dependable and Secure
  Computing}} \bibinfo{volume}{22}, \bibinfo{number}{2} (\bibinfo{year}{2025}),
  \bibinfo{pages}{1195--1209}.
\newblock
\href{https://doi.org/10.1109/TDSC.2024.3429574}{doi:\nolinkurl{10.1109/TDSC.2024.3429574}}


\bibitem[Cao et~al\mbox{.}(2023)]%
        {cao2023earace}
\bibfield{author}{\bibinfo{person}{Yetong Cao}, \bibinfo{person}{Chao Cai},
  \bibinfo{person}{Anbo Yu}, \bibinfo{person}{Fan Li}, {and}
  \bibinfo{person}{Jun Luo}.} \bibinfo{year}{2023}\natexlab{}.
\newblock \showarticletitle{{EarAcE}: Empowering Versatile Acoustic Sensing via
  Earable Active Noise Cancellation Platform}.
\newblock \bibinfo{journal}{\emph{Proceedings of the ACM on Interactive,
  Mobile, Wearable and Ubiquitous Technologies}} \bibinfo{volume}{7},
  \bibinfo{number}{2} (\bibinfo{date}{June} \bibinfo{year}{2023}),
  \bibinfo{pages}{47:1--47:23}.
\newblock
\href{https://doi.org/10.1145/3596242}{doi:\nolinkurl{10.1145/3596242}}


\bibitem[Cao et~al\mbox{.}(2025b)]%
        {cao2025espiro}
\bibfield{author}{\bibinfo{person}{Yetong Cao}, \bibinfo{person}{Dong Ma},
  \bibinfo{person}{Wentao Xie}, \bibinfo{person}{Qian Zhang}, {and}
  \bibinfo{person}{Jun Luo}.} \bibinfo{year}{2025}\natexlab{b}.
\newblock \showarticletitle{{ESPIRO}: Natural Pulmonary Function Monitoring via
  Earphone-Acquired Speech}. In \bibinfo{booktitle}{\emph{Proceedings of the
  31st Annual International Conference on Mobile Computing and Networking}}.
  \bibinfo{publisher}{Association for Computing Machinery},
  \bibinfo{address}{New York, NY, USA}, \bibinfo{pages}{329--344}.
\newblock
\href{https://doi.org/10.1145/3680207.3723477}{doi:\nolinkurl{10.1145/3680207.3723477}}


\bibitem[Causon et~al\mbox{.}(2020)]%
        {Causon2020_AR}
\bibfield{author}{\bibinfo{person}{Andrew Causon}, \bibinfo{person}{Kevin~J.
  Munro}, \bibinfo{person}{Christopher~J. Plack}, {and}
  \bibinfo{person}{Garreth Prendergast}.} \bibinfo{year}{2020}\natexlab{}.
\newblock \showarticletitle{The Role of the Clinically Obtained Acoustic Reflex
  as a Research Tool for Subclinical Hearing Pathologies}.
\newblock \bibinfo{journal}{\emph{Trends in Hearing}}  \bibinfo{volume}{24}
  (\bibinfo{year}{2020}), \bibinfo{pages}{2331216520972860}.
\newblock
\href{https://doi.org/10.1177/2331216520972860}{doi:\nolinkurl{10.1177/2331216520972860}}
\newblock
\shownote{BBN elicitors tested at 63–88 dB SPL; AR threshold/growth
  analyses}.


\bibitem[Chan et~al\mbox{.}(2022a)]%
        {chan2022off}
\bibfield{author}{\bibinfo{person}{Justin Chan}, \bibinfo{person}{Nada Ali},
  \bibinfo{person}{Ali Najafi}, \bibinfo{person}{Anna Meehan},
  \bibinfo{person}{Lisa~R Mancl}, \bibinfo{person}{Emily Gallagher},
  \bibinfo{person}{Randall Bly}, {and} \bibinfo{person}{Shyamnath Gollakota}.}
  \bibinfo{year}{2022}\natexlab{a}.
\newblock \showarticletitle{An off-the-shelf otoacoustic-emission probe for
  hearing screening via a smartphone}.
\newblock \bibinfo{journal}{\emph{Nature biomedical engineering}}
  \bibinfo{volume}{6}, \bibinfo{number}{11} (\bibinfo{year}{2022}),
  \bibinfo{pages}{1203--1213}.
\newblock


\bibitem[Chan et~al\mbox{.}(2023)]%
        {chan2023wireless}
\bibfield{author}{\bibinfo{person}{Justin Chan}, \bibinfo{person}{Antonio
  Glenn}, \bibinfo{person}{Malek Itani}, \bibinfo{person}{Lisa~R. Mancl},
  \bibinfo{person}{Emily Gallagher}, \bibinfo{person}{Randall Bly},
  \bibinfo{person}{Shwetak Patel}, {and} \bibinfo{person}{Shyamnath
  Gollakota}.} \bibinfo{year}{2023}\natexlab{}.
\newblock \showarticletitle{Wireless earbuds for low-cost hearing screening}.
  In \bibinfo{booktitle}{\emph{Proceedings of the 21st Annual International
  Conference on Mobile Systems, Applications and Services}}.
  \bibinfo{publisher}{Association for Computing Machinery},
  \bibinfo{address}{New York, NY, USA}, \bibinfo{pages}{84--95}.
\newblock
\href{https://doi.org/10.1145/3581791.3596856}{doi:\nolinkurl{10.1145/3581791.3596856}}


\bibitem[Chan and Gollakota(2022)]%
        {chan2022cochlea}
\bibfield{author}{\bibinfo{person}{Justin Chan} {and}
  \bibinfo{person}{Shyamnath Gollakota}.} \bibinfo{year}{2022}\natexlab{}.
\newblock \showarticletitle{Inner-ear cochlea testing with earphones}. In
  \bibinfo{booktitle}{\emph{Proceedings of the 20th Annual International
  Conference on Mobile Systems, Applications and Services}}.
  \bibinfo{publisher}{Association for Computing Machinery},
  \bibinfo{address}{New York, NY, USA}, \bibinfo{pages}{609--610}.
\newblock
\href{https://doi.org/10.1145/3498361.3538657}{doi:\nolinkurl{10.1145/3498361.3538657}}


\bibitem[Chan et~al\mbox{.}(2022b)]%
        {chan2022performing}
\bibfield{author}{\bibinfo{person}{Justin Chan}, \bibinfo{person}{Ali Najafi},
  \bibinfo{person}{Mallory Baker}, \bibinfo{person}{Julie Kinsman},
  \bibinfo{person}{Lisa~R Mancl}, \bibinfo{person}{Susan Norton},
  \bibinfo{person}{Randall Bly}, {and} \bibinfo{person}{Shyamnath Gollakota}.}
  \bibinfo{year}{2022}\natexlab{b}.
\newblock \showarticletitle{Performing tympanometry using smartphones}.
\newblock \bibinfo{journal}{\emph{Communications medicine}}
  \bibinfo{volume}{2}, \bibinfo{number}{1} (\bibinfo{year}{2022}),
  \bibinfo{pages}{57}.
\newblock


\bibitem[Chen et~al\mbox{.}(2023)]%
        {chen2023hearing}
\bibfield{author}{\bibinfo{person}{Fei Chen}, \bibinfo{person}{Hui Xue},
  \bibinfo{person}{Meng Wang}, \bibinfo{person}{Zhiling Cai}, {and}
  \bibinfo{person}{Shipeng Zhu}.} \bibinfo{year}{2023}\natexlab{}.
\newblock \showarticletitle{Hearing care: safe listening method and system for
  personal listening devices}.
\newblock \bibinfo{journal}{\emph{International Journal of Environmental
  Research and Public Health}} \bibinfo{volume}{20}, \bibinfo{number}{3}
  (\bibinfo{year}{2023}), \bibinfo{pages}{2161}.
\newblock


\bibitem[Chen et~al\mbox{.}(2024)]%
        {chen2024exploring}
\bibfield{author}{\bibinfo{person}{Tao Chen}, \bibinfo{person}{Yongjie Yang},
  \bibinfo{person}{Xiaoran Fan}, \bibinfo{person}{Xiuzhen Guo},
  \bibinfo{person}{Jie Xiong}, {and} \bibinfo{person}{Longfei Shangguan}.}
  \bibinfo{year}{2024}\natexlab{}.
\newblock \showarticletitle{Exploring the Feasibility of Remote Cardiac
  Auscultation Using Earphones}. In \bibinfo{booktitle}{\emph{Proceedings of
  the 30th Annual International Conference on Mobile Computing and
  Networking}}. \bibinfo{publisher}{Association for Computing Machinery},
  \bibinfo{address}{New York, NY, USA}, \bibinfo{pages}{357--372}.
\newblock
\href{https://doi.org/10.1145/3636534.3649366}{doi:\nolinkurl{10.1145/3636534.3649366}}


\bibitem[Dai et~al\mbox{.}(2018)]%
        {dai2018mmhumidity}
\bibfield{author}{\bibinfo{person}{Qinglang Dai}, \bibinfo{person}{Yongzhi
  Huang}, \bibinfo{person}{Lu Wang}, \bibinfo{person}{Rukhsana Ruby}, {and}
  \bibinfo{person}{Kaishun Wu}.} \bibinfo{year}{2018}\natexlab{}.
\newblock \showarticletitle{{mm-Humidity}: Fine-Grained Humidity Sensing with
  Millimeter Wave Signals}. In \bibinfo{booktitle}{\emph{2018 IEEE 24th
  International Conference on Parallel and Distributed Systems (ICPADS)}}.
  \bibinfo{publisher}{IEEE}, \bibinfo{pages}{204--211}.
\newblock
\href{https://doi.org/10.1109/PADSW.2018.8644907}{doi:\nolinkurl{10.1109/PADSW.2018.8644907}}


\bibitem[Fan et~al\mbox{.}(2023)]%
        {fan2023apg}
\bibfield{author}{\bibinfo{person}{Xiaoran Fan}, \bibinfo{person}{David Pearl},
  \bibinfo{person}{Richard Howard}, \bibinfo{person}{Longfei Shangguan}, {and}
  \bibinfo{person}{Trausti Thormundsson}.} \bibinfo{year}{2023}\natexlab{}.
\newblock \showarticletitle{{APG}: Audioplethysmography for Cardiac Monitoring
  in Hearables}. In \bibinfo{booktitle}{\emph{Proceedings of the 29th Annual
  International Conference on Mobile Computing and Networking}}.
  \bibinfo{publisher}{Association for Computing Machinery},
  \bibinfo{address}{New York, NY, USA}, Article \bibinfo{articleno}{67},
  \bibinfo{numpages}{15}~pages.
\newblock
\href{https://doi.org/10.1145/3570361.3613281}{doi:\nolinkurl{10.1145/3570361.3613281}}


\bibitem[Fan et~al\mbox{.}(2021)]%
        {fan2021headfi}
\bibfield{author}{\bibinfo{person}{Xiaoran Fan}, \bibinfo{person}{Longfei
  Shangguan}, \bibinfo{person}{Siddharth Rupavatharam},
  \bibinfo{person}{Yanyong Zhang}, \bibinfo{person}{Jie Xiong},
  \bibinfo{person}{Yunfei Ma}, {and} \bibinfo{person}{Richard Howard}.}
  \bibinfo{year}{2021}\natexlab{}.
\newblock \showarticletitle{{HeadFi}: bringing intelligence to all headphones}.
  In \bibinfo{booktitle}{\emph{Proceedings of the 27th Annual International
  Conference on Mobile Computing and Networking}}.
  \bibinfo{publisher}{Association for Computing Machinery},
  \bibinfo{address}{New York, NY, USA}, \bibinfo{pages}{147--159}.
\newblock
\href{https://doi.org/10.1145/3447993.3448624}{doi:\nolinkurl{10.1145/3447993.3448624}}


\bibitem[Feeney(2023)]%
        {Feeney2023_WBART}
\bibfield{author}{\bibinfo{person}{M.~Patrick Feeney}.}
  \bibinfo{year}{2023}\natexlab{}.
\newblock \showarticletitle{Wideband Acoustic Reflex Measurement}.
\newblock \bibinfo{journal}{\emph{Seminars in Hearing}} \bibinfo{volume}{44},
  \bibinfo{number}{1} (\bibinfo{year}{2023}), \bibinfo{pages}{84--92}.
\newblock
\href{https://doi.org/10.1055/s-0042-1758752}{doi:\nolinkurl{10.1055/s-0042-1758752}}


\bibitem[Feeney et~al\mbox{.}(2017a)]%
        {Feeney2017_WAI_Norms}
\bibfield{author}{\bibinfo{person}{M.~P. Feeney} {et~al\mbox{.}}}
  \bibinfo{year}{2017}\natexlab{a}.
\newblock \showarticletitle{Normative Wideband Reflectance, Equivalent
  Admittance at the Tympanic Membrane, and Acoustic Stapedius Reflex Threshold
  in Adults}.
\newblock \bibinfo{journal}{\emph{Ear and Hearing}}  \bibinfo{volume}{38}
  (\bibinfo{year}{2017}), \bibinfo{pages}{e84--e98}.
\newblock
\href{https://doi.org/10.1097/AUD.0000000000000399}{doi:\nolinkurl{10.1097/AUD.0000000000000399}}


\bibitem[Feeney et~al\mbox{.}(2017b)]%
        {feeney2017normative}
\bibfield{author}{\bibinfo{person}{M~Patrick Feeney},
  \bibinfo{person}{Douglas~H Keefe}, \bibinfo{person}{Lisa~L Hunter},
  \bibinfo{person}{Denis~F Fitzpatrick}, \bibinfo{person}{Angela~C Garinis},
  \bibinfo{person}{Daniel~B Putterman}, {and} \bibinfo{person}{Garnett~P
  McMillan}.} \bibinfo{year}{2017}\natexlab{b}.
\newblock \showarticletitle{Normative wideband reflectance, equivalent
  admittance at the tympanic membrane, and acoustic stapedius reflex threshold
  in adults}.
\newblock \bibinfo{journal}{\emph{Ear and hearing}} \bibinfo{volume}{38},
  \bibinfo{number}{3} (\bibinfo{year}{2017}), \bibinfo{pages}{e142--e160}.
\newblock


\bibitem[Fernandez et~al\mbox{.}(2015)]%
        {Fernandez2015_AgingAfterNoise}
\bibfield{author}{\bibinfo{person}{Katharine~A. Fernandez},
  \bibinfo{person}{Penelope W.~C. Jeffers}, \bibinfo{person}{Kumud Lall},
  \bibinfo{person}{M.~Charles Liberman}, {and} \bibinfo{person}{Sharon~G.
  Kujawa}.} \bibinfo{year}{2015}\natexlab{}.
\newblock \showarticletitle{Aging after Noise Exposure: Acceleration of
  Cochlear Synaptopathy in "Recovered" Ears}.
\newblock \bibinfo{journal}{\emph{Journal of Neuroscience}}
  \bibinfo{volume}{35}, \bibinfo{number}{19} (\bibinfo{year}{2015}),
  \bibinfo{pages}{7509--7520}.
\newblock
\href{https://doi.org/10.1523/JNEUROSCI.5138-14.2015}{doi:\nolinkurl{10.1523/JNEUROSCI.5138-14.2015}}


\bibitem[Goel et~al\mbox{.}(2016)]%
        {goel2016spirocall}
\bibfield{author}{\bibinfo{person}{Mayank Goel}, \bibinfo{person}{Elliot Saba},
  \bibinfo{person}{Maia Stiber}, \bibinfo{person}{Eric Whitmire},
  \bibinfo{person}{Josh Fromm}, \bibinfo{person}{Eric~C. Larson},
  \bibinfo{person}{Gaetano Borriello}, {and} \bibinfo{person}{Shwetak~N.
  Patel}.} \bibinfo{year}{2016}\natexlab{}.
\newblock \showarticletitle{{SpiroCall}: Measuring Lung Function over a Phone
  Call}. In \bibinfo{booktitle}{\emph{Proceedings of the 2016 CHI Conference on
  Human Factors in Computing Systems}}. \bibinfo{publisher}{Association for
  Computing Machinery}, \bibinfo{address}{New York, NY, USA},
  \bibinfo{pages}{5675--5685}.
\newblock
\href{https://doi.org/10.1145/2858036.2858401}{doi:\nolinkurl{10.1145/2858036.2858401}}


\bibitem[Gorga and Stelmachowicz(1983)]%
        {Gorga1983_temporal}
\bibfield{author}{\bibinfo{person}{Michael~P. Gorga} {and}
  \bibinfo{person}{Patricia~G. Stelmachowicz}.}
  \bibinfo{year}{1983}\natexlab{}.
\newblock \showarticletitle{Temporal Characteristics of the Acoustic Reflex}.
\newblock \bibinfo{journal}{\emph{Audiology}} \bibinfo{volume}{22},
  \bibinfo{number}{2} (\bibinfo{year}{1983}), \bibinfo{pages}{120--127}.
\newblock
\href{https://doi.org/10.3109/00206098309072774}{doi:\nolinkurl{10.3109/00206098309072774}}


\bibitem[GRAS(2020a)]%
        {EarSimulator}
\bibfield{author}{\bibinfo{person}{GRAS}.} \bibinfo{year}{2020}\natexlab{a}.
\newblock \bibinfo{title}{GRAS 43AC-S1 CCP Ear Simulator Kit According to IEC
  60318-4}.
\newblock
\newblock
\shownote{\url{https://www.grasacoustics.com/products/ear-simulator-kit/product/235-43ac-s1}}.


\bibitem[GRAS(2020b)]%
        {EarSimulator2}
\bibfield{author}{\bibinfo{person}{GRAS}.} \bibinfo{year}{2020}\natexlab{b}.
\newblock \bibinfo{title}{GRAS 45BB-15 KEMAR with Hi-Res Ear Simulators for
  Test of Ear}.
\newblock
\newblock
\shownote{\url{https://www.grasacoustics.com/products/head-torso-simulators-kemar/kemar-for-ear-headphone-test-2-ch/product/811-45bb-15}}.


\bibitem[Hill et~al\mbox{.}(2016)]%
        {Hill2016_HairCellSynaptopathy}
\bibfield{author}{\bibinfo{person}{Kayla Hill}, \bibinfo{person}{Hu Yuan},
  \bibinfo{person}{Xianren Wang}, {and} \bibinfo{person}{Su-Hua Sha}.}
  \bibinfo{year}{2016}\natexlab{}.
\newblock \showarticletitle{Noise-Induced Loss of Hair Cells and Cochlear
  Synaptopathy Are Mediated by the Activation of {AMPK}}.
\newblock \bibinfo{journal}{\emph{The Journal of Neuroscience}}
  \bibinfo{volume}{36}, \bibinfo{number}{28} (\bibinfo{year}{2016}),
  \bibinfo{pages}{7497--7510}.
\newblock
\href{https://doi.org/10.1523/JNEUROSCI.0782-16.2016}{doi:\nolinkurl{10.1523/JNEUROSCI.0782-16.2016}}


\bibitem[Huang(2023)]%
        {huang_patent_assistive_eyeglasses_2023}
\bibfield{author}{\bibinfo{person}{Yongzhi Huang}.}
  \bibinfo{year}{2023}\natexlab{}.
\newblock \bibinfo{title}{Sensory-substitution assistive eyeglasses for blind
  users}.
\newblock \bibinfo{howpublished}{Chinese Utility Model CN218979652U}.
\newblock
\urldef\tempurl%
\url{https://patents.google.com/patent/CN218979652U/zh}
\showURL{%
\tempurl}
\newblock
\shownote{In Chinese}.


\bibitem[Huang(2025)]%
        {huang_patent_glasses_2025}
\bibfield{author}{\bibinfo{person}{Yongzhi Huang}.}
  \bibinfo{year}{2025}\natexlab{}.
\newblock \bibinfo{title}{Glasses}.
\newblock \bibinfo{howpublished}{Chinese Utility Model CN222723306U}.
\newblock
\urldef\tempurl%
\url{https://patents.google.com/patent/CN222723306U/zh}
\showURL{%
\tempurl}
\newblock
\shownote{Granted April 8, 2025. In Chinese}.


\bibitem[Huang(2026)]%
        {huang2026smartphone_arxiv}
\bibfield{author}{\bibinfo{person}{Yongzhi Huang}.}
  \bibinfo{year}{2026}\natexlab{}.
\newblock \bibinfo{title}{Smartphone-Based Identification of Unknown Liquids
  via Active Vibration Sensing}.
\newblock
\showeprint[arxiv]{2603.28787}~[eess.SP]
\href{https://doi.org/10.48550/arXiv.2603.28787}{doi:\nolinkurl{10.48550/arXiv.2603.28787}}
\newblock
\shownote{arXiv preprint}.


\bibitem[Huang et~al\mbox{.}(2018)]%
        {huang2018oinput}
\bibfield{author}{\bibinfo{person}{Yongzhi Huang}, \bibinfo{person}{Shaotian
  Cai}, \bibinfo{person}{Lu Wang}, {and} \bibinfo{person}{Kaishun Wu}.}
  \bibinfo{year}{2018}\natexlab{}.
\newblock \showarticletitle{{Oinput}: A Bone-Conductive {QWERTY} Keyboard
  Recognition for Wearable Device}. In \bibinfo{booktitle}{\emph{2018 IEEE 24th
  International Conference on Parallel and Distributed Systems (ICPADS)}}.
  \bibinfo{publisher}{IEEE}, \bibinfo{pages}{946--953}.
\newblock
\href{https://doi.org/10.1109/PADSW.2018.8644590}{doi:\nolinkurl{10.1109/PADSW.2018.8644590}}


\bibitem[Huang et~al\mbox{.}(2021a)]%
        {huang2021vi}
\bibfield{author}{\bibinfo{person}{Yongzhi Huang}, \bibinfo{person}{Kaixin
  Chen}, \bibinfo{person}{Yandao Huang}, \bibinfo{person}{Lu Wang}, {and}
  \bibinfo{person}{Kaishun Wu}.} \bibinfo{year}{2021}\natexlab{a}.
\newblock \showarticletitle{Vi-liquid: unknown liquid identification with your
  smartphone vibration}. In \bibinfo{booktitle}{\emph{Proceedings of the 27th
  Annual International Conference on Mobile Computing and Networking}} (New
  Orleans, Louisiana) \emph{(\bibinfo{series}{MobiCom '21})}.
  \bibinfo{publisher}{Association for Computing Machinery},
  \bibinfo{address}{New York, NY, USA}, \bibinfo{pages}{174--187}.
\newblock
\showISBNx{9781450383424}
\href{https://doi.org/10.1145/3447993.3448621}{doi:\nolinkurl{10.1145/3447993.3448621}}


\bibitem[Huang et~al\mbox{.}(2023)]%
        {huang2021portable}
\bibfield{author}{\bibinfo{person}{Yongzhi Huang}, \bibinfo{person}{Kaixin
  Chen}, \bibinfo{person}{Yandao Huang}, \bibinfo{person}{Lu Wang}, {and}
  \bibinfo{person}{Kaishun Wu}.} \bibinfo{year}{2023}\natexlab{}.
\newblock \showarticletitle{A portable and convenient system for unknown liquid
  identification with smartphone vibration}.
\newblock \bibinfo{journal}{\emph{IEEE Transactions on Mobile Computing}}
  \bibinfo{volume}{22}, \bibinfo{number}{4} (\bibinfo{year}{2023}),
  \bibinfo{pages}{1894--1911}.
\newblock
\href{https://doi.org/10.1109/TMC.2021.3121332}{doi:\nolinkurl{10.1109/TMC.2021.3121332}}


\bibitem[Huang et~al\mbox{.}(2021b)]%
        {huang2021li}
\bibfield{author}{\bibinfo{person}{Yongzhi Huang}, \bibinfo{person}{Kaixin
  Chen}, \bibinfo{person}{Lu Wang}, \bibinfo{person}{Yinying Dong},
  \bibinfo{person}{Qianyi Huang}, {and} \bibinfo{person}{Kaishun Wu}.}
  \bibinfo{year}{2021}\natexlab{b}.
\newblock \showarticletitle{Lili: liquor quality monitoring based on light
  signals}. In \bibinfo{booktitle}{\emph{Proceedings of the 27th Annual
  International Conference on Mobile Computing and Networking}} (New Orleans,
  Louisiana) \emph{(\bibinfo{series}{MobiCom '21})}.
  \bibinfo{publisher}{Association for Computing Machinery},
  \bibinfo{address}{New York, NY, USA}, \bibinfo{pages}{256--268}.
\newblock
\showISBNx{9781450383424}
\href{https://doi.org/10.1145/3447993.3483246}{doi:\nolinkurl{10.1145/3447993.3483246}}


\bibitem[Huang et~al\mbox{.}(2024)]%
        {huang2023beverage}
\bibfield{author}{\bibinfo{person}{Yongzhi Huang}, \bibinfo{person}{Kaixin
  Chen}, \bibinfo{person}{Jiayi Zhao}, \bibinfo{person}{Lu Wang}, {and}
  \bibinfo{person}{Kaishun Wu}.} \bibinfo{year}{2024}\natexlab{}.
\newblock \showarticletitle{Beverage deterioration monitoring based on surface
  tension dynamics and absorption spectrum analysis}.
\newblock \bibinfo{journal}{\emph{IEEE Transactions on Mobile Computing}}
  \bibinfo{volume}{23}, \bibinfo{number}{5} (\bibinfo{year}{2024}),
  \bibinfo{pages}{3722--3740}.
\newblock
\href{https://doi.org/10.1109/TMC.2023.3279837}{doi:\nolinkurl{10.1109/TMC.2023.3279837}}


\bibitem[Huang et~al\mbox{.}(2025)]%
        {huang2025reconstructing}
\bibfield{author}{\bibinfo{person}{Yongzhi Huang}, \bibinfo{person}{Jiayi
  Zhao}, {and} \bibinfo{person}{Kaishun Wu}.} \bibinfo{year}{2025}\natexlab{}.
\newblock \showarticletitle{Reconstructing Ear Canal Channels for Fine-Grained
  Detection of Tympanic Membrane Changes}.
\newblock \bibinfo{journal}{\emph{Proc. ACM Interact. Mob. Wearable Ubiquitous
  Technol.}} \bibinfo{volume}{9}, \bibinfo{number}{3}, Article
  \bibinfo{articleno}{89} (\bibinfo{date}{Sept.} \bibinfo{year}{2025}),
  \bibinfo{numpages}{48}~pages.
\newblock
\href{https://doi.org/10.1145/3749516}{doi:\nolinkurl{10.1145/3749516}}


\bibitem[Huang et~al\mbox{.}(2026)]%
        {huang2026earlog}
\bibfield{author}{\bibinfo{person}{Yongzhi Huang}, \bibinfo{person}{Jiayi
  Zhao}, {and} \bibinfo{person}{Kaishun Wu}.} \bibinfo{year}{2026}\natexlab{}.
\newblock \showarticletitle{{EarLog}: Opportunistic Ear Physiology Sensing
  toward Personalized Hearing Risk Awareness}. In \bibinfo{booktitle}{\emph{The
  32nd Annual International Conference on Mobile Computing and Networking}}
  (Austin, TX, USA) \emph{(\bibinfo{series}{MobiCom '26})}.
  \bibinfo{publisher}{Association for Computing Machinery},
  \bibinfo{address}{New York, NY, USA}.
\newblock
\href{https://doi.org/10.1145/3795866.3844456}{doi:\nolinkurl{10.1145/3795866.3844456}}


\bibitem[Interacoustics(2020)]%
        {InteracousticsTitan}
\bibfield{author}{\bibinfo{person}{Interacoustics}.}
  \bibinfo{year}{2020}\natexlab{}.
\newblock \bibinfo{title}{Titan - A Modular and Portable Tympanometer}.
\newblock
\newblock
\shownote{\url{https://www.interacoustics.com/tympanometers/titan}}.


\bibitem[Jin et~al\mbox{.}(2022)]%
        {jin2022earhealth}
\bibfield{author}{\bibinfo{person}{Yincheng Jin}, \bibinfo{person}{Yang Gao},
  \bibinfo{person}{Xiaotao Guo}, \bibinfo{person}{Jun Wen},
  \bibinfo{person}{Zhengxiong Li}, {and} \bibinfo{person}{Zhanpeng Jin}.}
  \bibinfo{year}{2022}\natexlab{}.
\newblock \showarticletitle{{EarHealth}: an earphone-based acoustic otoscope
  for detection of multiple ear diseases in daily life}. In
  \bibinfo{booktitle}{\emph{Proceedings of the 20th Annual International
  Conference on Mobile Systems, Applications and Services}}.
  \bibinfo{publisher}{Association for Computing Machinery},
  \bibinfo{address}{New York, NY, USA}, \bibinfo{pages}{397--408}.
\newblock
\href{https://doi.org/10.1145/3498361.3538935}{doi:\nolinkurl{10.1145/3498361.3538935}}


\bibitem[Knoetze et~al\mbox{.}(2021)]%
        {knoetze2021sound}
\bibfield{author}{\bibinfo{person}{Megan Knoetze}, \bibinfo{person}{Faheema
  Mahomed-Asmail}, \bibinfo{person}{Vinaya Manchaiah}, {and}
  \bibinfo{person}{De~Wet Swanepoel}.} \bibinfo{year}{2021}\natexlab{}.
\newblock \showarticletitle{Sound-level monitoring earphones with smartphone
  feedback as an intervention to promote healthy listening behaviors in young
  adults}.
\newblock \bibinfo{journal}{\emph{Ear and Hearing}} \bibinfo{volume}{42},
  \bibinfo{number}{5} (\bibinfo{year}{2021}), \bibinfo{pages}{1173--1182}.
\newblock


\bibitem[Kujawa and Liberman(2009a)]%
        {Kujawa2009Adding}
\bibfield{author}{\bibinfo{person}{S. Kujawa} {and} \bibinfo{person}{M.
  Liberman}.} \bibinfo{year}{2009}\natexlab{a}.
\newblock \showarticletitle{Adding Insult to Injury: Cochlear Nerve
  Degeneration after “Temporary” Noise-Induced Hearing Loss}.
\newblock \bibinfo{journal}{\emph{The Journal of Neuroscience}}
  \bibinfo{volume}{29} (\bibinfo{year}{2009}), \bibinfo{pages}{14077 -- 14085}.
\newblock
\href{https://doi.org/10.1523/JNEUROSCI.2845-09.2009}{doi:\nolinkurl{10.1523/JNEUROSCI.2845-09.2009}}


\bibitem[Kujawa and Liberman(2009b)]%
        {Kujawa2009_TTSsynap}
\bibfield{author}{\bibinfo{person}{Sharon~G. Kujawa} {and}
  \bibinfo{person}{M.~Charles Liberman}.} \bibinfo{year}{2009}\natexlab{b}.
\newblock \showarticletitle{Adding Insult to Injury: Cochlear Nerve
  Degeneration after “Temporary” Noise-Induced Hearing Loss}.
\newblock \bibinfo{journal}{\emph{Journal of Neuroscience}}
  \bibinfo{volume}{29}, \bibinfo{number}{45} (\bibinfo{year}{2009}),
  \bibinfo{pages}{14077--14085}.
\newblock
\href{https://doi.org/10.1523/JNEUROSCI.2845-09.2009}{doi:\nolinkurl{10.1523/JNEUROSCI.2845-09.2009}}


\bibitem[Li et~al\mbox{.}(2025a)]%
        {li2025fedwmsam_arxiv}
\bibfield{author}{\bibinfo{person}{Tianle Li}, \bibinfo{person}{Yongzhi Huang},
  \bibinfo{person}{Linshan Jiang}, \bibinfo{person}{Chang Liu},
  \bibinfo{person}{Qipeng Xie}, \bibinfo{person}{Wenfeng Du},
  \bibinfo{person}{Lu Wang}, {and} \bibinfo{person}{Kaishun Wu}.}
  \bibinfo{year}{2025}\natexlab{a}.
\newblock \bibinfo{title}{A Fast and Flat Federated Learning Method via
  Weighted Momentum and Sharpness-Aware Minimization}.
\newblock
\showeprint[arxiv]{2511.22080}~[cs.LG]
\href{https://doi.org/10.48550/arXiv.2511.22080}{doi:\nolinkurl{10.48550/arXiv.2511.22080}}
\newblock
\shownote{arXiv preprint}.


\bibitem[Li et~al\mbox{.}(2025b)]%
        {li2025fedwmsam}
\bibfield{author}{\bibinfo{person}{Tianle Li}, \bibinfo{person}{Yongzhi Huang},
  \bibinfo{person}{Linshan Jiang}, \bibinfo{person}{Chang Liu},
  \bibinfo{person}{Qipeng Xie}, \bibinfo{person}{Wenfeng Du},
  \bibinfo{person}{Lu Wang}, {and} \bibinfo{person}{Kaishun Wu}.}
  \bibinfo{year}{2025}\natexlab{b}.
\newblock \showarticletitle{FedWMSAM: Fast and Flat Federated Learning via
  Weighted Momentum and Sharpness-Aware Minimization}. In
  \bibinfo{booktitle}{\emph{Advances in Neural Information Processing
  Systems}}, \bibfield{editor}{\bibinfo{person}{D.~Belgrave},
  \bibinfo{person}{C.~Zhang}, \bibinfo{person}{H.~Lin},
  \bibinfo{person}{R.~Pascanu}, \bibinfo{person}{P.~Koniusz},
  \bibinfo{person}{M.~Ghassemi}, {and} \bibinfo{person}{N.~Chen}} (Eds.),
  Vol.~\bibinfo{volume}{38, Main Conference}. \bibinfo{publisher}{Curran
  Associates, Inc.}, \bibinfo{pages}{4425--4462}.
\newblock
\href{https://doi.org/10.52202/085713-0157}{doi:\nolinkurl{10.52202/085713-0157}}


\bibitem[Li et~al\mbox{.}(2025c)]%
        {li2025fedwcm}
\bibfield{author}{\bibinfo{person}{Tianle Li}, \bibinfo{person}{Yongzhi Huang},
  \bibinfo{person}{Linshan Jiang}, \bibinfo{person}{Qipeng Xie},
  \bibinfo{person}{Chang Liu}, \bibinfo{person}{Wenfeng Du},
  \bibinfo{person}{Lu Wang}, {and} \bibinfo{person}{Kaishun Wu}.}
  \bibinfo{year}{2025}\natexlab{c}.
\newblock \showarticletitle{{FedWCM}: Unleashing the Potential of
  Momentum-based Federated Learning in Long-Tailed Scenarios}. In
  \bibinfo{booktitle}{\emph{Proceedings of the 54th International Conference on
  Parallel Processing}}. \bibinfo{publisher}{Association for Computing
  Machinery}, \bibinfo{address}{New York, NY, USA}, \bibinfo{pages}{73--82}.
\newblock
\href{https://doi.org/10.1145/3754598.3754657}{doi:\nolinkurl{10.1145/3754598.3754657}}


\bibitem[Li et~al\mbox{.}(2025d)]%
        {li2025fedwcm_arxiv}
\bibfield{author}{\bibinfo{person}{Tianle Li}, \bibinfo{person}{Yongzhi Huang},
  \bibinfo{person}{Linshan Jiang}, \bibinfo{person}{Qipeng Xie},
  \bibinfo{person}{Chang Liu}, \bibinfo{person}{Wenfeng Du},
  \bibinfo{person}{Lu Wang}, {and} \bibinfo{person}{Kaishun Wu}.}
  \bibinfo{year}{2025}\natexlab{d}.
\newblock \bibinfo{title}{{FedWCM}: Unleashing the Potential of Momentum-based
  Federated Learning in Long-Tailed Scenarios}.
\newblock
\showeprint[arxiv]{2507.14980}
\href{https://doi.org/10.48550/arXiv.2507.14980}{doi:\nolinkurl{10.48550/arXiv.2507.14980}}
\newblock
\shownote{arXiv preprint}.


\bibitem[Liberman et~al\mbox{.}(2016)]%
        {Liberman2016Differential}
\bibfield{author}{\bibinfo{person}{M.~Charles Liberman},
  \bibinfo{person}{Michael~J. Epstein}, \bibinfo{person}{Sandra~S. Cleveland},
  \bibinfo{person}{Haobing Wang}, {and} \bibinfo{person}{St{\'e}phane~F.
  Maison}.} \bibinfo{year}{2016}\natexlab{}.
\newblock \showarticletitle{Toward a Differential Diagnosis of Hidden Hearing
  Loss in Humans}.
\newblock \bibinfo{journal}{\emph{PLOS ONE}} \bibinfo{volume}{11},
  \bibinfo{number}{9} (\bibinfo{year}{2016}), \bibinfo{pages}{e0162726}.
\newblock
\href{https://doi.org/10.1371/journal.pone.0162726}{doi:\nolinkurl{10.1371/journal.pone.0162726}}


\bibitem[Qiu and Stucker(1998a)]%
        {Qiu1998_ARlatency}
\bibfield{author}{\bibinfo{person}{W.~W. Qiu} {and} \bibinfo{person}{F.
  Stucker}.} \bibinfo{year}{1998}\natexlab{a}.
\newblock \showarticletitle{Characterization of Acoustic Reflex Latency in
  Normal-Hearing Subjects}.
\newblock \bibinfo{journal}{\emph{Scandinavian Audiology}}
  \bibinfo{volume}{27}, \bibinfo{number}{1} (\bibinfo{year}{1998}),
  \bibinfo{pages}{43--49}.
\newblock
\href{https://doi.org/10.1080/010503998419696}{doi:\nolinkurl{10.1080/010503998419696}}


\bibitem[Qiu and Stucker(1998b)]%
        {Qiu1998Characterization}
\bibfield{author}{\bibinfo{person}{W.~W. Qiu} {and} \bibinfo{person}{F.
  Stucker}.} \bibinfo{year}{1998}\natexlab{b}.
\newblock \showarticletitle{Characterization of acoustic reflex latency in
  normal-hearing subjects.}
\newblock \bibinfo{journal}{\emph{Scandinavian audiology}}  \bibinfo{volume}{27
  1} (\bibinfo{year}{1998}), \bibinfo{pages}{43--9}.
\newblock
\href{https://doi.org/10.1080/010503998419696}{doi:\nolinkurl{10.1080/010503998419696}}


\bibitem[Reynolds and Bielefeld(2023)]%
        {reynolds2023music}
\bibfield{author}{\bibinfo{person}{Alison Reynolds} {and}
  \bibinfo{person}{Eric~C Bielefeld}.} \bibinfo{year}{2023}\natexlab{}.
\newblock \showarticletitle{Music as a unique source of noise-induced hearing
  loss}.
\newblock \bibinfo{journal}{\emph{Hearing Research}}  \bibinfo{volume}{430}
  (\bibinfo{year}{2023}), \bibinfo{pages}{108706}.
\newblock


\bibitem[R{\"o}ddiger et~al\mbox{.}(2021)]%
        {roddiger2021earrumble}
\bibfield{author}{\bibinfo{person}{Tobias R{\"o}ddiger},
  \bibinfo{person}{Christopher Clarke}, \bibinfo{person}{Daniel Wolffram},
  \bibinfo{person}{Matthias Budde}, {and} \bibinfo{person}{Michael Beigl}.}
  \bibinfo{year}{2021}\natexlab{}.
\newblock \showarticletitle{{EarRumble}: Discreet Hands- and Eyes-Free Input by
  Voluntary Tensor Tympani Muscle Contraction}. In
  \bibinfo{booktitle}{\emph{Proceedings of the 2021 CHI Conference on Human
  Factors in Computing Systems}}. \bibinfo{publisher}{Association for Computing
  Machinery}, \bibinfo{address}{New York, NY, USA}, Article
  \bibinfo{articleno}{743}, \bibinfo{numpages}{14}~pages.
\newblock
\href{https://doi.org/10.1145/3411764.3445205}{doi:\nolinkurl{10.1145/3411764.3445205}}


\bibitem[{Samsung Electronics}(2016)]%
        {Samsung}
\bibfield{author}{\bibinfo{person}{{Samsung Electronics}}.}
  \bibinfo{year}{2016}\natexlab{}.
\newblock \bibinfo{title}{{Samsung EHS64 Wired Stereo Headset with Remote and
  Mic}}.
\newblock \bibinfo{howpublished}{Samsung product page}.
\newblock
\urldef\tempurl%
\url{https://www.samsung.com/in/mobile-accessories/headphone-ehs64-ehs64avfbecinu/}
\showURL{%
Retrieved August 27, 2026 from \tempurl}
\newblock
\shownote{Product model EHS64AVFBECINU; product launch date: May 1, 2016}.


\bibitem[Song et~al\mbox{.}(2020)]%
        {song2020spirosonic}
\bibfield{author}{\bibinfo{person}{Xingzhe Song}, \bibinfo{person}{Boyuan
  Yang}, \bibinfo{person}{Ge Yang}, \bibinfo{person}{Ruirong Chen},
  \bibinfo{person}{Erick Forno}, \bibinfo{person}{Wei Chen}, {and}
  \bibinfo{person}{Wei Gao}.} \bibinfo{year}{2020}\natexlab{}.
\newblock \showarticletitle{{SpiroSonic}: monitoring human lung function via
  acoustic sensing on commodity smartphones}. In
  \bibinfo{booktitle}{\emph{Proceedings of the 26th Annual International
  Conference on Mobile Computing and Networking}}.
  \bibinfo{publisher}{Association for Computing Machinery},
  \bibinfo{address}{New York, NY, USA}, Article \bibinfo{articleno}{52},
  \bibinfo{numpages}{14}~pages.
\newblock
\href{https://doi.org/10.1145/3372224.3419209}{doi:\nolinkurl{10.1145/3372224.3419209}}


\bibitem[Sun et~al\mbox{.}(2022)]%
        {sun2022earmonitor}
\bibfield{author}{\bibinfo{person}{Xue Sun}, \bibinfo{person}{Jie Xiong},
  \bibinfo{person}{Chao Feng}, \bibinfo{person}{Wenwen Deng},
  \bibinfo{person}{Xudong Wei}, \bibinfo{person}{Dingyi Fang}, {and}
  \bibinfo{person}{Xiaojiang Chen}.} \bibinfo{year}{2022}\natexlab{}.
\newblock \showarticletitle{Earmonitor: In-ear Motion-resilient Acoustic
  Sensing Using Commodity Earphones}.
\newblock \bibinfo{journal}{\emph{Proceedings of the ACM on Interactive,
  Mobile, Wearable and Ubiquitous Technologies}} \bibinfo{volume}{6},
  \bibinfo{number}{4} (\bibinfo{year}{2022}), \bibinfo{pages}{182:1--182:22}.
\newblock
\href{https://doi.org/10.1145/3569472}{doi:\nolinkurl{10.1145/3569472}}


\bibitem[{Titan | Handheld tympanometer}(2024)]%
        {titan-manual}
\bibfield{author}{\bibinfo{person}{{Titan | Handheld tympanometer}}.}
  \bibinfo{year}{2024}\natexlab{}.
\newblock \bibinfo{title}{Instructions for Use -- Titan}.
\newblock
\newblock
\shownote{Accessed: August 30, 2024. Available at:
  \url{https://www.interacoustics.com/images/files/manuals/en/8100759_D-0100608-M_-_2023_12_Instructions_for_Use_Titan_EN.pdf}}.


\bibitem[Trevi{\~n}o et~al\mbox{.}(2023)]%
        {Trevino2023_MEMR}
\bibfield{author}{\bibinfo{person}{Monica Trevi{\~n}o}, \bibinfo{person}{Andie
  Zang}, {and} \bibinfo{person}{Edward Lobarinas}.}
  \bibinfo{year}{2023}\natexlab{}.
\newblock \showarticletitle{The Middle Ear Muscle Reflex: Current and Future
  Role in Assessing Noise-Induced Cochlear Damage}.
\newblock \bibinfo{journal}{\emph{The Journal of the Acoustical Society of
  America}} \bibinfo{volume}{153}, \bibinfo{number}{1} (\bibinfo{year}{2023}),
  \bibinfo{pages}{436--466}.
\newblock
\href{https://doi.org/10.1121/10.0016853}{doi:\nolinkurl{10.1121/10.0016853}}


\bibitem[Turchet et~al\mbox{.}(2024)]%
        {turchet2024hearing}
\bibfield{author}{\bibinfo{person}{Luca Turchet}, \bibinfo{person}{Simone
  Luiten}, \bibinfo{person}{Tjebbe Treub}, \bibinfo{person}{Marloes van~der
  Burgt}, \bibinfo{person}{Costanza Siani}, {and} \bibinfo{person}{Alberto
  Boem}.} \bibinfo{year}{2024}\natexlab{}.
\newblock \showarticletitle{Hearing loss prevention at loud music events via
  real-time visuo-haptic feedback}.
\newblock \bibinfo{journal}{\emph{Journal on Multimodal User Interfaces}}
  \bibinfo{volume}{18}, \bibinfo{number}{1} (\bibinfo{year}{2024}),
  \bibinfo{pages}{43--53}.
\newblock


\bibitem[Valero et~al\mbox{.}(2016)]%
        {Valero2016_MEMR}
\bibfield{author}{\bibinfo{person}{Michelle~D. Valero},
  \bibinfo{person}{Kelsey~E. Hancock}, {and} \bibinfo{person}{M.~Charles
  Liberman}.} \bibinfo{year}{2016}\natexlab{}.
\newblock \showarticletitle{The Middle Ear Muscle Reflex in the Diagnosis of
  Cochlear Neuropathy}.
\newblock \bibinfo{journal}{\emph{Hearing Research}}  \bibinfo{volume}{332}
  (\bibinfo{year}{2016}), \bibinfo{pages}{29--38}.
\newblock
\href{https://doi.org/10.1016/j.heares.2015.11.005}{doi:\nolinkurl{10.1016/j.heares.2015.11.005}}


\bibitem[Wang et~al\mbox{.}(2021)]%
        {huang_patent_action_recognition_2021}
\bibfield{author}{\bibinfo{person}{Lu Wang}, \bibinfo{person}{Yongzhi Huang},
  \bibinfo{person}{Runping Lu}, {and} \bibinfo{person}{Kaishun Wu}.}
  \bibinfo{year}{2021}\natexlab{}.
\newblock \bibinfo{title}{Action recognition method, system and application}.
\newblock \bibinfo{howpublished}{Chinese Patent Application CN113516086A}.
\newblock
\urldef\tempurl%
\url{https://patents.google.com/patent/CN113516086A/zh}
\showURL{%
\tempurl}
\newblock
\shownote{In Chinese}.


\bibitem[Wang et~al\mbox{.}(2025)]%
        {huang_patent_multikeyboard_2025}
\bibfield{author}{\bibinfo{person}{Lu Wang}, \bibinfo{person}{Jiayi Zhao},
  \bibinfo{person}{Yongzhi Huang}, {and} \bibinfo{person}{Kaishun Wu}.}
  \bibinfo{year}{2025}\natexlab{}.
\newblock \bibinfo{title}{Multi-keyboard mixed key sound identification method,
  device, equipment and storage medium}.
\newblock \bibinfo{howpublished}{Chinese Patent CN116415166B}.
\newblock
\urldef\tempurl%
\url{https://patents.google.com/patent/CN116415166B/en}
\showURL{%
\tempurl}
\newblock
\shownote{Granted November 21, 2025. In Chinese}.


\bibitem[Wang and Ren(2012)]%
        {WangRen2012Repeated}
\bibfield{author}{\bibinfo{person}{Yong Wang} {and} \bibinfo{person}{Chongyu
  Ren}.} \bibinfo{year}{2012}\natexlab{}.
\newblock \showarticletitle{Effects of Repeated ``Benign'' Noise Exposures in
  Young {CBA} Mice: Shedding Light on Age-Related Hearing Loss}.
\newblock \bibinfo{journal}{\emph{Journal of the Association for Research in
  Otolaryngology}} \bibinfo{volume}{13}, \bibinfo{number}{4}
  (\bibinfo{year}{2012}), \bibinfo{pages}{505--515}.
\newblock
\href{https://doi.org/10.1007/s10162-012-0329-0}{doi:\nolinkurl{10.1007/s10162-012-0329-0}}


\bibitem[Withnell and Gowdy(2013)]%
        {withnell2013analysis}
\bibfield{author}{\bibinfo{person}{Robert~H Withnell} {and}
  \bibinfo{person}{Lauren~E Gowdy}.} \bibinfo{year}{2013}\natexlab{}.
\newblock \showarticletitle{An analysis of the acoustic input impedance of the
  ear}.
\newblock \bibinfo{journal}{\emph{Journal of the Association for Research in
  Otolaryngology}}  \bibinfo{volume}{14} (\bibinfo{year}{2013}),
  \bibinfo{pages}{611--622}.
\newblock


\bibitem[{World Health Organization} and {International Telecommunication
  Union}(2019)]%
        {WHOITU2019_SafeListening}
\bibfield{author}{\bibinfo{person}{{World Health Organization}} {and}
  \bibinfo{person}{{International Telecommunication Union}}.}
  \bibinfo{year}{2019}\natexlab{}.
\newblock \bibinfo{title}{Safe Listening Devices and Systems: A WHO–ITU
  Standard}.
\newblock
  \bibinfo{howpublished}{\url{https://www.who.int/publications/i/item/9789241515276}}.
\newblock
\newblock
\shownote{Mode 1: 80 dB for 40 h/week (dose-based guidance)}.


\bibitem[Wu et~al\mbox{.}(2022a)]%
        {huang_patent_capillary_2022}
\bibfield{author}{\bibinfo{person}{Kaishun Wu}, \bibinfo{person}{Yinying Dong},
  \bibinfo{person}{Yongzhi Huang}, \bibinfo{person}{Lu Wang}, {and}
  \bibinfo{person}{Ruixiong Yang}.} \bibinfo{year}{2022}\natexlab{a}.
\newblock \bibinfo{title}{Liquid detection method and system based on
  structured-light measurement of surface capillary waves}.
\newblock \bibinfo{howpublished}{Chinese Patent CN110749528B}.
\newblock
\urldef\tempurl%
\url{https://patents.google.com/patent/CN110749528B/zh}
\showURL{%
\tempurl}
\newblock
\shownote{Granted January 25, 2022. In Chinese}.


\bibitem[Wu et~al\mbox{.}(2020)]%
        {huang_patent_humidity_2020}
\bibfield{author}{\bibinfo{person}{Kaishun Wu}, \bibinfo{person}{Yongzhi
  Huang}, \bibinfo{person}{Qinglang Dai}, {and} \bibinfo{person}{Lu Wang}.}
  \bibinfo{year}{2020}\natexlab{}.
\newblock \bibinfo{title}{Humidity sensing and detection method and system
  based on millimeter-wave signals}.
\newblock \bibinfo{howpublished}{Chinese Patent CN109374652B}.
\newblock
\urldef\tempurl%
\url{https://patents.google.com/patent/CN109374652B/zh}
\showURL{%
\tempurl}
\newblock
\shownote{Granted March 20, 2020. In Chinese}.


\bibitem[Wu et~al\mbox{.}(2021)]%
        {huang_patent_bone_input_2021}
\bibfield{author}{\bibinfo{person}{Kaishun Wu}, \bibinfo{person}{Yongzhi
  Huang}, \bibinfo{person}{Lu Wang}, \bibinfo{person}{Shaotian Cai}, {and}
  \bibinfo{person}{Jianhao Zhang}.} \bibinfo{year}{2021}\natexlab{}.
\newblock \bibinfo{title}{Intelligent input method and system based on bone
  conduction}.
\newblock \bibinfo{howpublished}{Chinese Patent CN109933202B}.
\newblock
\urldef\tempurl%
\url{https://patents.google.com/patent/CN109933202B/zh}
\showURL{%
\tempurl}
\newblock
\shownote{Granted November 30, 2021. In Chinese}.


\bibitem[Wu et~al\mbox{.}(2022b)]%
        {huang_patent_sensory_substitution_2022}
\bibfield{author}{\bibinfo{person}{Kaishun Wu}, \bibinfo{person}{Yongzhi
  Huang}, \bibinfo{person}{Lu Wang}, {and} \bibinfo{person}{Kaixin Chen}.}
  \bibinfo{year}{2022}\natexlab{b}.
\newblock \bibinfo{title}{Auxiliary sensing method and system based on sensory
  substitution}.
\newblock \bibinfo{howpublished}{U.S. Patent US11310620B2}.
\newblock
\urldef\tempurl%
\url{https://patents.google.com/patent/US11310620B2/en}
\showURL{%
\tempurl}
\newblock
\shownote{Granted April 19, 2022}.


\bibitem[Wu et~al\mbox{.}(2025)]%
        {huang_patent_lora_isac_2025}
\bibfield{author}{\bibinfo{person}{Kaishun Wu}, \bibinfo{person}{Yongzhi
  Huang}, \bibinfo{person}{Lu Wang}, \bibinfo{person}{Xiaoshen Li},
  \bibinfo{person}{Hong Liu}, \bibinfo{person}{Li Li}, {and}
  \bibinfo{person}{Min Sun}.} \bibinfo{year}{2025}\natexlab{}.
\newblock \bibinfo{title}{Low-power wide area network integrated sensing and
  communication method based on channel sensing and reinforcement learning}.
\newblock \bibinfo{howpublished}{U.S. Patent US12238543B2}.
\newblock
\urldef\tempurl%
\url{https://patents.google.com/patent/US12238543B2/en}
\showURL{%
\tempurl}
\newblock
\shownote{Granted February 25, 2025}.


\bibitem[Wu et~al\mbox{.}(2018)]%
        {huang_patent_wifi_fire_2018}
\bibfield{author}{\bibinfo{person}{Kaishun Wu}, \bibinfo{person}{Yongzhi
  Huang}, \bibinfo{person}{Lu Wang}, \bibinfo{person}{Shuxin Zhong}, {and}
  \bibinfo{person}{Hailiang Yang}.} \bibinfo{year}{2018}\natexlab{}.
\newblock \bibinfo{title}{Method and system for indoor fire detection and alarm
  based on wireless signal transmission}.
\newblock \bibinfo{howpublished}{Chinese Patent CN107025751B}.
\newblock
\urldef\tempurl%
\url{https://patents.google.com/patent/CN107025751B/en}
\showURL{%
\tempurl}
\newblock
\shownote{Granted May 8, 2018. In Chinese}.


\bibitem[Xie et~al\mbox{.}(2024)]%
        {xie2024litecrypt}
\bibfield{author}{\bibinfo{person}{Qipeng Xie}, \bibinfo{person}{Weizheng
  Wang}, \bibinfo{person}{Yongzhi Huang}, \bibinfo{person}{Mengyao Zheng},
  \bibinfo{person}{Shuai Shang}, \bibinfo{person}{Linshan Jiang},
  \bibinfo{person}{Salabat Khan}, {and} \bibinfo{person}{Kaishun Wu}.}
  \bibinfo{year}{2024}\natexlab{}.
\newblock \showarticletitle{{LiteCrypt}: Enhancing {IoMT} Security with
  Optimized {HE} and Lightweight Dual-Authorization}. In
  \bibinfo{booktitle}{\emph{2024 IEEE 30th International Conference on Parallel
  and Distributed Systems (ICPADS)}}. \bibinfo{publisher}{IEEE},
  \bibinfo{pages}{166--175}.
\newblock
\href{https://doi.org/10.1109/ICPADS63350.2024.00031}{doi:\nolinkurl{10.1109/ICPADS63350.2024.00031}}


\bibitem[Xie et~al\mbox{.}(2023)]%
        {xie2023earspiro}
\bibfield{author}{\bibinfo{person}{Wentao Xie}, \bibinfo{person}{Qingyong Hu},
  \bibinfo{person}{Jin Zhang}, {and} \bibinfo{person}{Qian Zhang}.}
  \bibinfo{year}{2023}\natexlab{}.
\newblock \showarticletitle{EarSpiro: Earphone-based Spirometry for Lung
  Function Assessment}.
\newblock \bibinfo{journal}{\emph{Proceedings of the ACM on Interactive,
  Mobile, Wearable and Ubiquitous Technologies}} \bibinfo{volume}{6},
  \bibinfo{number}{4} (\bibinfo{year}{2023}), \bibinfo{pages}{1--27}.
\newblock
\href{https://doi.org/10.1145/3569480}{doi:\nolinkurl{10.1145/3569480}}


\bibitem[Xu et~al\mbox{.}(2025)]%
        {xu2025easyspiro}
\bibfield{author}{\bibinfo{person}{Chi Xu}, \bibinfo{person}{Wentao Xie},
  \bibinfo{person}{Baichen Yang}, \bibinfo{person}{Yizhen Zhang},
  \bibinfo{person}{Yanbin Gong}, \bibinfo{person}{Jin Zhang},
  \bibinfo{person}{Wei Li}, \bibinfo{person}{Shifang Yang}, {and}
  \bibinfo{person}{Qian Zhang}.} \bibinfo{year}{2025}\natexlab{}.
\newblock \showarticletitle{{EasySpiro}: Assessing Lung Function via Arbitrary
  Exhalations on Commodity Earphones}. In \bibinfo{booktitle}{\emph{Proceedings
  of the 31st Annual International Conference on Mobile Computing and
  Networking}}. \bibinfo{publisher}{Association for Computing Machinery},
  \bibinfo{address}{New York, NY, USA}, \bibinfo{pages}{513--528}.
\newblock
\href{https://doi.org/10.1145/3680207.3723489}{doi:\nolinkurl{10.1145/3680207.3723489}}


\bibitem[Yin et~al\mbox{.}(2022)]%
        {yin2022aware}
\bibfield{author}{\bibinfo{person}{Xiangyu Yin}, \bibinfo{person}{Kai Huang},
  \bibinfo{person}{Erick Forno}, \bibinfo{person}{Wei Chen},
  \bibinfo{person}{Heng Huang}, {and} \bibinfo{person}{Wei Gao}.}
  \bibinfo{year}{2022}\natexlab{}.
\newblock \showarticletitle{Out-Clinic Pulmonary Disease Evaluation via
  Acoustic Sensing and Multi-Task Learning on Commodity Smartphones}. In
  \bibinfo{booktitle}{\emph{Proceedings of the 20th ACM Conference on Embedded
  Networked Sensor Systems}}. \bibinfo{publisher}{Association for Computing
  Machinery}, \bibinfo{address}{New York, NY, USA},
  \bibinfo{pages}{1182--1188}.
\newblock
\href{https://doi.org/10.1145/3560905.3568437}{doi:\nolinkurl{10.1145/3560905.3568437}}


\bibitem[Yin et~al\mbox{.}(2023)]%
        {yin2023ptease}
\bibfield{author}{\bibinfo{person}{Xiangyu Yin}, \bibinfo{person}{Kai Huang},
  \bibinfo{person}{Erick Forno}, \bibinfo{person}{Wei Chen},
  \bibinfo{person}{Heng Huang}, {and} \bibinfo{person}{Wei Gao}.}
  \bibinfo{year}{2023}\natexlab{}.
\newblock \showarticletitle{{PTEase}: Objective Airway Examination for
  Pulmonary Telemedicine using Commodity Smartphones}. In
  \bibinfo{booktitle}{\emph{Proceedings of the 21st Annual International
  Conference on Mobile Systems, Applications and Services}}.
  \bibinfo{publisher}{Association for Computing Machinery},
  \bibinfo{address}{New York, NY, USA}, \bibinfo{pages}{110--123}.
\newblock
\href{https://doi.org/10.1145/3581791.3596854}{doi:\nolinkurl{10.1145/3581791.3596854}}


\bibitem[Zhao et~al\mbox{.}(2024)]%
        {zhao2024chameleon}
\bibfield{author}{\bibinfo{person}{Jiayi Zhao}, \bibinfo{person}{Yongzhi
  Huang}, \bibinfo{person}{Qipeng Xie}, \bibinfo{person}{Weizheng Wang},
  \bibinfo{person}{Lu Wang}, {and} \bibinfo{person}{Kaishun Wu}.}
  \bibinfo{year}{2024}\natexlab{}.
\newblock \showarticletitle{Chameleon: An Adaptive System for Overlapping
  Keystroke Signal Separation and Identification}. In
  \bibinfo{booktitle}{\emph{2024 IEEE 30th International Conference on Parallel
  and Distributed Systems (ICPADS)}}. IEEE, \bibinfo{pages}{60--67}.
\newblock
\href{https://doi.org/10.1109/ICPADS63350.2024.00018}{doi:\nolinkurl{10.1109/ICPADS63350.2024.00018}}


\end{thebibliography}

\appendix

\section{Clinical Background and Illustrative Content Confounding}
\label{app:clinical_background}
\subsection{Scope of This Appendix}
\label{app:clinical_scope}

The argument in Sec.~\ref{sec:why_formulation} rests on two observations. First, a commodity in-ear path can capture small reflection changes, but it does not inherit the stable absolute criterion of a clinical system. Second, arbitrary playback changes which ear-canal paths are excited, so the observed reflection changes with content even when the ear does not. The following sections explain these two effects. Appendix~\ref{app:residual_derivations} then derives the residual cues used for AR detection.

\subsection{Clinical Acoustic-Admittance Background}
\label{app:clinical_admittance}

Clinical AR assessment~\citep{Qiu1998Characterization,Kujawa2009Adding} commonly describes the middle-ear response through acoustic admittance, the reciprocal of acoustic impedance:
\[
Y = \frac{1}{Z}.
\]
With controlled single-frequency stimulation, a low-level probe tone gives a reference reflection coefficient \(R_{\mathrm{ref}}\). A higher-level elicited state gives a detection coefficient \(R_{\mathrm{det}}\). Their acoustic-admittance change is
\[
\Delta Y
=
\frac{1-R_{\mathrm{ref}}}{Z_1(1+R_{\mathrm{ref}})}
-
\frac{1-R_{\mathrm{det}}}{Z_2(1+R_{\mathrm{det}})},
\]
where \(Z_1\) and \(Z_2\) are calibrated impedances for the same ear.

\paragraph{Controlled Titan--COTS comparison.}
The comparison in Sec.~\ref{sec:cost} uses an Interacoustics Titan clinical system~\citep{titan-manual,InteracousticsTitan} and a matched COTS path built from wired Samsung EHS64 headphones~\citep{Samsung} and an ICS-40720 analog differential-output MEMS microphone. Titan presents a $500\,\mathrm{Hz}$ elicitor and monitors a $226\,\mathrm{Hz}$ probe tone. The COTS path uses SPL-matched playback and records the reflected signal under the corresponding stimulus condition.

Figure~\ref{fig:titan_mic} reports pairwise percentage contrasts relative to the original playback. The pink curve compares the human ear with the silicone artificial ear. The blue curve compares the silicone ear with the IEC 60318-4 ear simulator. The orange line is the Titan admittance-change criterion, the green bands are Titan-detected AR ranges, and the dashed green curve fits the Titan measurements. The experiment includes five volunteers, a silicone artificial ear~\citep{EarSimulator2}, and an IEC 60318-4-compliant ear simulator~\citep{EarSimulator}. This comparison is separate from the contralateral music-validation protocol in Sec.~\ref{sec:setup}.

With dedicated probe and elicitor tones and stable instrumentation, Titan attributes small changes in reflected sound power to AR onset. It begins to detect small responses near $65\,\mathrm{dB}$ SPL, and the response grows markedly above $75\,\mathrm{dB}$ SPL. The two COTS contrasts also rise overall, but they vary irregularly with SPL. Neither passive fixture can generate a physiological AR. Their different geometries and acoustic impedances can still create a nonzero offset, and that offset changes with level. The passive-fixture contrast even crosses the Titan criterion. An absolute value calibrated at one level therefore cannot serve as a global AR threshold on the COTS path.

\subsection{An Illustrative Multipath-Content Model under Arbitrary Playback}
\label{app:content_confound}

Absolute reflection criteria face a second problem under arbitrary playback. Unlike a clinical probe tone, music is nonstationary and ranges from sparse tonal passages to broadband mixtures. Each spectrum excites a different combination of ear-canal paths. We use a factorized model to show how temporal overlap and frequency-dependent absorption enter the observed reflection.

The reverberant field in the canal decays over time. We denote its envelope by
\[
L_p(t) = L_{p0} - \beta t / T_{\mathrm{floor}},
\]
where \(L_{p0}\) is the initial sound-pressure level, \(\beta\) controls the decay rate, and \(T_{\mathrm{floor}}\) is the time required to approach the ambient-noise floor. The envelope \(g_{\mathrm{mix}}(t)\) summarizes the cumulative effect of overlapping reflections.

The canal wall, surrounding tissues, and canal morphology also shape frequency-dependent absorption and scattering. We collect these effects in \(g_{\mathrm{abs}}(f,x)\), where \(f\) is frequency and \(x\) is propagation distance along the canal.

The resulting illustrative effective reflection coefficient is
\[
R_{\mathrm{eff}}(f,t,x)
=
R_{\mathrm{ideal}}(f)\, g_{\mathrm{mix}}(t)\, g_{\mathrm{abs}}(f,x),
\]
where \(R_{\mathrm{ideal}}(f)\) is the idealized reflection coefficient under simplified conditions.

This factorization exposes why a fixed reflectance threshold fails. The overlap and absorption terms are weighted by the current playback spectrum and gain, so the microphone observes a different composite response as the content changes. \earlog{} therefore predicts the no-AR response for the current playback and tests the residual, rather than comparing the raw reflection with one absolute reference.

\section{Residual-Cue Derivations and Online AR Decision}
\label{app:residual_derivations}

Section~\ref{sec:pLDS} defines the estimator, and Secs.~\ref{sec:roughness} and~\ref{sec:LDA} define the residual descriptors and online decision. This appendix derives the residual relation and describes pLDS window construction, regularization details, training-time feature normalization, and the hysteresis parameters.

\subsection{From Middle-Ear Load Change to LDS}
\label{app:lds_derivation}

AR changes the effective middle-ear load seen from the ear canal~\citep{feeney2017normative,withnell2013analysis,baricevich2025frequency}. This change modifies the acoustic boundary condition and becomes observable through the ear-canal standing-wave response in a subset of frequency bands.

In analysis band \(b\), let \(S_b(t)\) denote the short-time root-mean-square energy of the decoded playback signal and let \(R_b(t)\) denote the corresponding inward-microphone energy. We define the source-side log loudness and the band-wise log transmission ratio as
\[
L_{s,b}(t)=\log S_b(t),
\qquad
\tau_b(t)=\log R_b(t)-\log S_b(t).
\]

Within a short window, we approximate the headphone gain in band \(b\) by \(G_b\). We represent ear-canal reverberation by \(H_b^{\mathrm{rev}}\) and the additional middle-ear transfer change caused by AR by \(H_b^{\mathrm{AR}}\). The transmission ratio can then be written as
\[
\tau_b(t)
\approx
\log\left|G_bH_b^{\mathrm{rev}}(f_b,t)\right|
+
\log\left|H_b^{\mathrm{AR}}(L_{s,b}(t),t)\right|.
\]

The first term describes the device and ear-canal response that the no-AR baseline seeks to predict. During AR, we write
\[
H_b^{\mathrm{AR}}(L_{s,b})
=
1+\varepsilon_b(L_{s,b}),
\]
where \(\varepsilon_b\) denotes the loudness-dependent transfer perturbation associated with the effective middle-ear load change \(\Delta Z_L(f_b,L_{s,b})\). Applying the chain rule gives
\[
\frac{\partial\tau_b}{\partial L_{s,b}}
\approx
\frac{\partial\log|H_b^{\mathrm{AR}}|}{\partial Z_L}
\frac{\partial Z_L}{\partial L_{s,b}}.
\]

Without AR, the no-AR baseline accounts for the ordinary level-dependent response and the remaining loudness sensitivity stays close to zero. When AR engages, the effective load becomes loudness-dependent and the first factor above can be amplified near ear-canal resonance bands.

The main pipeline computes this quantity after baseline subtraction. Let \(\tau_b^{\mathrm{base}}(t)\) denote the predicted no-AR transmission ratio for the same playback condition. The residual is
\[
r_b(t)
=
\tau_b(t)-\tau_b^{\mathrm{base}}(t).
\]
The LDS used in the main text is therefore
\[
\mathrm{LDS}_b
\approx
\frac{\partial r_b(t)}
{\partial L_{s,b}(t)}.
\]
This expression links the observed residual slope to the loudness-dependent middle-ear transfer change while suppressing the dominant device and playback response.

\subsection{pLDS Window Construction and Regularization Details}
\label{app:plds_estimation}

A single-band LDS can absorb variation transferred from neighboring bands. For the short estimation window \(t=t_1,\ldots,t_W\), \earlog{} arranges the residual and source-side loudness observations as
\[
\begin{aligned}
\mathbf r(t)
&=
[r_1(t),\ldots,r_B(t)]^\top,
\\
\mathbf L_s(t)
&=
[L_{s,1}(t),\ldots,L_{s,B}(t)]^\top,
\end{aligned}
\]
which supply the per-frame observations for the multiband regression in Eq.~\ref{eq:plds_ridge}. Dynamic audio often produces strongly correlated loudness trajectories across bands, so ordinary least squares can yield an unstable coefficient matrix. The ridge term governed by \(\lambda_A\) regularizes that collinearity. The estimator still relies on the excitation present in the current window, which is why Sec.~\ref{sec:pLDS} excludes low-energy bands.

\subsection{Verifier Training and Hysteresis Parameters}
\label{app:verifier_runtime}

Equations~\ref{eq:verifier_features} and~\ref{eq:verifier_score} define the feature vector and online score. The z-score mean and variance are computed from the training participants and then fixed for evaluation. Covariance shrinkage is applied when fitting the LDA discriminant to reduce instability under the available labels.

The runtime parameters are the smoothing coefficient \(\lambda_s\), entry and exit thresholds \(\gamma_{\mathrm{hi}}\) and \(\gamma_{\mathrm{lo}}\), and consecutive-window counts \(K_+\) and \(K_-\). The two thresholds prevent rapid switching near the decision boundary, while the consecutive-window counts require sustained evidence from valid windows. Their state semantics and event aggregation are defined in Sec.~\ref{sec:LDA}.

\section{Details of FIR Identification and Volume-Dependent Extrapolation}
\label{ap:fir_details_main}

The three subsections describe bucketed identification, low-order baseline fitting, and runtime construction used by Sec.~\ref{sec:baseline}. The coefficient update itself is defined in Sec.~\ref{sec:Sensitivity}.

\subsection{Bucketed LMS Identification}
\label{ap:fir_lms}

For each band-energy bucket \(V\), Sec.~\ref{sec:Parameter} defines a local no-AR FIR on the tap support selected by Sec.~\ref{sec:Sensitivity}. Equation~\ref{eq:sparse_lms_update} is the authoritative coefficient update. After convergence, the supported coefficient vector provides the local no-AR impulse response \(\mathbf{h}^{(V)}\) used by the bucket-level baseline model.

\subsection{Band-Wise Polynomial Extrapolation Across Loudness Buckets}
\label{ap:volume_extrapolation}

The gammatone filterbank maps each bucket-specific response into the \(B\) analysis bands used throughout the main pipeline. Section~\ref{sec:baseline_construction} defines the band-wise polynomial model for \(\theta_b(V)\). Here, \(V\) is the bucket-level playback condition, and the model coefficients are fitted from the corresponding low-volume calibration buckets. The low-order fit captures the smooth variation observed across those buckets and provides a baseline for playback states without an exact stored match.

\subsection{Online Construction of the No-AR Baseline}
\label{ap:baseline_runtime}

For each analysis window, \earlog{} computes the current playback state \(V_t\), evaluates the band-wise models from Sec.~\ref{ap:volume_extrapolation}, and reconstructs \(H_{\mathrm{baseline}}(f,V_t)\). Section~\ref{sec:baseline_construction} defines the resulting log-magnitude residual. This stage predicts the no-AR baseline but does not make an AR decision; the residual is passed to the LDS, pLDS, and roughness calculations in Sec.~\ref{sec:verification}.
\section{Calibration Transfer and Baseline Refresh}
\label{app:baseline_transfer}

This appendix expands the calibration-transfer and refresh mechanisms in Sec.~\ref{sec:baseline_maintenance}. The main text defines the anchor-matching and online transfer objectives. Here, we describe paired-difference and drift-basis construction, anchor-set construction, support-window formation, frequency-domain transport, and the consistency failure that requires a new low-volume baseline.

\subsection{Building the Drift Basis}
\label{app:drift_basis}

Let \(\boldsymbol{\theta}_i^{(a)}\in\mathbb{R}^{B}\) and \(\boldsymbol{\theta}_i^{(b)}\in\mathbb{R}^{B}\) denote band-wise no-AR baseline parameters from the same playback bucket under two fit or environmental conditions. Both observations come from valid low-volume windows used to form the no-AR reference. Their difference is
\[
\Delta\boldsymbol{\theta}_i
=
\boldsymbol{\theta}_i^{(b)}
-
\boldsymbol{\theta}_i^{(a)}.
\]
Stacking \(N\) paired differences gives
\[
\mathbf{\Delta}
=
\left[
\Delta\boldsymbol{\theta}_1,\ldots,
\Delta\boldsymbol{\theta}_N
\right]
\in\mathbb{R}^{B\times N}.
\]

\earlog{} applies singular value decomposition:
\[
\mathbf{\Delta}
=
\mathbf U\mathbf\Sigma\mathbf R^\top.
\]
The leading \(r\) left singular vectors form the drift basis
\[
\mathbf{\Psi}
=
\mathbf U_{[:,1:r]}
\in\mathbb{R}^{B\times r}.
\]
Its columns capture the dominant band-wise response changes observed across refitting, seal, humidity, and related non-thermal conditions.

\subsection{Anchor-Band Matching}
\label{app:anchor_matching}

Bands that are weakly affected by the drift basis provide a stable reference for matching the current condition. \earlog{} defines the anchor set
\[
\mathcal A
=
\left\{
b\in\{1,\ldots,B\}:
\|\mathbf{\Psi}_{b,:}\|_2\leq\tau_A
\right\},
\]
where \(\mathbf{\Psi}_{b,:}\) is row \(b\) of the drift basis. Let \(\mathbf P_{\mathcal A}\) denote the diagonal selector that preserves these bands.

The online support set \(\mathcal S_t\) contains the first valid low-volume support windows available under the current condition. Their band-wise average forms \(\mathbf q_t\) in the same log-transfer representation as the stored baseline parameters. Equation~\ref{eq:calibration_anchor_matching} matches this support to the stored bucket that agrees most closely on the anchor bands.

\subsection{Online Transfer Estimation}
\label{app:transfer_estimation}

Equation~\ref{eq:calibration_online_transfer} estimates the current drift coefficients after anchor matching. At the transfer-function level,
\[
\begin{aligned}
\log|
\tilde H_{\mathrm{baseline}}(f,V_t,u_t)
|
&=
\log|
H_{\mathrm{baseline}}(f,V_t)
|\\
&\quad+
\sum_{j=1}^{r}
\hat{\alpha}_{t,j}\psi_j(f),
\end{aligned}
\]
where \(\psi_j(f)\) is the frequency-domain profile of drift component \(j\). The residual passed to Sec.~\ref{sec:verification} is
\[
\Delta\log\tilde H(f,t)
=
\log|H_{\mathrm{obs}}(f,t)|
-
\log|
\tilde H_{\mathrm{baseline}}(f,V_t,u_t)
|.
\]

\subsection{Consistency Failure and Baseline Refresh}
\label{app:baseline_refresh}

The transferred baseline is checked on valid low-volume support windows. \earlog{} monitors the residual energy and the low-volume LDS, pLDS, and roughness values produced by the transported baseline. After a transient failure, the validity gate in Sec.~\ref{sec:guardrails} rejects the affected window. Persistent failure across consecutive support windows suspends AR detection and triggers a low-volume refresh of the band-energy buckets and baseline parameters.

Listening exposure and signal-quality status remain active during refresh. The refreshed baseline must pass the same consistency test before AR detection resumes. This fallback covers shifts that are not represented by the learned drift basis without introducing a second AR classifier or recursively transporting a previous estimate.

\section{Ear-Canal Length Estimation and Sparse FIR Tap Placement}
\label{ap:FIR_Tap}

This appendix gives the resonance-based propagation prior used to select the sparse FIR support in Sec.~\ref{sec:Sensitivity}. We first obtain a coarse effective length \(L\) from the standing-wave condition
\[
2\pi fL/c=(2n+1)\pi/2,
\]
where \(c\) is the speed of sound and \(n\) is a nonnegative integer. For the fundamental mode,
\[
L=c/(4f).
\]

In our implementation, \(L\) is constrained to \(15\)--\(45\,\mathrm{mm}\). The lower bound accounts for the shorter effective propagation distance introduced by headphone insertion relative to the theoretical \(25\)--\(45\,\mathrm{mm}\) ear-canal range. The upper bound corresponds to a minimum candidate resonance frequency of approximately \(1906\,\mathrm{Hz}\).

A single resonance may be shifted by multipath. \earlog{} therefore forms a relative-gain curve by dividing the measured frequency-response magnitude by the source magnitude and identifies prominent resonance peaks within the allowed frequency range. For a detected resonance \(f_n\), the corresponding effective length is
\[
L
=
\frac{c}{4f_n}(2n+1).
\]

The resulting one-way propagation delay is \(t_0=L/c\). \earlog{} places nonzero FIR taps near the direct arrival and the round-trip delays
\[
k_i\approx2it_0f_s,
\qquad i=0,1,\ldots,N,
\]
where \(f_s\) is the sampling rate and \(i\) indexes successive reflections. This propagation prior reduces the number of free FIR coefficients and supports stable response estimation within the short analysis window.
\section{Details of Observation Screening}
\label{app:guardrails}

This appendix complements Sec.~\ref{sec:guardrails} by defining the validity indicators and window-routing rules. It specifies how windows are assigned to valid, invalid, or overload-only states and how persistent baseline inconsistency is handed to baseline maintenance.

\subsection{Window Validity and Routing}
\label{app:validity_gate}

Let \(y_{\mathrm{raw}}(n)\) denote the inward-microphone samples in an analysis window of length \(N\). \earlog{} identifies samples that approach the calibrated saturation boundary of the analog front end:
\[
\begin{aligned}
\mathcal C
&=
\left\{
\begin{gathered}
n\in[1,N]:\\
|y_{\mathrm{raw}}(n)|
\geq
\eta_{\mathrm{sat}}
\end{gathered}
\right\},
\end{aligned}
\]
where \(\eta_{\mathrm{sat}}\) is obtained during front-end calibration. The clipped-sample ratio is
\[
r_{\mathrm{clip}}
=
\frac{|\mathcal C|}{N}.
\]
A calibrated operating range separates hard saturation from mild clipping. Hard-saturated windows are logged as \emph{overload-only}; mildly clipped windows retain enough response structure to remain eligible for the constrained repair path summarized in Sec.~\ref{sec:guardrails}.

The other three indicators examine different parts of the playback--response relation. The leakage ratio \(\lambda_{\mathrm{leak}}(t)\) compares outer- and inner-microphone energy to reveal seal loss or abnormal external-sound leakage. The source-coherence score \(c_{\mathrm{src}}(t)\) tests whether the inward observation remains aligned with the known playback. The baseline-consistency score \(m_{\mathrm{fit}}(t)\) measures whether the current response remains compatible with the recent no-AR baseline.

A window remains valid only while all four indicators stay within their calibrated ranges. The gate changes state after either a violation or a recovery persists across consecutive windows, avoiding rapid switching around a boundary. Saturated windows are routed to \emph{overload-only}; other failures are routed to \emph{invalid}. Rejected windows do not update the AR state or event count, but they remain in the listening-exposure and signal-quality records. Persistent baseline inconsistency is passed to Sec.~\ref{sec:baseline_maintenance} for baseline adaptation or refresh.

\section{Implementation Details}
\label{app:Implementation}

This appendix expands Sec.~\ref{Module} with the analog front end, thermal compensation, electromagnetic shielding, deployment geometry, runtime, and power measurements.

\begin{figure}[!tbp]
  \centering
  \includegraphics[width=0.78\linewidth]{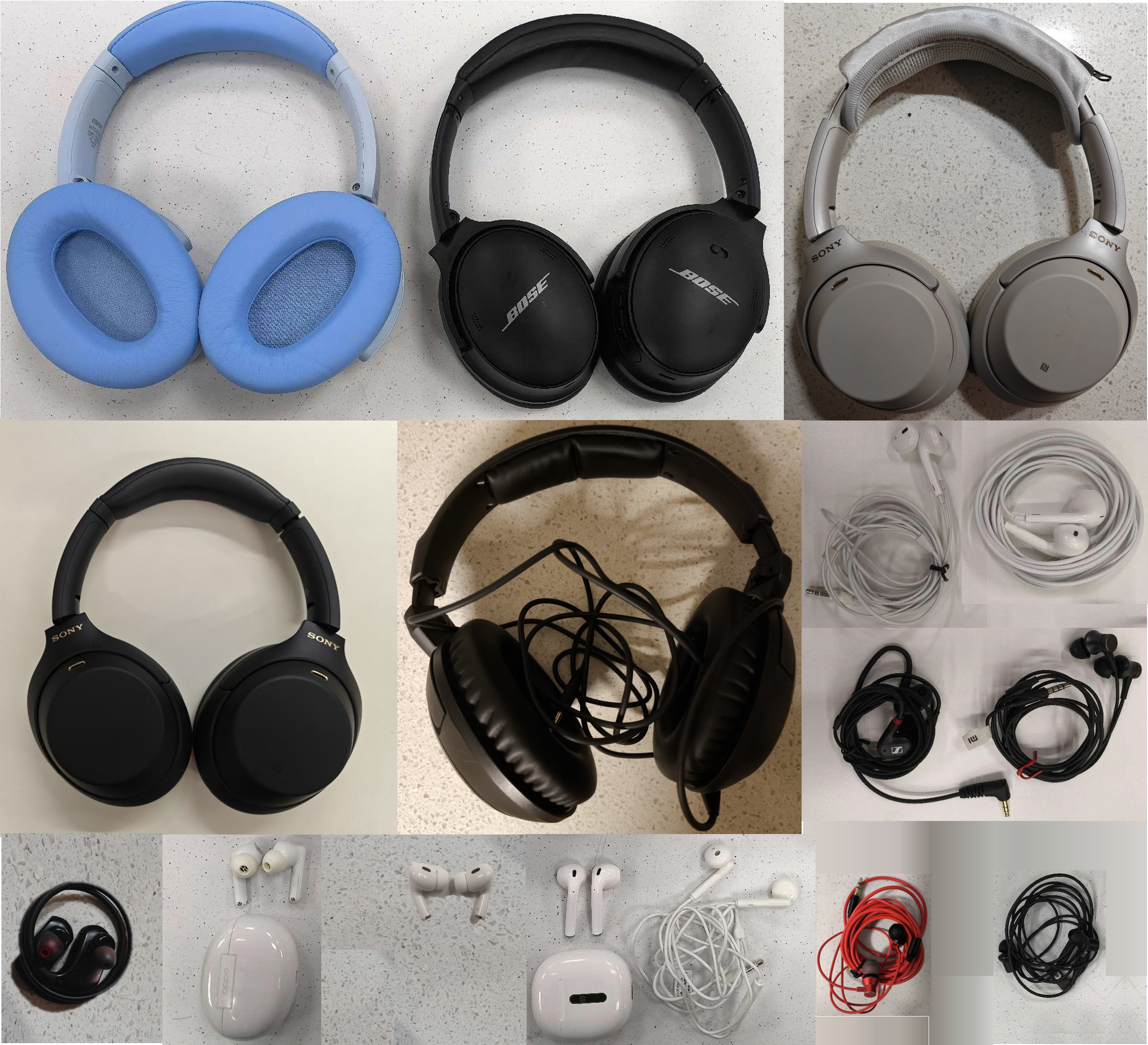}
  \caption{Representative headphone models used in the robustness study.}
  \Description{A collage of representative over-ear, in-ear, and semi-in-ear headphone models used in the robustness study.}
  \label{fig:app_headphones}
\end{figure}

\subsection{Analog Front End and Thermal Compensation}
\label{app:analog_frontend}

The inner and outer ICS-40720 microphones are conditioned through separate analog paths and digitized on independent ADC channels. A resistor-capacitor anti-alias filter is followed by active high-pass and low-pass stages that confine the analog response to 50~Hz--20~kHz. Final band limiting and equalization are performed digitally.

Let \(G_0(T)\) denote the uncompensated microphone-to-ADC gain at temperature \(T\), and let \(C_{\mathrm{NTC}}(T)\) denote the gain correction introduced by the NTC feedback network. The feedback topology is selected so that
\[
\begin{aligned}
\partial_T\log C_{\mathrm{NTC}}(T)
&\approx
-\partial_T\log G_0(T),\\
\partial_T\log
\left[
G_0(T)C_{\mathrm{NTC}}(T)
\right]
&\approx 0
\end{aligned}
\]
over the calibrated operating range. The analog network removes the dominant first-order thermal gain drift before the ADC, and digital normalization absorbs the smaller residual offset.

\subsection{Shielding and Deployment}
\label{app:hardware_deployment}

The printed circuit board sits close to the headphone driver, where it is exposed to low-frequency magnetic fields from coil current and high-frequency electromagnetic interference from driver wiring. The circuit is mounted in a \(0.5\,\mathrm{mm}\) polylactic-acid carrier. An inner \(0.5\,\mathrm{mm}\) 1J85 permalloy shell diverts low-frequency magnetic fields, while an outer grounded \(200\,\mathrm{\mu m}\) copper mesh attenuates high-frequency interference.

Figure~\ref{fig:headphone_display} shows the in-ear and over-ear placement. In an in-ear assembly, the module lies between the driver and the canal entrance. In an over-ear assembly, the electronics remain at least \(1\,\mathrm{cm}\) from the driver magnet to reduce magnetic coupling to the MEMS front end. Figure~\ref{fig:app_headphones} shows representative devices from the robustness study.

\subsection{Runtime and Power}
\label{app:implementation_power}

The fixed-point implementation runs on an Arduino Portenta H7 with an STM32H747 processor. The Cortex-M7 core runs at 480~MHz and the Cortex-M4 core at 240~MHz. The complete pipeline processes the two 48-kHz microphone streams at \(1.05\times\) real time.

The prototype development board draws \(0.45\,\mathrm{W}\) during continuous inference and \(0.40\,\mathrm{W}\) during a fallback low-volume refresh. These values include the general-purpose MCU board and its peripheral circuitry. Relative to an audio-only baseline on the same platform, the \earlog{}-specific incremental overhead is \(16\,\mathrm{mW}\), corresponding to \(0.016\,\mathrm{Wh}\) per listening hour. This is approximately \(8\%\) of a representative \(0.20\,\mathrm{Wh}\) earbud battery. The whole-board and incremental measurements describe different quantities and should not be compared as product-level power figures. Clock alignment, SPL matching, and the evaluation platform are summarized in Appendix~\ref{app:eval_runtime}.

\section{Evaluation Details}
\label{app:eval_details}

This appendix provides participant statistics, stimulus schedules, hardware profiling, and the detailed robustness protocols supporting Secs.~\ref{sec:setup} and~\ref{sec:Evaluation}. Section~\ref{sec:setup} is the authoritative description of the contralateral Titan reference, independent label generation, window states, event aggregation, and metric denominators.

\begin{figure*}[!t]
    \centering
    \begin{minipage}[t]{0.60\textwidth}
        \vspace{0pt}
        \centering
        \includegraphics[height=32mm,keepaspectratio]{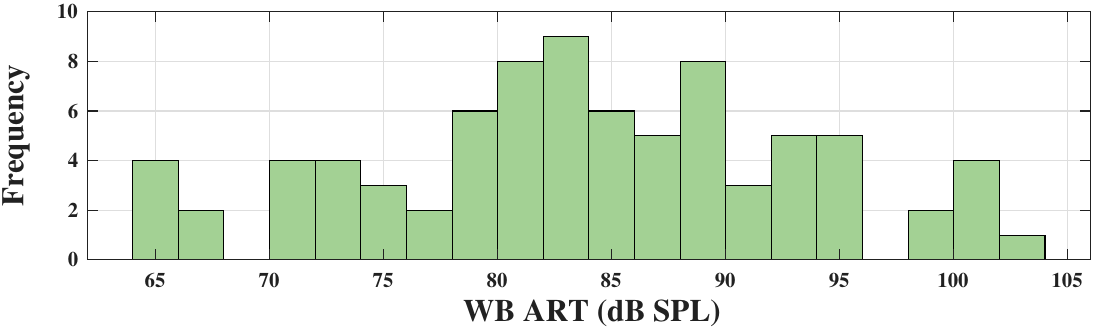}
        \caption{Distribution of evaluation ears across WB-ART values.}
        \Description{Distribution of the 81 evaluation ears across the three listener-specific WB-ART groups.}
        \label{fig:app_distribution}
    \end{minipage}\hfill\begin{minipage}[t]{0.38\textwidth}
        \vspace{0pt}
        \centering
        \includegraphics[height=32mm,keepaspectratio]{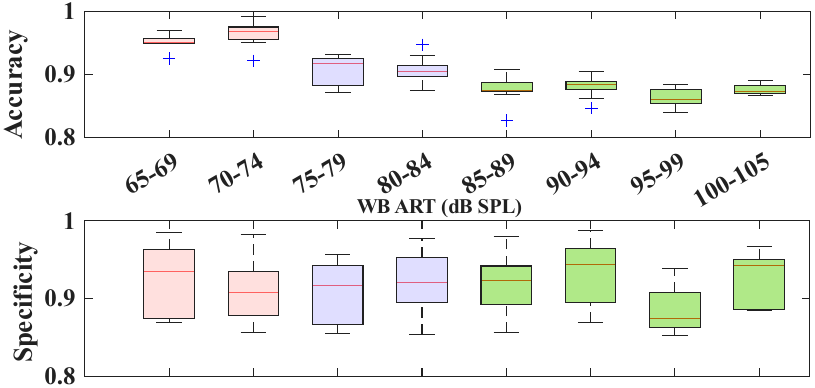}
        \caption{\earlog{} maintains broad cross-threshold performance under wideband, piano, and instrument playback in a noisy environment.}
        \Description{Cross-threshold performance for wideband, piano, and instrument playback in a noisy environment, compared across the listener-specific WB-ART groups.}
        \label{fig:app_wbart_groups}
    \end{minipage}
\end{figure*}

\subsection{Participant Statistics and WB-ART Stratification}
\label{app:eval_participants}

The cohort contains 81 participants, with one evaluation ear from each participant. For subgroup analysis, the evaluation ears are stratified by WB-ART into ART-1 (65--75~dB, \(n=16\)), ART-2 (75--85~dB, \(n=29\)), and ART-3 (above 85~dB, \(n=36\)). Figure~\ref{fig:app_distribution} shows the cohort distribution.

\subsection{Stimulus Schedule, Control Conditions, and Held-Out Content}
\label{app:eval_scope}

The complete corpus spans controlled measurement, robustness testing, and ordinary-listening deployment. A 35-mm straight tube provides a non-physiological control. Playback includes broadband white noise over 20~Hz--20~kHz, piano tones, instrument sounds, and music excerpts at 50--90~dB SPL. The total recording duration exceeds 400 hours. The clinically referenced 145,800-window benchmark and unilateral music validation are defined separately in Sec.~\ref{sec:setup}.

The robustness study spans 16 headphone models and ambient-noise conditions at 45, 55, 65, and 70~dB. These conditions represent a quiet dormitory, classroom discussion, office conversation, and roadside noise. The headphone set includes over-ear, in-ear, and semi-in-ear form factors, together with wired and wireless devices. Figure~\ref{fig:app_headphones} shows representative models.

Unless stated otherwise, supervised training uses wideband stimuli. Piano notes and instrument timbres are held out for cross-content evaluation without fine-tuning. The unilateral music experiment keeps the same wideband-trained classifier fixed and uses the synchronized contralateral Titan reference defined in Sec.~\ref{sec:setup}.

\subsection{Hardware, Clocking, and Deployment Profiling}
\label{app:eval_runtime}

Training and offline processing use a computer with an Intel Core i7-9750H processor and 8~GB RAM. The model is trained in MATLAB R2023b and deployed on the Arduino Portenta H7. The inner and outer microphone streams are captured on separate ADC channels with 16-bit resolution at 48~kHz. The microphone streams are aligned to the digital playback by cross-correlation, and playback levels are matched to the clinical setup with a calibrated ear simulator. The fixed-point implementation runs at \(1.05\times\) real time.

\subsection{Confidence Intervals}
\label{app:eval_splits}

Confidence intervals, when reported, are computed by participant-level bootstrapping. Window-level resampling is not used because adjacent windows from the same participant are correlated.

\subsection{Recognition Across WB-ART Groups Under Cross-Content Playback}
\label{app:eval_wbart_content}

Figure~\ref{fig:app_wbart_groups} complements Sec.~\ref{sec:eval:wbart} by combining wideband, piano, and instrument playback in a noisy environment. Each evaluation ear is tested repeatedly at 50--90~dB SPL. Accuracy reaches 0.959 for ART-1 ears around 65--70~dB WB-ART, ranges from 0.883 to 0.925 in the ART-2 transition range around 75~dB, and reaches 0.877 for ART-3 ears around 95--100~dB. The broader content set preserves the cross-threshold operating range established by the clinically referenced benchmark.

\subsection{Refitting and Everyday-Behavior Protocol}
\label{app:eval_refit}

For the refitting benchmark, each participant repeatedly removes and re-wears the same headphone model. After every re-wear, we record a short low-volume support segment and then play content at 10~dB below the participant's WB-ART. Synchronized Titan-inactive intervals define the AR-negative reference condition, so every positive \earlog{} output is counted as a false alarm.

For the everyday-behavior benchmark, participants walk, speak, chew, and smile during the same Titan-referenced below-threshold playback. Figure~\ref{fig:realistic_robustness_a} reports false-alarm rate among valid windows on the left axis and valid-window ratio on the right axis. The joint measurement verifies that \earlog{} controls erroneous event logging while retaining sensing coverage. Invalid and overload-only intervals remain in the exposure and signal-quality records.

\subsection{Cross-Day Drift and Reference Maintenance}
\label{app:eval_refresh}

The cross-day study follows multiple participants for seven days. Each participant uses the same evaluation ear and headphone model throughout the study. Every daily session contains a prespecified below-WB-ART support segment, a seated controlled-playback segment, and an above-threshold block for performance verification.

We compare two policies. The stale-reference policy reuses the initial baseline throughout the study. The full-system policy enables the calibration transfer, consistency check, and low-volume refresh described in Sec.~\ref{sec:baseline_maintenance}. Each thin trajectory in Fig.~\ref{fig:realistic_robustness_b} follows one participant under one policy. The aggregate curves report the participant-level median and interquartile range.

The refresh analysis records trigger count and accumulated low-volume playback time for every participant and day. These measurements distinguish the continuous contribution of calibration transfer from the occasional cost of rebuilding the no-AR baseline.

\balance
\section{Discussion}
\label{app:extended_discussion}

\earlog{} establishes a new paradigm for ear-health monitoring by recording AR recruitment during everyday listening over time. Using \earlog{}, we also found that listeners with normal hearing may still show roughly 20--30 AR activations per hour even when the volume is limited. These event logs may support future studies of how repeated high-load listening contributes to hearing damage and provide listener-specific evidence for long-term risk management.

We did not build \earlog{} around an end-to-end classifier because reliable AR labels are scarce. Like many physiological-sensing problems, repeatedly eliciting AR to collect a large positive dataset would be difficult to justify under safety and ethics constraints. Arbitrary dynamic audio also produces very different signal patterns, making it unrealistic to annotate every type of content window by window. Individual ear anatomy, headphone response, and fit further change how the same physiological event appears acoustically. With limited labels, a model trained directly on raw signals can easily learn playback content or fit instead of physiology. \earlog{} therefore avoids asking the model to learn every source of variation by itself.

Our experiments cover refitting, walking, speaking, chewing, smiling, ambient noise, and multi-day drift. When interference cannot be explained, \earlog{} leaves the affected windows out of the AR event log. More conditions still need to be evaluated. These disturbances are typically random and brief, while AR events occur independently and do not systematically coincide with them. Removing interference therefore takes priority over retaining every window.

The \earlog{} prototype recreates the inner- and outer-microphone arrangement commonly found in active-noise-cancelling headphones, although its implementation remains different from an actual ANC signal chain. It still requires synchronized access to playback and microphone data, together with a stable analog front end. The NTC network compensates for temperature-dependent front-end gain and reduces baseline drift and low-volume refreshes. This component is optional. Software calibration can handle the same drift, but it requires additional computation and energy.

Using ultrasound instead of everyday playback as the probe is also feasible. Ultrasound offers a more stable excitation, but it requires wider-band transducers and acquisition front ends, an ultrasonic microphone, and software support for synchronizing the probe signal. These additions increase power consumption, hardware compatibility requirements, and the risk of intermodulation distortion, making direct deployment on ordinary headphones less practical.

\earlog{} may also extend beyond AR to other ear-health states and conditions. Voluntary middle-ear muscle activity, small pressure transients, and slowly changing ear states may also alter the in-ear response. Distinguishing these processes will require longer temporal context, binaural evidence, or clinical information, opening an important direction for future work.

\end{document}